\documentclass[reprint,amsmath,amssymb,aps,prc,superscriptaddress,nofootinbib]{revtex4-2}
\usepackage[utf8]{inputenc}

\usepackage{bm}
\usepackage[normalem]{ulem}
\usepackage{multirow}

\usepackage{pgfplots}
\usepgfplotslibrary{groupplots,dateplot}
\usetikzlibrary{patterns,shapes.arrows}
\pgfplotsset{compat=newest}
\definecolor{color0}{rgb}{0.12156862745098,0.466666666666667,0.705882352941177}
\definecolor{color1}{rgb}{0.172549019607843,0.627450980392157,0.172549019607843}
\definecolor{color2}{rgb}{1,0.498039215686275,0.0549019607843137}

\pgfplotsset{error bar legend/.style={%
    /pgfplots/legend image code/.prefix code={%
      \pgfkeysgetvalue{/pgfplots/error bars/error mark}{\pgfplotserrorbarsmark}%
      \draw[%
        /pgfplots/every error bar, 
        mark=\pgfplotserrorbarsmark, 
        /pgfplots/error bars/error mark options, 
        sharp plot,
        ##1
      ] plot coordinates {(0.3cm, -0.15cm) (0.3cm, 0.15cm)};%
    }
  }
}

\usepackage{graphicx}% Include figure files
\usepackage{dcolumn}% Align table columns on decimal point
\usepackage{bm}% bold math
\usepackage{hyperref}

\newcommand{\apjl}{Astrophysical Journal, Letters}

\newcommand{\aap}{Astronomy \& Astrophysics}
\newcommand{\mnras}{Monthly Notices of the Royal Astronomical Society}
\newcommand{\pasa}{Publications of the Astronomical Society of Australia}

\begin{document}

\preprint{APS/PRC}

%\title{Determining the strength of the cosmic hand-break - \texorpdfstring{$^{66}\mathrm{Ni}(n,\gamma)^{67}\mathrm{Ni}$}{66Ni(n,g)67Ni}}
%\title{Can we measure nuclear level densities and $\gamma$-ray strength functions of neutron rich nuclei with the Oslo method and inverse kinematics?}
%\title{\texorpdfstring{The Oslo Method With Radioactive Beam - Nuclear Level Density and $\gamma$-ray Strength Function of $^{67}\mathrm{Ni}$}{The Oslo Method With Radioactive Beam - Nuclear Level Density and Gamma-ray Strength Function of Ni-67}}

\title{\texorpdfstring{Nuclear Level Density and $\gamma$-ray Strength Function of $^{67}\mathrm{Ni}$ and the impact on the $i$-process}{Nuclear Level Density and Gamma-ray Strength Function of Ni-67 and the impact on the $i$-process}}

\author{V.~W.~Ingeberg}\email{vetlewi@fys.uio.no}
\author{S.~Siem}\email{sunniva.siem@fys.uio.no}
\affiliation{Department of Physics, University of Oslo, NO-0316 Oslo, Norway}
\author{M.~Wiedeking}%\email{mwiedeking@lbl.gov}
\affiliation{Nuclear Science Division, Lawrence Berkeley National Laboratory, Berkeley, CA 94720, USA}
\affiliation{SSC Laboratory, iThemba LABS, P.O. Box 722, 7129 Somerset West, South Africa}
\affiliation{School of Physics, University of the Witwatersrand, 2050 Johannesburg, South Africa}
\author{A.~Choplin}
\affiliation{Institut d’Astronomie et d’Astrophysique, Universit\'e Libre de Bruxelles (ULB), Campus de la Plaine, CP 226, 1050 Brussels, Belgium}
\author{S.~Goriely}
\affiliation{Institut d’Astronomie et d’Astrophysique, Universit\'e Libre de Bruxelles (ULB), Campus de la Plaine, CP 226, 1050 Brussels, Belgium}
\author{L.~Siess}
\affiliation{Institut d’Astronomie et d’Astrophysique, Universit\'e Libre de Bruxelles (ULB), Campus de la Plaine, CP 226, 1050 Brussels, Belgium}
\author{K.~J.~Abrahams}
\affiliation{Department of Physics, University of the Western Cape, P/B X17 Bellville 7535, South Africa}
\author{K.~Arnswald}
\affiliation{Institut für Kernphysik, University of Cologne, D-50937 Cologne, Germany}
\author{F.~Bello Garrote}
\altaffiliation[Present address: ]{Oslo University Hospital, NO-0424 Oslo, Norway}
\affiliation{Department of Physics, University of Oslo, NO-0316 Oslo, Norway}
%\author{T.~Berry}
%\affiliation{Department of Physics, University of Surrey, Guildford GU2 7XH, United Kingdom}
\author{D.~L.~Bleuel}
\affiliation{Lawrence Livermore National Laboratory, 7000 East Avenue, Livermore, California, 94550-9234, USA}
\author{J.~Cederkäll}
\affiliation{Physics Department, University of Lund, Box-118, SE-22100, Lund, Sweden}
\affiliation{ISOLDE, EP Department, CERN, CH-1211 Geneva, Switzerland}
\author{T.~L.~Christoffersen}
\affiliation{Department of Physics, University of Oslo, NO-0316 Oslo, Norway}
\author{D.~M.~Cox}
\affiliation{Department of Physics, University of Jyväskylä, P.O. Box 35, FIN-40014 Jyväskylä, Finland}
\affiliation{Helsinki Institute of Physics, University of Helsinki, P.O. Box 64, FIN-00014, Helsinki, Finland}
%\author{L.~Crespo~Campo}
%\altaffiliation[Present address: ]{Bayer AS, NO-0283 Oslo, Norway}
%\affiliation{Department of Physics, University of Oslo, NO-0316 Oslo, Norway}
\author{H.~De~Witte}
\affiliation{Instituut voor Kern- en Stralingsfysica, K.U. Leuven, Celestijnenlaan 200D, B-3001 Leuven, Belgium}
\author{L.~P.~Gaffney}
\altaffiliation[Present address: ]{Department of Physics, University of Liverpool, L69 7ZE, United Kingdom}
\affiliation{ISOLDE, EP Department, CERN, CH-1211 Geneva, Switzerland}
\author{A.~Görgen}
\affiliation{Department of Physics, University of Oslo, NO-0316 Oslo, Norway}
\author{C.~Henrich}
\affiliation{Institut für Kernphysik, Technische Universität Darmstadt, D-64289 Darmstadt, Germany}
\author{A.~Illana}
\altaffiliation[Present address: ]{Departamento de Estructura de la Materia, Física Térmica y Electrónica and IPARCOS, Universidad Complutense de Madrid, E-28040 Madrid, Spain}
\affiliation{Instituut voor Kern- en Stralingsfysica, K.U. Leuven, Celestijnenlaan 200D, B-3001 Leuven, Belgium}
\author{P.~Jones}
\affiliation{SSC Laboratory, iThemba LABS, P.O. Box 722, 7129 Somerset West, South Africa}
\author{B.~V.~Kheswa}
\altaffiliation[Present address: ]{Department of Physics, University of Johannesburg, P.O. Box 524, Auckland Park 2006, South Africa}
\affiliation{Department of Physics, University of Oslo, NO-0316 Oslo, Norway}
\author{T.~Kröll}
\affiliation{Institut für Kernphysik, Technische Universität Darmstadt, D-64289 Darmstadt, Germany}
\author{S.~N.~T.~Majola}
\altaffiliation[Present address: ]{Department of Physics, University of Johannesburg, P.O. Box 524, Auckland Park 2006, South Africa}
\affiliation{SSC Laboratory, iThemba LABS, P.O. Box 722, 7129 Somerset West, South Africa}
\author{K.~L.~Malatji}
\affiliation{SSC Laboratory, iThemba LABS, P.O. Box 722, 7129 Somerset West, South Africa}
\affiliation{Department of Physics, Stellenbosch University, Private Bag X1, Matieland, 7602, South Africa}
%\author{T.~Nogwanya}
%\affiliation{iThemba LABS, P.O. Box 722, 7129 Somerset West, South Africa}
%\affiliation{Department of Physics, University of the Western Cape, P/B X17 Bellville 7535, South Africa}
\author{J.~Ojala}
\altaffiliation[Present address: ]{Department of Physics, University of Liverpool, L69 7ZE, United Kingdom}
\author{J.~Pakarinen}
\affiliation{Department of Physics, University of Jyväskylä, P.O. Box 35, FIN-40014 Jyväskylä, Finland}
\affiliation{Helsinki Institute of Physics, University of Helsinki, P.O. Box 64, FIN-00014, Helsinki, Finland}
\author{G.~Rainovski}
\affiliation{Faculty of Physics, St. Kliment Ohridski University of Sofia, BG-1164 Sofia, Bulgaria}
\author{P.~Reiter}
%\author{D.~Rosiak}
\affiliation{Institut für Kernphysik, University of Cologne, D-50937 Cologne, Germany}
\author{M.~von~Schmid}
\affiliation{Institut für Kernphysik, Technische Universität Darmstadt, D-64289 Darmstadt, Germany}
\author{M.~Seidlitz}
%\author{B.~Siebeck}
\affiliation{Institut für Kernphysik, University of Cologne, D-50937 Cologne, Germany}
%\author{J.~Snäll}
%\affiliation{Physics Department, University of Lund, Box-118, SE-22100, Lund, Sweden}
%\author{K.~Sowazi}
%\affiliation{iThemba LABS, P.O. Box 722, 7129 Somerset West, South Africa}
%\affiliation{Department of Physics, University of the Western Cape, P/B X17 Bellville 7535, South Africa}
\author{G.~M.~Tveten}
%\altaffiliation[Present address: ]{Expert Analytics AS, NO-0160 Oslo, Norway}
\affiliation{Department of Physics, University of Oslo, NO-0316 Oslo, Norway}
\author{N.~Warr}
\affiliation{Institut für Kernphysik, University of Cologne, D-50937 Cologne, Germany}
\author{F.~Zeiser}
%\altaffiliation[Present address: ]{Statskraft, P.O. Box 200 Lilleaker, NO-0216 Oslo, Norway}
\affiliation{Department of Physics, University of Oslo, NO-0316 Oslo, Norway}

\collaboration{The ISOLDE Collaboration}

%\date{February 21, 2022}
\date{\today}

\begin{abstract}
Proton-$\gamma$ coincidences from $(\mathrm{d},\mathrm{p})$ reactions between a $^{66}\mathrm{Ni}$ beam and a deuterated polyethylene target have been analyzed with the inverse-Oslo method to find the nuclear level density (NLD) and $\gamma$-ray strength function ($\gamma$SF) of $^{67}\mathrm{Ni}$. The $^{66}\mathrm{Ni}(n,\gamma)$ capture cross section has been calculated using the Hauser-Feshbach model in \verb+TALYS+ using the measured NLD and $\gamma$SF as constraints. The results confirm that the $^{66}\mathrm{Ni}(n,\gamma)$ reaction acts as a bottleneck when relying on one-zone nucleosynthesis calculations. However, the impact of this reaction is strongly dampened in multi-zone models of low-metallicity AGB stars experiencing i-process nucleosynthesis.
\end{abstract}

\maketitle

\section{Introduction}
The origin and production mechanism of the heavy elements has been under investigation for a long time. The first direct observations of nucleosynthesis of heavy elements was reported by P. Merrill in 1952 \cite{Merrill1952}. Soon after, the mechanism behind the production of heavy elements through neutron capture was outlined in the famous B$^2$FH paper in 1957 \cite{RevModPhys.29.547}. 

Nucleosynthesis beyond the Fe/Ni mass region is due to the slow neutron capture (s-process), rapid neutron capture (r-process), photo-disintegration processes, and more recent results call for an intermediate neutron capture process (i-process) \cite{Cowan1977,Roederer2016}. These processes are sensitive to nuclear properties such as the nuclear level densities (NLD) and $\gamma$-ray strength functions ($\gamma$SF). The availability and uncertainties in nuclear data affect the ability to reliably calculate reaction rates, in particular for the r- and i-processes as they involve neutron-rich nuclei for which data is non-existing or at the very least sparse. This can have significant impact on the neutron capture rates and on the final abundance distribution of elements and their isotopes predicted by models. Currently, the vast majority of nuclei for which NLDs and $\gamma$SFs have been measured lie at, or very near, the line of stability \cite{GorielyCRP2019}. Experimentally measured NLD and $\gamma$SF can provide significant constraints on calculated neutron capture rates which are relevant for nucleosynthesis models \cite{Larsen2019, Wiedeking24}.

Away from stability, directly measured $(n,\gamma)$ cross sections are not available and NLDs and $\gamma$SFs can provide the necessary constraints.  
The Oslo method is particularly suitable as it extracts NLD and $\gamma$SF simultaneously \cite{OsloMethodNIM, Midtb2019}. These quantities, together with the optical model potentials, are the main ingredients of the \emph{Hauser-Feshbach} theory \cite{PhysRev.87.366} to calculate neutron capture cross-sections and reaction rates. The Oslo Method has been extended to include total absorption spectroscopy following $\beta$-decay leading to the $\beta$-Oslo method \cite{PhysRevLett.113.232502} and has demonstrated the versatility of the Oslo method and its applicability even for nuclei far from stability \cite{PhysRevLett.113.232502,PhysRevC.97.054329,lewis19,PhysRevLett.116.242502,Liddick2019BenchmarkingMethod}. The $\beta$-Oslo method requires specific $\beta$-Q value conditions for the mother-daughter pair placing a limit on the nuclei which can be measured with this technique.

Another approach is to use inverse kinematics with radioactive beams. Here, the beam intensity is the main limiting factor.
Applying the Oslo method to experimental data from an inverse kinematics experiment was first demonstrated in Ref. \cite{Ingeberg2020} extracting the NLD and $\gamma$SF of $^{87}\mathrm{Kr}$ following a $\mathrm{d}(^{86}\mathrm{Kr},\mathrm{p})^{87}\mathrm{Kr}$ pick-up reaction.
In this paper, we have applied this inverse-Oslo method to data from a radioactive ion beam experiment to extract the NLD and $\gamma$SF of $^{67}\mathrm{Ni}$. Based on these results, the neutron capture rate of $^{66}\mathrm{Ni}$ was constrained using the Hauser-Feshbach theory, which is of particular importance for our understanding of the weak i-process. In the weak i-process the neutron density is large ($\sim 10^{15}$), but exposure is low due to a quick and abrupt termination of the neutron production. In such a case only the first n-capture peak elements are produced significantly \cite{McKay2020}. This was studied in Ref. \cite{McKay2020}, where it was found that the $^{66}\mathrm{Ni}(n,\gamma)$ reaction have a major impact on the overall production rate of heavier elements, essentially acting as a bottleneck for the entire weak i-process. Constraining the $^{66}\mathrm{Ni}(n,\gamma)$ reaction could improve our understanding of the i-process nucleosynthesis.

\begin{figure*}
\centering
\includegraphics[width=0.9\textwidth]{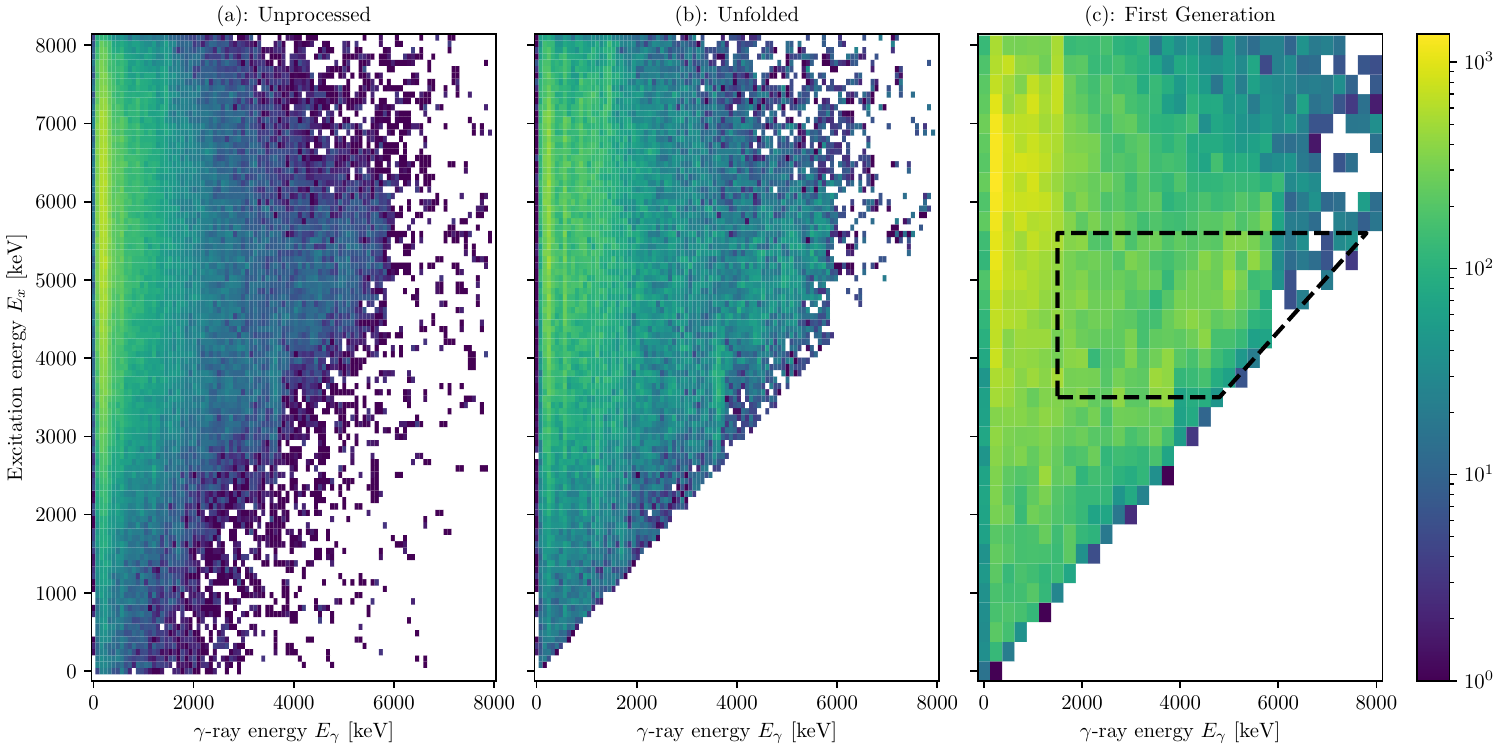}
\caption{\label{fig:all_matrices}(a) Unprocessed excitation-$\gamma$ coincidence and (b) unfolded matrices with $80$ keV/bin on both axes. The first-generation matrix (c) was re-binned from $80$ to $250$ keV/bin to reduce fluctuations. The dashed lines in (c) indicate the region for which the NLD and $\gamma$SF are extracted.}
\end{figure*}

\section{Experiment and analysis}
The experiment was performed using the HIE-ISOLDE facility at CERN \cite{Kadi18}, where 1.4 GeV proton beam from the PSBooster bombarded a uranium carbide target inducing fission.
The evaporated fission products were ionized by the resonance ionization laser ion source (RILIS) \cite{Fedoseyev2000} and the $^{66}\mathrm{Ni}$ ions were separated by the General Purpose Separator (GPS) to be collected and cooled by a penning trap (REXTRAP) \cite{AMES200517}. The ions were then charge bread to a $16^+$ charge state by an electron-beam ion source (REXEBIS) \cite{WENANDER2002528,Wenander_2010}. The highly ionized $^{66}\mathrm{Ni}$ ions were re-accelerated by the HIE-ISOLDE linear accelerator to an energy of 4.47(1) MeV/u. The $\approx 3.5\times 10^6$ pps intense $^{66}\mathrm{Ni}$ beam impinged on a $\approx 670$ $\mu$g/cm$^2$ thick secondary deuterated polyethylene target (99\% enrichment in deuterium) \cite{kheswa_production_2020} placed at the Miniball experimental station and produced the desired $\mathrm{d}(^{66}\mathrm{Ni},\mathrm{p})^{67}\mathrm{Ni}$ reactions. The total time of beam on target was approximately $140$ hours. The same reaction has previously been measured at ISOLDE by Ref. \cite{diriken_study_2014}, but at a much lower beam energy of $2.5$ MeV/u. That data-set was not considered as it only reached excitation energies up to 3.5 MeV, much less than what is required for the Oslo Method.

Protons from the reaction were measured with the C-REX particle detector array \cite{hellgartner_axial_2023,Hellgartner15}, while coincident $\gamma$-rays were measured using the Miniball detector array \cite{EBERTH2001389,Warr2013}. Only the downstream particle array was considered due to low yields in the upstream array. The particle detectors considered covered angles between $25^\circ$ and $49^\circ$ in the laboratory frame (CM angles $40^\circ$ to $76^\circ$). Two of the eight Miniball high-purity germanium (HPGe) clusters \cite{Warr2013} were replaced by six large-volume ($3.5\times 8$ inch) LaBr$_3$:Ce detectors to increase the high $\gamma$-ray energy efficiency of the array. The placement of each detector is given in table \ref{tab:DetPlace}. Signals from the C-REX detectors were processed by an analog data acquisition system (DAQ) which fed four 32-channel analogue-to-digital converter (ADC) modules from Mesytec. The signals from the $\gamma$-ray detectors were processed by a digital DAQ consisting of XIA Digital Gamma Finders (DGF-4C). The two systems were synchronized, read out simultaneously, and the data written to disk. 

\begin{table}
    \centering
    \caption{Positioning of the $\gamma$-ray detectors in the experimental setup.\label{tab:DetPlace}}
    \begin{tabular}{c|c|c|c}
      Detector & $\phi$  & $\theta$  &  Target distance \\ 
             & [deg] & [deg] & [cm] \\ 
      \hline
      LaBr$_{3}$:Ce 1 & 136  & 147 & 25\\
      LaBr$_{3}$:Ce 2 & 184  & 147 & 25\\
      LaBr$_{3}$:Ce 3 & 210  & 147 & 25\\
      LaBr$_{3}$:Ce 4 & 32  & 64 & 25\\
      LaBr$_{3}$:Ce 5 & 358  & 64 & 25\\
      LaBr$_{3}$:Ce 6 & 330  & 64 & 25\\
      Miniball 1 & 148 & 64 & 12\\
      Miniball 2 & 228 & 64 & 12\\
      Miniball 3 & 188 & 101 & 12\\
      Miniball 4 & 358 & 101 & 12\\
      Miniball 5 & 312 & 56 & 12\\
      Miniball 6 & 32 & 56 & 12\\
    \end{tabular}
\end{table}

Prompt particle-$\gamma$ coincidences were selected from the data by applying time gates on the time difference between the detected $\gamma$-rays and particles, while the $(\mathrm{d},\mathrm{p})$ reaction was selected by applying the appropriate energy cut on the $\Delta \text{E} - \text{E}$ matrix. Background events were selected in a similar fashion by placing time gates of half the size as the prompt gate on either sides of the prompt peak. This resulted in a total of $3.2 \times 10^{5}$ and $1.1 \times 10^6$ background subtracted proton-$\gamma$ coincidences with the LaBr$_3$:Ce detectors and the Miniball clusters, respectively. The low number of proton-$\gamma$ coincidences with the LaBr$_3$:Ce compared to those with the Miniball detectors is due to the lower geometric efficiency of the LaBr$_3$:Ce detectors which were located considerably further from the target. Despite having only a quarter of the Miniball coincidences, the more efficient LaBr$_3$:Ce detectors measured the majority of high-energy $\gamma$-ray transitions. For this reason, the results presented here are based on the analysis of proton-$\gamma$ coincident events from the LaBr$_3$:Ce detectors only.

Due to the considerable residual kinetic energy of $^{67}\mathrm{Ni}$ the $\gamma$-ray energies were Doppler corrected. The maximum deflection of the residual $^{67}\mathrm{Ni}$ was less than $1.5^\circ$ and the maximum spread in velocity was less than $4.2$\%. The correction was performed using a constant factor based only on the $\gamma$-detector angle relative to the beam axis.

For each proton-$\gamma$ event the excitation energy was calculated from the  kinematic reconstruction of the two-body reaction. The proton-$\gamma$ coincidences were collected in an excitation-$\gamma$ energy matrix shown in Fig. \ref{fig:all_matrices}(a). Due to the Lorentz boost associated with inverse-kinematics there were no well-resolved levels in the particle spectra suitable for an in-beam calibration and the FWHM of the excitation energy was $\approx 1.1$ MeV.

The first step to obtain NLDs and $\gamma$SFs from excitation-$\gamma$ matrices is to correct for the detector response using the unfolding method \cite{UnfoldingNIM} by using a detector response deduced from a \verb+Geant4+ \cite{AGOSTINELLI2003250} simulation of the experimental setup \cite{Ingeberg2022d}. This resulted in the \emph{unfolded} matrix shown in Fig. \ref{fig:all_matrices}(b).

Next, an iterative subtraction method \cite{FirstGenerationNIM} is applied to the unfolded matrix in order to deduce the distribution of primary $\gamma$-rays emitted from each excitation bin, resulting in the \emph{first generation} matrix shown in Fig. \ref{fig:all_matrices}(c).

The first-generation matrix is proportional to the $\gamma$-ray transmission coefficients $\mathcal{T}(E_\gamma)$ and the NLD $\rho(E_f = E_x - E_\gamma)$ at the final excitation energy $E_f$ following the first $\gamma$ decay of the cascade \cite{OsloMethodNIM}
\begin{equation}
    \Gamma(E_x, E_\gamma) \propto \mathcal{T}(E_\gamma)\rho(E_f = E_x - E_\gamma). \label{eq:FirstGenProp}
\end{equation}
The NLD and transmission coefficients are extracted by normalizing the first-generation $\gamma$-ray spectra for each excitation bin and by fitting the experimental matrix ($\Gamma(E_\gamma, E_x)$) with a theoretical matrix \cite{OsloMethodNIM}
\begin{equation}
    \Gamma_\text{th}(E_x, E_\gamma) = \frac{\mathcal{T}(E_\gamma) \rho(E_x - E_\gamma)}{\sum\limits_{E_\gamma = E_\gamma^\text{min}}^{E_x} \mathcal{T}(E_\gamma) \rho(E_x - E_\gamma)} \label{eq:fgTheo}
\end{equation}
by minimizing
\begin{equation}
    \chi^2 = \sum\limits_{E_x, E_\gamma} \left( \frac{\Gamma(E_\gamma, E_x) - \Gamma_\text{th}(E_\gamma, E_x)}{\Delta \Gamma(E_\gamma, E_x)} \right)^2, \label{eq:ChiSqMin}
\end{equation}
where $\mathcal{T}(E_\gamma)$ and $\rho(E_f = E_x - E_\gamma)$ are treated as free parameters for each $\gamma$-ray energy $E_\gamma$ and final energy $E_f$, respectively. The denominator, $\Delta \Gamma (E_\gamma, E_x)$, represents the experimental uncertainty in the first-generation matrix. Considering the experimental resolution and to ensure that only primary transitions, originating in the quasi-continuum, are considered a minimum $\gamma$-ray energy of $1.5$ MeV was selected. Similarly, only excitation energy bins between $3.5$ and $5.6$ MeV were included in the fit. These $E_{\gamma}$ and $E_x$ energy regions have been chosen based on the guidelines of Ref. \cite{Larsen2011}. 

The resulting $\gamma$-ray transmission coefficients are converted to $\gamma$SF by
\begin{equation*}
    f(E_\gamma) = \frac{\mathcal{T}(E_\gamma)}{2\pi E_\gamma^3}
\end{equation*}
under the assumption that the transmission coefficients are dominated by dipole transitions.

The theoretical first-generation matrix, \eqref{eq:fgTheo}, is invariant under transformation of NLD and $\gamma$SF \cite{OsloMethodNIM}
\begin{equation}
\begin{split}
    \widetilde{\rho}(E_f) = A \rho(E_f) e^{\alpha E_f} \\
    \widetilde{f}(E_\gamma) = B f(E_\gamma) e^{\alpha E_\gamma},
\end{split} \label{eq:TransfNLD_GSF}
\end{equation}
where $A$, $B$ and $\alpha$ are transformation parameters. This means that the extracted NLD and $\gamma$SF do not represent the physical NLD and $\gamma$SF, but contain information on the functional shape. To deduce the absolute values of the NLD and $\gamma$SF the extracted NLD and $\gamma$SF will have to be normalized to auxiliary data. 

\section{\texorpdfstring{Normalization of NLD and $\gamma$SF}{Normalization of NLD and gamma-SF}}\label{sec:Normalization}
Many nuclei along the island of stability have auxiliary data available with which the normalization of the NLDs and $\gamma$SFs is achieved. Typical data used for this purpose are the level density from resolved discrete states, the level density at the neutron separation energy ($S_n = 5.8$ MeV) from s- and/or p-wave neutron resonance spacing data, and the average radiative widths of s-wave neutron resonances. 
When applying the Oslo method, NLDs do not extend to $S_n$ as the method probes the NLD at the final excitation energy $E_f = E_x^\text{max} - E_\gamma$, the maximum energy possible to be considered in the first-generation matrix is $E_x^\text{max}$, and is typically the neutron separation energy (depending on reaction kinematics) \cite{Larsen2011}, but in this case slightly below at $5.6$ MeV, which corresponds to a final and highest excitation energy of 4.1 MeV. Hence, the measured NLDs are typically interpolated to the NLDs from resonance data at $S_n$.

The impact on the uncertainty of NLDs and $\gamma$SF on the availability of neutron resonance data has recently been demonstrated \cite{Ingeberg2022b}. In the absence of neutron resonance data, the resonance spacing or radiative widths of neutron resonances are generally estimated from systematics of neighboring nuclei, see for instance Refs.\cite{Larsen2016, Brits2019, Kheswa2015GalacticProperties}. Similarly, for $\beta$-Oslo measurements various combinations of solutions have been used ranging from systematic information of nearby nuclei \cite{PhysRevLett.113.232502, PhysRevC.109.054305}, model calculations \cite{PhysRevLett.116.242502, PhysRevLett.132.202701, PhysRevC.109.054305}, or the Shape method \cite{PhysRevLett.132.202701, PhysRevC.109.054305, PhysRevC.107.L011602, wiedeking2020independent}. Neither for $^{67}\mathrm{Ni}$ nor for surrounding nuclei are neutron resonance data available. The situation is further aggravated by an incomplete level scheme at low excitation energies resulting in an unreliable level density of resolved discrete states which, together with the low experimental resolution, makes the application of the Shape method not ideal. With the absence of any reliable normalization points the only practical solution for normalizing the NLD and $\gamma$SF is the reliance on model estimates.

\subsection{Bayesian analysis}
Despite the lack of auxiliary data for normalization it is still possible to constrain the experimental NLD and $\gamma$SF as the true values are expected to fall within the range of commonly accepted models. For example, the NLD in statistical nuclei grows exponentially and are often well reproduced by large shell model calculations \cite{Brown2014,PhysRevC.97.054329} at low energies and the constant temperature \cite{guttormsen_experimental_2015}, Back-shifted Fermi Gas \cite{PhysRevC.96.014312} or microscopic Hartree-Fock \cite{PhysRevC.93.064302} calculations at high energies. Similarly, the $\gamma$SF is expected to follow the tail of the GDR and typically features resonances such as the upbend and/or a PDR. This is typically well described by the phenomenological Simple Modified Lorentzian model (SMLO) or the microscopic mean field plus quasi-particle random phase approximation both of which being the recommended global models to describe the PSF \cite{GorielyCRP2019}. By assuming a combination of NLD and $\gamma$SF models one can constrain the possible values for the normalization parameters, $A$, $B$ and $\alpha$ in Eq. \eqref{eq:TransfNLD_GSF}.

This can be taken a step further by considering several models for NLD and $\gamma$SF to get a range of acceptable normalization parameters. To perform such an analysis we apply Bayesian statistics. Analyzing Oslo method data within a Bayesian statistics framework has previously been demonstrated by Midtbø \textit{et al.} \cite{Midtb2019} and our analysis is based on the framework presented therein. The starting point of the normalization analysis is Bayes' theorem
\begin{equation}
    P(\bm{\theta}|\{ \rho_i \}, \{ f_i \}) = \frac{\mathcal{L}(\{ \rho_i \}, \{ f_i \} | \bm{\theta}) P(\bm{\theta}) }{P(\{ \rho_i \}, \{ f_i \})}, \label{eq:BayesTheorem}
\end{equation}
where $P(\bm{\theta}|\{ \rho_i \}, \{ f_i \})$ is the posterior probability of a set of normalization and model parameters $\bm{\theta} = (A, B, \alpha, \bm{\theta}_\text{NLD}, \bm{\theta}_{\gamma\text{SF}})$, given a set of non-normalized measured NLD points $\{ \rho_i \}$ and $\gamma$SF points $\{ f_i \}$. The likelihood $\mathcal{L}(\{ \rho_i \}, \{ f_i \} | \bm{\theta})$ is the probability of measuring non-normalized NLD and $\gamma$SF given the set of parameters $\bm{\theta}$. $P(\bm{\theta})$ is the prior probability of a set of normalization and model parameters. The prior will be further discussed in section \ref{subsec:priors}, while the physical models and their parameters are presented in section \ref{subsec:Models} and summarized in table \ref{tab:ModelParam}. The evidence $P(\{ \rho_i \}, \{ f_i \})$ is the probability of measuring the non-normalized NLD and $\gamma$SF, and can be treated as a normalization factor. With the assumption of Gaussian distributed uncertainties of the measured non-normalized NLD and $\gamma$SF leads to the likelihood
\begin{equation}
    \mathcal{L}(\bm{\theta}) = \prod_i \mathcal{L}_i(\bm{\theta}),\label{eq:tot_likelihood}
\end{equation}
with $\mathcal{L}_i(\bm{\theta})$ corresponding to
\begin{align}
\begin{split}
    \ln \mathcal{L}_\text{discrete} &= \sum_j \ln\frac{1}{\sqrt{2\pi \sigma_{\rho_j,\text{Oslo}}(\bm{\theta})}} \\
    &- \frac{1}{2}\sum_j \left(\frac{\rho_{j,\text{discrete}} - \rho_{j,\text{Oslo}}(\bm{\theta})}{\sigma_{\rho_j,\text{Oslo}}(\bm{\theta})} \right)^2, \label{eq:discrete_like}
\end{split} \\
\begin{split}
    \ln \mathcal{L}_\text{NLD} &= \sum_j \ln\frac{1}{\sqrt{2\pi \sigma_{\rho_j,\text{Oslo}}(\bm{\theta})}} \\
    &- \frac{1}{2}\sum_j \left(\frac{\rho_{j,\text{model}}(\bm{\theta}) - \rho_{j,\text{Oslo}}(\bm{\theta})}{\sigma_{\rho_j,\text{Oslo}}(\bm{\theta})} \right)^2, \label{eq:CT_like}
\end{split} \\
%\ln \mathcal{L}_{\rho_{S_n}} &= \left( \frac{\rho_{S_n} - \rho_{S_n,\text{model}}(\bm{\theta})}{\sigma_{\rho_{S_n}}} \right)^2,  \label{eq:rhoSn_like} \\
\begin{split}
    \ln \mathcal{L}_{\gamma\text{SF}} &= \sum_j \ln\frac{1}{\sqrt{2\pi \sigma_{f_j,\text{Oslo}}(\bm{\theta})}} \\
    &- \frac{1}{2}\sum_j \left( \frac{f_{j, \text{model}}(\bm{\theta}) - f_{j,\text{Oslo}}(\bm{\theta})}{\sigma_{f_j,\text{Oslo}}(\bm{\theta})} \right)^2, \label{eq:gsf_like}
\end{split} \\
    \ln \mathcal{L}_{^{68}\mathrm{Ni}} &= \sum_j \ln\frac{1}{\sqrt{2\pi \sigma_{j, ^{68}\mathrm{Ni}}}} \\
    &-\frac{1}{2} \sum_j \left( \frac{f_{j, \text{model}}(\bm{\theta}) - f_{j,^{68}\mathrm{Ni}}(\bm{\theta})}{\sigma_{j,^{68}\mathrm{Ni}}} \right)^2. \label{eq:gsf_68Ni}
\end{align}
Here, $\mathcal{L}_\text{discrete}$ represents the likelihood of measuring a set of NLDs given the known level density from discrete levels $\rho_{j,\text{discrete}}$, $\mathcal{L}_\text{NLD}$ is the likelihood given a specific NLD model and $\mathcal{L}_{\gamma\text{SF}}$ is the likelihood given a $\gamma$SF model. The NLD between excitation energies between $1.0$ and $2.4$ MeV are considered in Eq. \eqref{eq:discrete_like} while NLD between $2.6$ and $3.8$ MeV are considered in Eq. \eqref{eq:CT_like}. All measured $\gamma$SF energies are included in Eq. \eqref{eq:gsf_like}. The $\mathcal{L}_{^{68}\mathrm{Ni}}$ likelihood is included to further constrain the $\gamma$SF using the measured $\gamma$SF in $^{68}\mathrm{Ni}$ as the $\gamma$SF above the neutron separation energy are typically very similar in neighboring nuclei. %Here the contribution from the term ($\ln\left(1/\sqrt{2\pi\sigma_{j,^{68}\mathrm{Ni}}}\right)$) is omitted as it is constant and will not affect the overall result. 
%Further discussion on models and parameter priors will follow in Sect. \ref{subsec:Models} and Sect. \ref{subsec:priors}, respectively.
$\rho_{j,\text{Oslo}}(\bm{\theta})$, ($f_{j, \text{Oslo}}(\bm{\theta})$) and $\sigma_{\rho_j,\text{Oslo}}((\bm{\theta})$ ($\sigma_{f_j,\text{Oslo}}(\bm{\theta})$) are the normalized NLD ($\gamma$SF) and standard deviation, given a set of normalization parameters $A, B, \alpha \in \bm{\theta}$, respectively (see Eq. \ref{eq:TransfNLD_GSF}).% Initial guesses for the values of the $A$, $B$ and $\alpha$ parameters will be obtained from the maximum likelihood estimation (MLE), $\bm{\hat{\theta}}$, while the prior probability distribution for the model parameters are given by systematics if available, or other physical justifications, this is further discussed in Sect. \ref{subsec:priors}. The MLE is obtained by maximizing Eq. \eqref{eq:tot_likelihood}.

Instead of NLDs from resolved discrete states we have considered level densities based on the counting of levels from large-scale shell model (SM) calculations utilizing the \verb+ca48mh1g+ interaction, see Refs.\cite{Midtbo2018,PhysRevC.97.054329} for details.

To perform the actual Bayesian analysis, model and normalization parameters were sampled using the Bayesian nested sampling algorithm \verb+MultiNest+ \cite{Feroz2008,Feroz2009,Feroz2019} in the \verb+PyMultiNest+ package \cite{Buchner2014} to obtain a posterior probability distributions of the normalization and model parameters $\bm{\theta}$, given as samples $\bm{\theta}_i$.
%In addition to sampling of the posterior probability distribution with the nested sampling algorithm, we will also utilize the maximum likelihood estimation, $\bm{\hat{\theta}}$ to obtain an initial guess for the normalization parameters $A$, $B$ and $\alpha$.

\subsection{Models\label{subsec:Models}}
A total of three NLD models and four $\gamma$SF models have been considered, making the total number of inferred posteriors twelve. In this section we will describe the models considered.

\subsubsection*{NLD}
For energies above the point where the model space of the shell model calculation is exhausted the NLD has been modeled with three different models. These include the Back-Shifted Fermi Gas (BSFG) model \cite{PhysRev.50.332,doi:10.1139/p65-139}
\begin{equation}
    \rho_\text{BSFG}(E_x) = \frac{\sqrt{\pi}}{12\sigma} \frac{\exp(2\sqrt{a(E_x - \delta)})}{a^{1/4}(E_x - \delta)^{5/4}}, \label{eq:BSFG}
\end{equation}
where $a$ is the level density parameter, $\delta$ is the energy shift and $\sigma$ is the spin-cutoff parameter. The Constant Temperature (CT) model \cite{ERICSON1959481} 
\begin{equation}
    \rho_\text{CT}(E_x) = \frac{1}{T}e^{\frac{E_x - \delta}{T}}, \label{eq:CT}
\end{equation}
where $T$ is the nuclear temperature and $\delta$ is an energy shift parameter. Lastly, tabulated Hartree-Fock-Bogoliubov (HFB) calculations \cite{PhysRevC.78.064307} were also considered, and allowed to be re-normalized through
\begin{equation}
    \rho_\text{HFB}(E_x) = \hat{\rho}_\text{HFB}(E_x - \delta) e^{c\sqrt{E_x - \delta}}, \label{eq:HFB}
\end{equation}
where again $\delta$ is used to denote an energy shift, $c$ is a slope parameter and $\hat{\rho}_\text{HFB}$ are the tabulated NLDs.
It should be noted that $\delta$ in eqs. \eqref{eq:BSFG}, \eqref{eq:CT}, and \eqref{eq:HFB} are not the same parameters and will be sub-scripted whenever there is ambiguity which parameter is being referenced.
The spin distribution for the BSFG and CT models was given by the Bethe/Ericson distribution \cite{PhysRev.50.332,ERICSON1959481}
\begin{equation}
    g(E_x, J) = \exp\left( -\frac{J^2}{2\sigma^2(E_x)}\right) - \exp\left( -\frac{(J + 1)^2}{2\sigma^2(E_x)} \right), \label{eq:EricsonDist}
\end{equation}
with the spin-cutoff parameterized by \cite{Capote2009,PhysRevC.96.024313}
\begin{equation}
    \sigma^2(E_x) = \begin{cases}
    \sigma_d^2 & \quad E < E_d \\
    \sigma_d^2 + \frac{E - E_d}{S_n - E_d}(\sigma^2(S_n) - \sigma^2_d) & \quad E \geq E_d,
    \end{cases}
    \label{eq:SpinCutParam}
\end{equation}
where $\sigma_d$ is the discrete spin-cutoff parameter, obtained from shell-model result as discussed in section \ref{subsec:priors}, with a constant spin distribution at energies below $E_d = 2.0$ MeV.

\subsubsection*{$\gamma$SF}
For the $\gamma$SF we have considered the Simplified version of the Modified Lorentzian model (SMLO) \cite{PhysRevC.99.014303} and the microscopic Gogny-HFB plus quasi-particle random phase approximation (Gogny-HFB+QRPA) with phenomenological corrections \cite{Goriely2018}, which are the recommended global models for the $\gamma$SF \cite{GorielyCRP2019}. The SMLO model describes the E1 $\gamma$SF with the giant dipole resonance (GDR) for photo-emission strength and is given by
\begin{equation}
\begin{split}
    f_\text{GDR}(E_\gamma) = \frac{1}{3\pi^2 \hbar^2 c^2} \frac{\sigma_\text{TRK}}{1 - \exp\left(-E_\gamma/T\right)} \\
    \times \frac{2}{\pi} \frac{E_\gamma \Gamma(E_\gamma, T)}{(E_\gamma^2 - E_\text{GDR}^2)^2 + E_\gamma^2 \Gamma^2(E_\gamma, T)},
\end{split} \label{eq:SMLO_func}
\end{equation}
where $E_\text{GDR}$ is the centroid of the GDR, $\sigma_\text{TRK} = 60\frac{NZ}{A}$ is the Thomas-Reiche-Kuhn sum rule \cite{PhysRevC.99.014303} and $T$ is the temperature at the final excitation energy. The width,
\begin{equation}
    \Gamma(E_\gamma, T) = \frac{\Gamma_\text{GDR}}{E_\text{GDR}}\left(E_\gamma + \frac{(2\pi T)^2}{E_\text{GDR}}\right),
\end{equation}
depends on the final temperature and the $\Gamma_\text{GDR}$ which is the width found in the photo-absorption strength function.
Within the SMLO model the M1 strength is parameterized as two standard Lorentzians (SLO)
\begin{equation}
    f_\text{SLO}(E_\gamma) = \frac{1}{3\pi^2 \hbar^2 c^2} \frac{E_\gamma \Gamma^2}{(E_\gamma^2-E_r^2)^2 + (E_\gamma \Gamma)^2} \label{eq:SLO}
\end{equation}
for the scissors and the spin-flip resonances. The low-energy upbend is modeled as a simple exponential \cite{PhysRevC.78.064307}.
From the shape of the non-normalized $\gamma$SF there are no obvious signs of a scissors resonance which is consistent with the relatively low estimated axial deformation of $\beta_2 = ~0.05$ \cite{Hilaire07} and has been omitted from further consideration in the model for the $\gamma$SF. The assumed M1 strength contributions are thus
\begin{equation}
    f_\text{M1}(E_\gamma) =f_\text{SLO}(E_\gamma, E_\text{sf}, \Gamma_\text{sf}, \sigma_\text{sf}) + C\exp\left(-\eta E_\gamma\right), \label{eq:SMLO_M1_strength}
\end{equation}
where $E_\text{sf}$ is the mean energy, $\Gamma_\text{sf}$ is the width and $\sigma_\text{sf}$ the cross-section of the spin-flip resonance. The parameters of the upbend, $\eta$ and $C$ have no physical interpretation and are treated as free parameters chosen to best describe the shape of the upbend. The assumption for the M1 multipolarity of the upbend is based on results from shell model calculations \cite{PhysRevLett.111.232504,Frauendorf22,Brown2014,PhysRevLett.119.052502} and from a measurement albeit not conclusively \cite{Jones2018}.

In the Gogny-HFB+QRPA model, the photo-emission E1 strength is given by
\begin{equation}
\begin{split}
    f_\text{E1}(E_\gamma) = c_\text{E1}f_\text{E1}^\text{QRPA}(E_\gamma - \delta_\text{E1}) \\
    + \frac{f_0 U}{1 + \exp\left(E_\gamma - \epsilon_0\right)},
\end{split}
\label{eq:QRPA}
\end{equation}
where $f_\text{E1}^\text{QRPA}(E_\gamma)$ is the strength found in Gogny-HFB+QRPA calculations. The phenomenological parameters $f_0$, $\epsilon_0$ are free parameters and $U$ is the excitation energy of the initial level. The strength is allowed to be scaled and shifted through the free parameters $c_\text{E1}$ and $\delta_\text{E1}$, respectively. The M1 strength was estimated within the SMLO model, Eq. \eqref{eq:SMLO_M1_strength}.

Data points from a Coulomb dissociation measurement of $^{68}\mathrm{Ni}$ are also considered \cite{PhysRevLett.111.242503}.
%Data points for the $\gamma$SF of $^{68}\mathrm{Ni}$ are also considered \cite{PhysRevLett.111.242503}.
This features a strong narrow Pygmy dipole resonance (PDR)  at $E_x \approx 9.5$ MeV which has been accounted for by including a SLO in the E1 strength. Since the extracted $\gamma$SF in $^{67}\mathrm{Ni}$ only extends to $5.5$ MeV we cannot determine if such a PDR also exists in $^{67}\mathrm{Ni}$. In order to account for either possibility, we have repeated the analysis both with and without including the PDR, observed in $^{68}\mathrm{Ni}$, in the $^{67}\mathrm{Ni}$ strength. In the latter case, the two experimental points at $E_\gamma = 8.4$ MeV and $E_\gamma = 9.5$ MeV of $^{68}\mathrm{Ni}$ were excluded and removed from consideration in Eq. \eqref{eq:gsf_68Ni}. A list of all model combinations and their parameters are listed in table \ref{tab:ModelParam}.

\subsection{Parameter priors\label{subsec:priors}}
For all model parameters, the probability distribution were assumed to be Gaussian or truncated Gaussian probability density functions (PDF). The latter used for parameters where there are physical justifications for why the values cannot have certain values (e.g. no negative temperature). 

The normalization parameters required to transform the NLD and $\gamma$SF to the physical solution are arbitrary and we have no a priori knowledge other than that $A$ and $B$ has to be positive. Ideally, flat prior probability distributions should be used for $A$, $B$ and $\alpha$, with the former two truncated to positive values. However, this is computationally expensive. Instead truncated Gaussian distributions were used for $A$ and $B$ and a Gaussian probability distribution for $\alpha$. To maintain an un-informed prior the standard deviation of the probability distribution for $A$ and $B$ were set to $10$ times the centroid, while the standard deviation of the prior probability distribution of $\alpha$ was set to $1$ 1/MeV. The latter was chosen due to the high sensitivity of the likelihood to this parameter. The centroid of the prior probability distributions were selected to be the $A$, $B$ and $\alpha$ values that maximized the likelihood Eq. \eqref{eq:tot_likelihood}. Numerical values for the centroids of $A$, $B$ and $\alpha$ PDFs are listed in Appendix \ref{app:appendix}.

\begin{table}
\centering
    \caption{List of all NLD and $\gamma$SF models considered together with their parameters. The $E1$ strength for the $\gamma$SF is model by either by SMLO (E1) or HFB-QRPA (E1), while the $M1$ strength parameters are referred to as M1. The PDR feature seen in $^{68}\mathrm{Ni}$ was modeled as E1 with the parameters denoted as PDR.}
    \begin{tabular}{c|c|c} \hline
                                        & Model & Parameters \\ \hline
    \multirow{3}{*}{$\bm{\theta}_\text{NLD}$} & CT & $T$, $\delta_\text{CT}$, $\sigma_d$, $\sigma(S_n)$ \\
                                        & BSFG & $a$, $\delta_\text{BSFG}$, $\sigma_d$, $\sigma(S_n)$ \\
                                        & HFB & $c$, $\delta_\text{HFB}$ \\ \hline
    \multirow{4}{*}{$\bm{\theta}_{\gamma\text{SF}}$} & SMLO (E1)    & $E_\text{GDR}$, $\Gamma_\text{GDR}$, $\sigma_\text{GDR}$ \\
                                                     & HFB-QRPA (E1)& $c_\text{E1}$, $\delta_\text{E1}$, $f_0$, $\epsilon_0$, $U$ \\
                                                     & PDR (E1)     & $E_\text{PDR}$, $\Gamma_\text{PDR}$, $\sigma_\text{PDR}$ \\
                                                     & M1      & $E_\text{sf}$, $\Gamma_\text{sf}$, $\sigma_\text{sf}$, $C$, $\eta$ \\ \hline
    \end{tabular}
    \label{tab:ModelParam}
\end{table}

For the NLD models fairly weak informed priors were used. In the case of the BSFG model the level density parameter $a$ mean value of $\mu=8.2$ MeV was selected as the approximate mean of the systematics from  Refs. \cite{doi:10.1139/p65-139,VONEGIDY1988189}, Refs. \cite{VonEgidy2005,PhysRevC.80.054310} and Ref. \cite{PhysRevC.80.054310} with width $\sigma=\mu$. The shift parameter $\delta_\text{BSFG}$ is a Gaussian with $\mu=0$ MeV and $\sigma=10$ MeV. Similarly the CT model priors were selected from the systematics of Ref. \cite{PhysRevC.80.054310} with widths of $\sigma=2$ MeV for the temperature and $\sigma=10$ MeV for the shift parameter $\delta_\text{CT}$. The slope and shift parameters $c$ and $\delta_\text{HFB}$ for the HFB was centered around $0$ MeV$^{-1/2}$ and $0$ MeV with widths of $1$ MeV$^{-1/2}$ and $1$ MeV, respectively. The discrete spin-cutoff parameter from eq. \ref{eq:SpinCutParam} was found to be $\sigma_d = 2.50(25)$ by examining the tabulated shell model levels while the spin-cutoff at the neutron separation energy is $\sigma(S_n) = 3.80(38)$ from an average of spin-cut off models in \cite{PhysRevC.78.064307,doi:10.1139/p65-139,VonEgidy2005}. In both cases an uncertainty of $10\%$ was assumed. The priors for all NLD model parameters are listed in Table \ref{tab:NLD_Priors}.

\begin{table}
    \centering
    \caption{Model parameter priors used for the extrapolation of the NLD between the experimental points and the neutron separation energy. The last column indicates whether a Gaussian ($\mathcal{N}$) or truncated Gaussian ($\mathcal{N}_\text{t}$) PDF are used as the prior probability distribution.}
    \begin{tabular}{c| c c c} \hline
        Parameter & $\mu$ & $\sigma$ & PDF \\ \hline
        $a$ & $8.2$ MeV & $8.2$ MeV & $\mathcal{N}_\text{t}$ \\
        $\delta$ (BSFG) & $0$ MeV & $10$ MeV & $\mathcal{N}$ \\ \hline
        $T$ & $0.958$ MeV & $2$ MeV & $\mathcal{N}_\text{t}$ \\
        $\delta$ (CT) & $-0.359$ MeV & $10$ MeV & $\mathcal{N}$ \\ \hline
        $c$ (HFB) & $0$ MeV$^{-1/2}$ & $1$ MeV$^{-1/2}$ & $\mathcal{N}$ \\
        $\delta$ (HFB) & $0$ MeV & $1$ MeV & $\mathcal{N}$ \\ \hline
        $\sigma_d$ & $2.5$ & $0.25$ & $\mathcal{N}_\text{t}$ \\
        $\sigma(S_n)$ & $3.8$ & $0.38$ & $\mathcal{N}_\text{t}$ \\ \hline
    \end{tabular}
    \label{tab:NLD_Priors}
\end{table}

The parameters for the SMLO model were taken from the systematics in \cite{PhysRevC.99.014303}. Unless an uncertainty was provided it was assumed to be $10\%$. The scale $c_\text{E1}$ and shift $\delta_\text{E1}$ for the Gogny-HFB+QRPA was assumed to be centered around $1$ and $0$ MeV, respectively, with a width of $0.5$ MeV for both. The phenomenological parameters $f_0$ and $\epsilon_0$ are taken from \cite{Goriely2018}, while the probability distribution of the initial excitation energy $U$ was modeled as a truncated Gaussian at the excitation range used to extract the NLD and $\gamma$SF ($3.5$ and $5.6$ MeV) with a mean of $4.5$ MeV and a standard deviation of $1.1$ MeV. The spin-flip and upbend parameters were taken from \cite{PhysRevC.99.014303} assuming $10\%$ uncertainty, except for $C$, where the width was assumed to be the same as the mean. Lastly, the PDR parameters were taken from \cite{PhysRevLett.111.242503}. All priors parameters related to the $\gamma$SF models are listed in Table \ref{tab:GSF_priors}.

\begin{table}
    \centering
    \caption{Model parameter priors used for the $\gamma$SF models extrapolated to the extracted $\gamma$SF of $^{67}\mathrm{Ni}$. The last column indicates whether a Gaussian ($\mathcal{N}$) or truncated Gaussian ($\mathcal{N}_\text{t}$) PDF are used as the prior probability distribution.}
    \begin{tabular}{c|c c c} \hline
        Parameter & $\mu$ & $\sigma$ & PDF \\ \hline
        $E_\text{GDR}$ & $17.68$ MeV & $0.19$ MeV  & $\mathcal{N}_\text{t}$ \\
        $\Gamma_\text{GDR}$ & $6.0$ MeV & $2.8$ MeV  & $\mathcal{N}_\text{t}$ \\
        $\sigma_\text{GDR}$ & $978$ mb & $98$ mb  & $\mathcal{N}_\text{t}$ \\ \hline
        $c_\text{E1}$ & $1$ & $0.5$ & $\mathcal{N}_\text{t}$ \\
        $\delta_\text{E1}$ & $0$ MeV & $0.5$ MeV & $\mathcal{N}$ \\
        $f_0$ & $2.5\times10^{-10}$ MeV$^{-4}$ & $2.5\times10^{-10}$ MeV$^{-4}$ & $\mathcal{N}_\text{t}$ \\
        $\epsilon_0$ & $4$ MeV & $1$ MeV & $\mathcal{N}$ \\
        $U$ & $4.45$ MeV & $1.1$ MeV & $\mathcal{N}_\text{t}$ \\ \hline
        $E_\text{PDR}$ & $9.55$ MeV & $0.17$ MeV & $\mathcal{N}_\text{t}$ \\
        $\Gamma_\text{PDR}$ & $1.02$ & $0.26$ MeV & $\mathcal{N}_\text{t}$ \\
        $\sigma_\text{PDR}$ & $27.4$ mb & $5.6$ mb & $\mathcal{N}_\text{t}$ \\ \hline
        $E_\text{sf}$ & $8.93$ MeV & $0.90$ MeV & $\mathcal{N}_\text{t}$ \\
        $\Gamma_\text{sf}$ & $4.0$ MeV & $0.4$ MeV & $\mathcal{N}_\text{t}$ \\
        $\sigma_\text{sf}$ & $1.0$ mb & $0.1$ mb & $\mathcal{N}_\text{t}$ \\
        $C$ & $3.5 \times 10^{-8}$ MeV$^{-3}$ & $3.5 \times 10^{-8}$ MeV$^{-3}$ & $\mathcal{N}_\text{t}$ \\
        $\eta$ & $0.8$ 1/MeV & $0.08$ 1/MeV & $\mathcal{N}$ \\
        \hline
    \end{tabular}
    \label{tab:GSF_priors}
\end{table}

\subsection{Hauser-Feshbach calculations and results}
\begin{figure}
    \centering
    \includegraphics[width=0.5\textwidth]{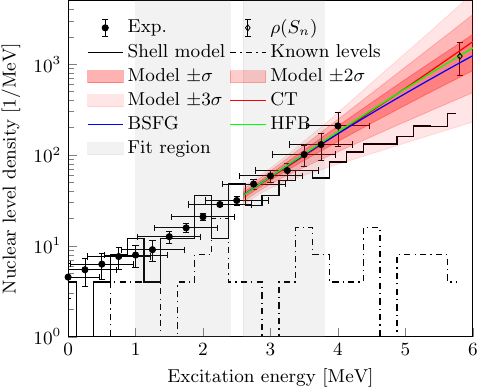}
    \caption{The measured nuclear level density of $^{67}\mathrm{Ni}$ is shown by the black circles while the black line shows the NLD from large scale SM calculations \cite{Midtbo2018}. The black dash-dotted line shows the NLD from known discrete levels \cite{JUNDE2005159,PhysRevC.91.054321}. The red, blue and green lines show the average for the CT, BSFG and HFB models, respectively. The red shaded area indicates the $\pm1\sigma$, $\pm2\sigma$ and $\pm3\sigma$ credibility intervals. The black open diamond represents the extrapolated NLD at the neutron separation energy. The gray shaded areas indicate the fitting regions to the discrete levels (lower energy) and to the extrapolated models normalized to the measured NLD (higher energy).}
    \label{fig:nld_67Ni}
\end{figure}

\begin{figure}
    \centering
    \includegraphics[width=0.5\textwidth]{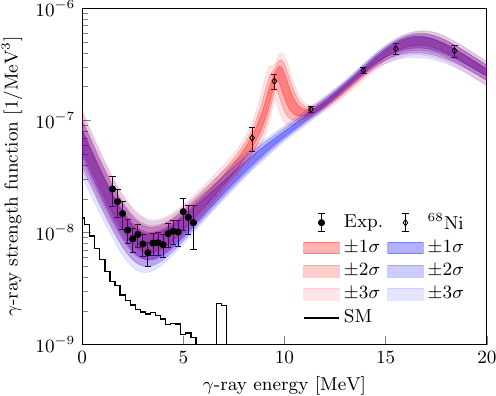}
    \caption[$\gamma$-ray strength function of $^{67}\mathrm{Ni}$]{The measured $\gamma$-ray strength function of $^{67}\mathrm{Ni}$ is shown by the black circles while the black line shows the M1-strength found in large scale SM calculations from \cite{Midtbo2018}.
    The red and blue shaded area indicates the $\pm1\sigma$, $\pm2\sigma$ and $\pm3\sigma$ credibility intervals for the models including the PDR and the models excluding the PDR, respectively. The black open diamonds are the $\gamma$SF data of $^{68}\mathrm{Ni}$ measured by Rossi \textit{et al.} \cite{PhysRevLett.111.242503}.}
    \label{fig:gsf_67Ni}
\end{figure}

\begin{figure}
    \centering
    \includegraphics[width=0.5\textwidth]{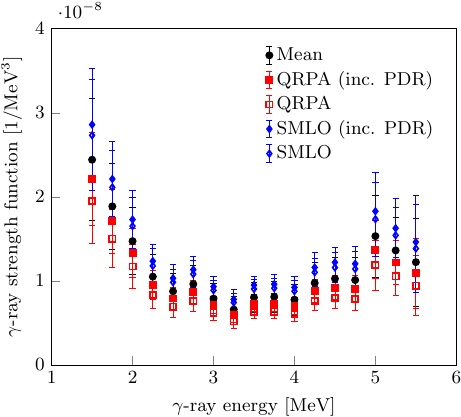}
    \caption{Comparison of the various $\gamma$SF normalizations including and excluding the PDR. The black circles show the average normalized $\gamma$SF.}
    \label{fig:gsf_sep_67Ni}
\end{figure}

For each sample $\bm{\theta}_i$ the normalized NLD and $\gamma$SF are calculated and the marginal posterior probability distribution for each NLD and $\gamma$SF point is obtained for each combination of NLD and $\gamma$SF models. Weighting each combination of models equally, the average normalized NLD and $\gamma$SF are shown in Fig. \ref{fig:nld_67Ni} and \ref{fig:gsf_67Ni}, respectively. The $68$, $95$ and $99.7\%$ credibility intervals (1, 2 and 3 standard deviations) of the model predictions are found by calculating the NLD and $\gamma$SF for each sample $\bm{\theta}_i$ and weighting each combination of models equally. The posterior mean values for all model and normalization parameters are listed in Appendix \ref{app:appendix}.
The impact of including the PDR structure in the models for the $\gamma$SF reported by \cite{PhysRevLett.111.242503} is shown in Fig. \ref{fig:gsf_sep_67Ni} to be negligible for the normalization of the $\gamma$SF. The NLD at the neutron separation energy is found to be $1.22(51)\times10^3$ MeV$^{-1}$ by averaging the NLD models.

\begin{figure}
    \centering
    %\ref{ng_xs_legend}
    \includegraphics[width=0.5\textwidth]{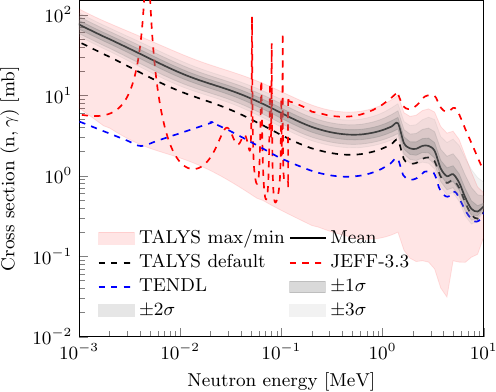}
    \caption{The mean calculated capture cross section from the sampled NLD and $\gamma$SF are given by the black line. The gray shaded area shows the $\pm1\sigma$, $\pm2\sigma$ and $\pm3\sigma$ credibility intervals. The black dashed line represents the capture cross section with TALYS default models (CT/Fermi-Gas+SMLO). The red shaded area shows the maximum and minimum limits based on calculations using the same NLD and $\gamma$SF model combinations as Ref. \cite{McKay2020}.
The red and blue dashed lines shows the JEFF-3.3 library \cite{Plompen2020TheJEFF-3.3} and the TENDL library \cite{Koning2019TENDL:Technology}, respectively, the former including a set of unobserved resonance added ad-hoc as explained in Ref. \cite{Rochman2020}.}
    \label{fig:ng_xs}
\end{figure}

\begin{figure}
    \centering
    \includegraphics[width=0.5\textwidth]{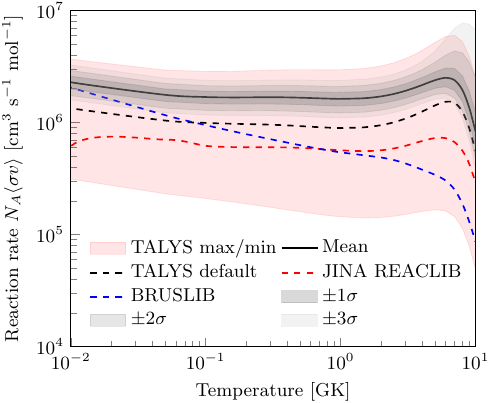}
    \caption{The mean calculated capture rate from the sampled NLD and $\gamma$SF are given by the black line. The gray filled area shows the $\pm1\sigma$, $\pm2\sigma$ and $\pm3\sigma$ credibility intervals. The black dashed line is the capture rate with TALYS default (CT/Fermi-Gas+SMLO). The red and blue dashed lines are the reaction rates from JINA REACLIB v1.1 \cite{cyburt_jina_2010} and BRUSLIB \cite{arnould06}, respectively. The red shaded area shows the maximum and minimum limits based on calculations using the same NLD and $\gamma$SF model combinations as Ref. \cite{McKay2020}.}
    \label{fig:ng_rate}
\end{figure}

The measured NLD and $\gamma$SF in $^{67}\mathrm{Ni}$ were used in Hauser-Feshbach calculations to constrain the neutron capture cross section of $^{66}\mathrm{Ni}$. From each posterior sample $\bm{\theta}_i$ of normalization and model parameters tabulated NLD and $\gamma$SF values were generated from linear interpolation and used as input to the \verb+TALYS+\footnote{Version 1.96} reaction code \cite{Koning2007}. Since the level scheme of $^{67}\mathrm{Ni}$ is far from complete only the first two discrete levels were included in the Hauser-Feshbach  calculations, while above those the provided tabulated NLD was used. Each realization of the capture cross section and reaction rate is weighted equally to obtain a mean cross section and reaction rate and the credibility bands shown in Figs. \ref{fig:ng_xs} and \ref{fig:ng_rate}. The impact of the optical model potential (OMP) has been tested by performing calculations with the phenomenological OMP of Ref. \cite{koning_local_2003} and the microscopic OMP of Ref. \cite{PhysRevC.63.024607} and found to have a negligible contribution to the uncertainty.

In the sensitivity study of Ref. \cite{McKay2020} the $^{66}\mathrm{Ni}(n,\gamma)$ reaction rates were taken from the JINA REACLIB v1.1 library \cite{cyburt_jina_2010}, indicated by the red dashed line in Fig. \ref{fig:ng_rate}. The maximum and minimum reaction rates considered by Ref. \cite{McKay2020} are based on the models listed in Table 1 of Ref. \cite{denissenkov_impact_2018}, and these results have been reproduced as indicated by the red band in Fig. \ref{fig:ng_xs} and \ref{fig:ng_rate}. For comparison, the capture cross section and rate with the default selection of NLD (CT+Fermi-Gas), $\gamma$SF (SMLO \cite{PhysRevC.99.014303}+LEE) models, and optical model potential (local phenomenological OMP) in \verb+TALYS+ version 1.96 are shown by the black dashed lines in Figs. \ref{fig:ng_xs} and \ref{fig:ng_rate}.

\section{Discussion}
The Oslo method relies on external nuclear data for the normalization. In the absence of those, additional uncertainties may be induced and model dependencies may become significant. This is apparent through the relatively large uncertainties toward $S_n$ on the measured NLD for $^{67}\mathrm{Ni}$.

The challenge specific to inverse kinematic reactions is the Lorentz boost which causes a strong angular dependence in the kinematic reconstruction of the residual excitation energy. This leads to an excitation-energy resolution which is limited by the angular opening of the particle telescope's active areas. 
The consequences are most apparent in the NLD where very little structures are visible. 

In contrast, the measured $\gamma$SF still retains noticeable features and clearly exhibits a well established enhancement for $E_x < 4$ MeV similar to those found in other $\mathrm{Ni}$ isotopes \cite{PhysRevC.98.054619,PhysRevC.94.044321,Ingeberg2022b,PhysRevC.96.014312,Spyrou2017,PhysRevC.97.054329}. Its observation indicates that the upbend is a structure which exists also away from stability. The upbend in $^{67}\mathrm{Ni}$ is predicted to be due to M1 strength, based on large-scale shell model calculations \cite{Midtbo2018}, and shown by the black solid line in Fig. \ref{fig:gsf_67Ni}. The significant strength in the  enhancement and the simultaneous absence of a measurable scissors resonance may be supportive of the suggested connection of the two structures \cite{Schwengner2017Low-EnergyNuclei,Frauendorf22}, although results on the $\gamma$SF in $^{142,144-151}\mathrm{Nd}$ seem to contradict this \cite{guttormsen22}.

The calculated $^{66}\mathrm{Ni}(n,\gamma)$ capture cross section in Fig. \ref{fig:ng_xs} features an uncertainty of $\approx 40\%$, constraining the cross section considerably. It is interesting to note that our cross section, lies consistently higher than the recommended values as provided in TALYS and TENDL but is smaller than those of JEFF 3.3 for $E_n > 100$ keV. These differences highlight the necessity for measurements of NLDs and $\gamma$SFs, especially for nuclei away from stability.

Ref. \cite{McKay2020} allows the capture rate to vary within the red band as indicated in Fig. \ref{fig:ng_rate}, with the nominal value taken from the JINA REACLIB v1.1 database \cite{cyburt_jina_2010}. Our results constrain the capture rate significantly and show that this rate is found at the high end of the range used by Ref. \cite{McKay2020}. This suggests a short exposure time for the weak i-process and could help pinpoint details in the stellar environment responsible to produce neutrons.

\section{Astrophysical implications}

\begin{figure}
    \centering
    \includegraphics[scale=0.38, trim = 0cm 0cm 0cm 0.cm]{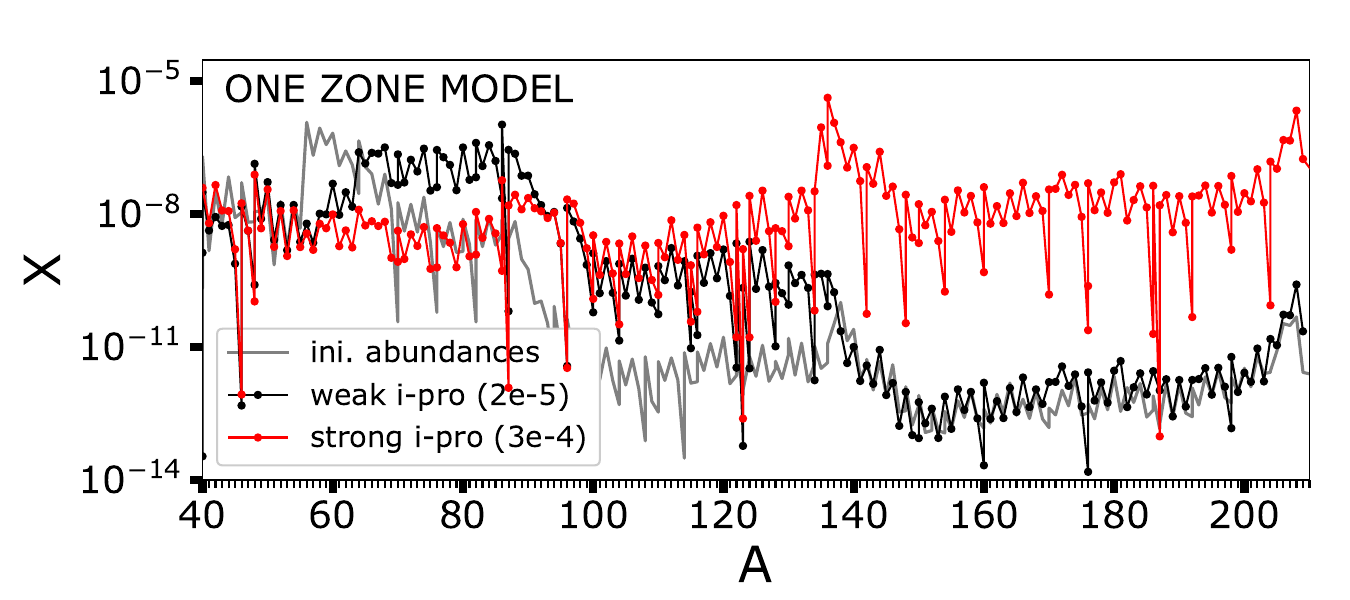}
    \caption{Final nuclei mass fractions (after decays) as a function of mass number after a weak (black) and a strong (red) i-process using the one-zone model. The initial proton mass fraction is indicated in parenthesis (see text for details). The grey pattern shows the initial abundances.}
    \label{fig:ipro}
\end{figure}

\begin{figure}
    \centering
    \includegraphics[scale=0.38, trim = 0cm 0cm 0cm 0.cm]{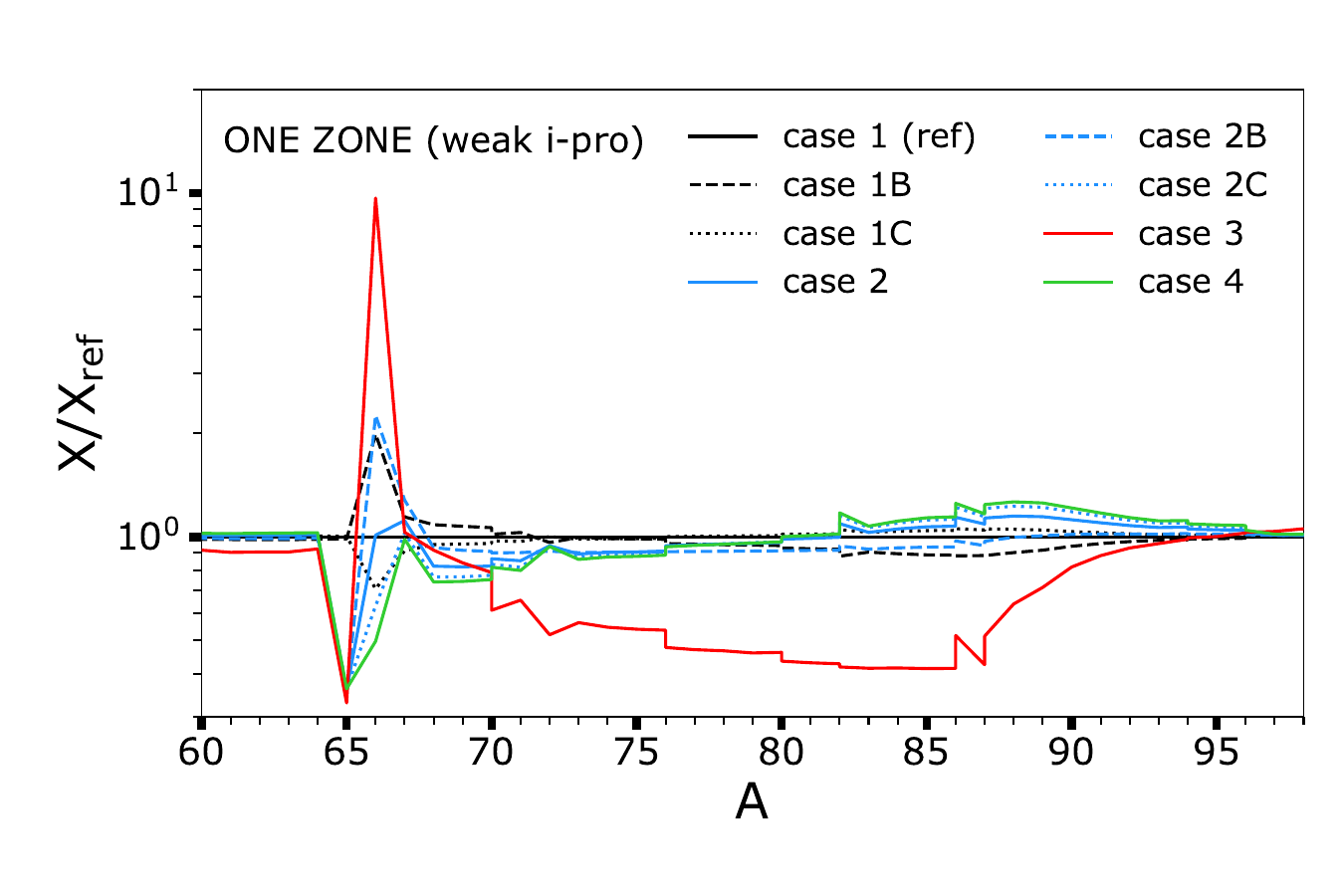}
    \includegraphics[scale=0.38, trim = 0cm 0cm 0cm 0.  cm]{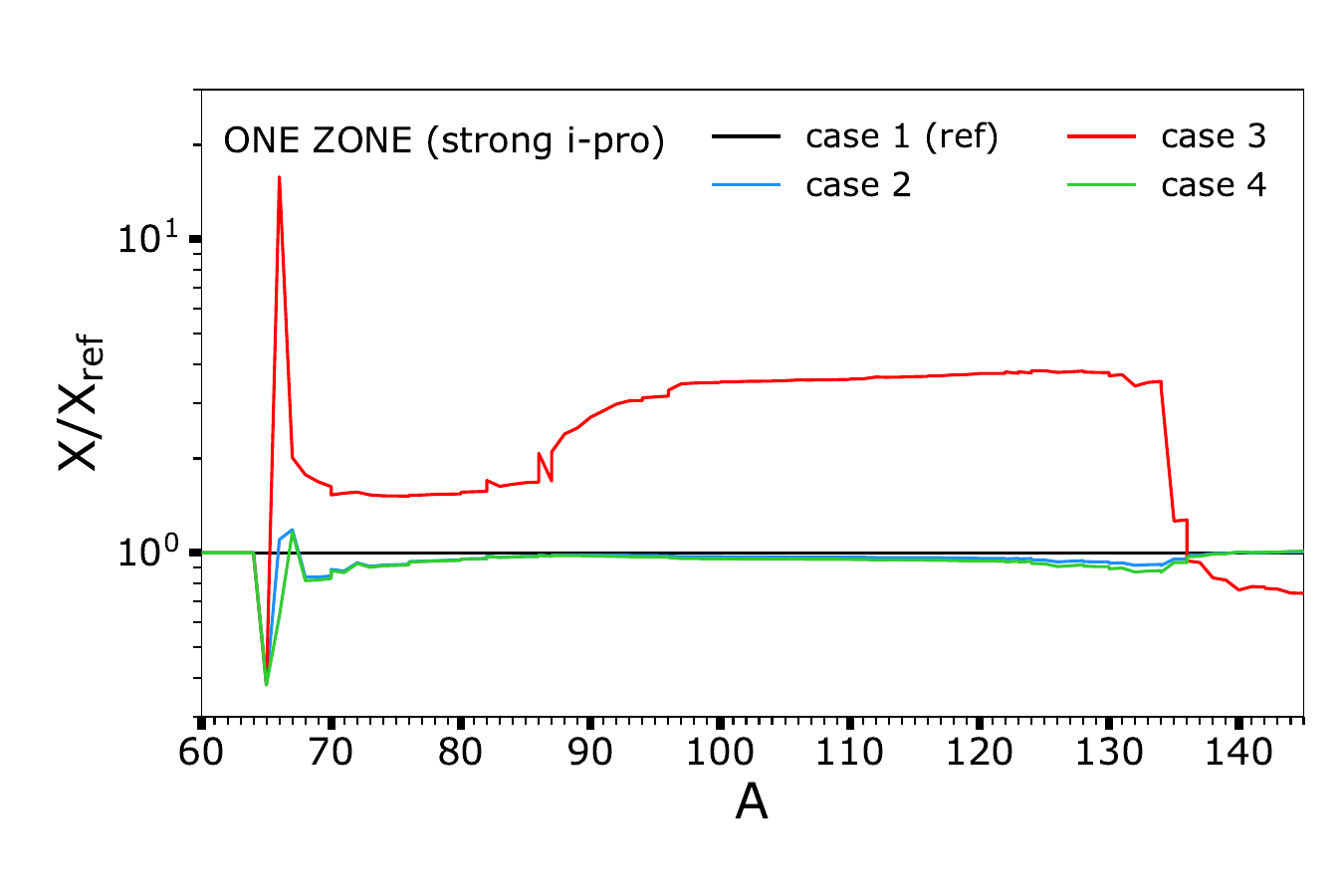}
    \caption{Final nuclei mass fractions (after decays) as a function of mass number after a weak (top panel) and a strong (bottom) i-process using the one-zone model. The abundances are normalized by the abundances of the case 1. The 8 cases correspond to the 8 different combinations of rates indicated in Table~\ref{tab:cases} (cases 1B, 1C, 2B and 2C are not shown in the bottom panel for clarity).}
    \label{fig:onezone}
\end{figure}

\begin{figure}
    \centering
    \includegraphics[scale=0.38, trim = 0cm 0cm 0cm 0.cm]{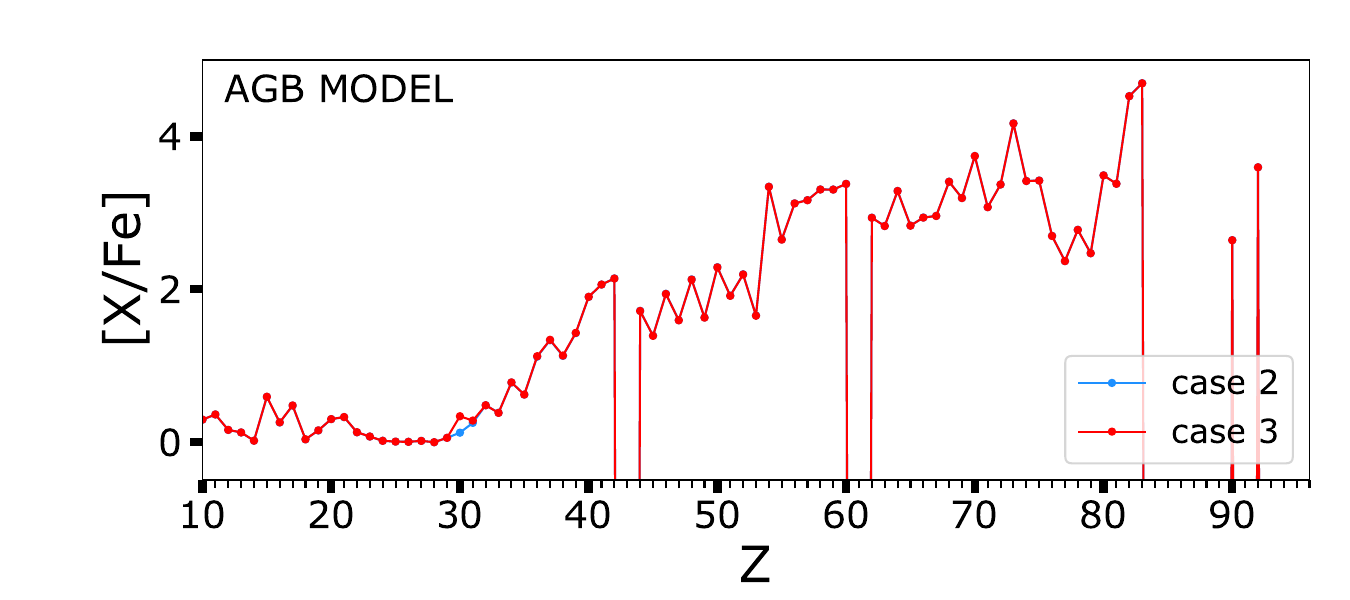}
    \includegraphics[scale=0.38, trim = 0cm 0cm 0cm 0.cm]{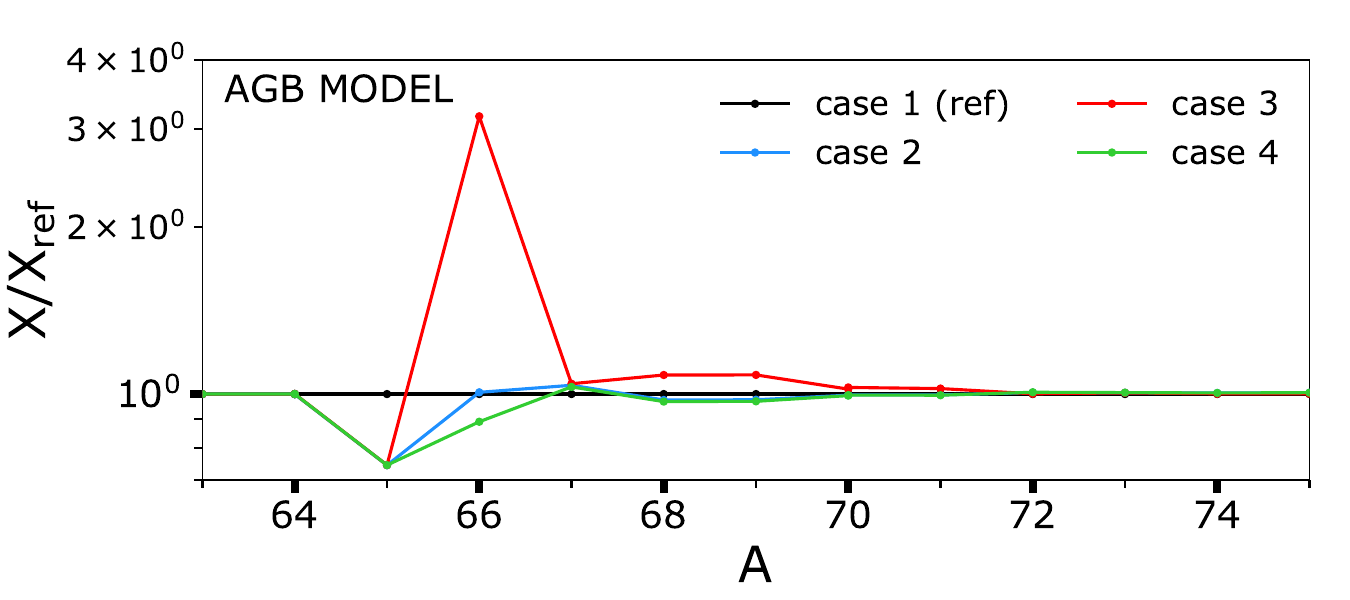}
    \caption{Final surface abundances (after decays) for a full AGB stellar model. Top: [X/Fe] ratios, defined as $[\textrm{X/Fe}] =\log_{10}(N_{\rm X} / N_{\rm Fe})_{\star} - \log_{10}(N_{\rm X} / N_{\rm Fe})_{\odot}$ with $N_{\rm X}$ the number density of an element X. The first and second $\log_{10}$ terms refer to the abundances of the model and the Sun, respectively. Bottom: same as Fig.~\ref{fig:onezone}. The 4 cases correspond to the 4 different combinations of rates indicated in Table~\ref{tab:cases}.}
    \label{fig:agb}
\end{figure}

The $^{66}\mathrm{Ni}(n,\gamma)$ capture rate was suggested to be of key importance for the overall production of heavy elements during the i-process nucleosynthesis taking place in Sakurai's object or rapidly accreting white dwarfs \cite{McKay2020}.
One possible astrophysical site with similar i-process conditions are low-metallicity low-mass stars during the early thermally pulsating Asymptotic Giant Branch (AGB) phase \cite{iwamoto04,cristallo09b,suda10,stancliffe11,choplin21}. 
During this stage, protons can be engulfed by the convective thermal pulse. As they are transported downward by convection on a timescale of about 1~hr, they are burnt rapidly by $^{12}\mathrm{C}(p,\gamma)^{13}\mathrm{N}$. After the decay of $^{13}\mathrm{N}$ to $^{13}\mathrm{C}$ in about 10~min, the reaction $^{13}\mathrm{C}(\alpha,n)$ is activated at the bottom of the thermal pulse and leads to neutron densities of up to $\sim 10^{15}$~cm$^{-3}$.
Stellar modeling was performed with the stellar evolution code STAREVOL \cite{siess00, siess06, goriely18}. A 1~$M_{\odot}$ AGB model at a metallicity of [Fe/H]~$=-2.5$ is considered. This model is discussed in detail in Ref. \cite{choplin21}.

We estimated the impact of the new $^{66}\mathrm{Ni}(n,\gamma)$ rate on the i-process nucleosynthesis in these objects. The effects are evaluated on both multi-zone stellar evolution models and in one-zone nucleosynthesis calculations. 
We used a reaction network including 1160 nuclei, linked through 2123 nuclear reactions ($n$-, $p$-, $\alpha$-captures and $\alpha$-decays) and weak interactions (electron captures, $\beta$-decays). Nuclear reaction rates were taken from BRUSLIB, the Nuclear Astrophysics Library of the Université Libre de Bruxelles\footnote{Available at http://www.astro.ulb.ac.be/bruslib/} \cite{arnould06} and the updated experimental and
theoretical rates from the NETGEN interface \cite{xu13}. Additional details can be found in Refs. \cite{choplin21,goriely21,choplin22}. 
In our sensitivity analysis, eight cases were investigated. We consider, for the $^{66}\mathrm{Ni}(n,\gamma)$ reaction, the recommended Maxwellian-averaged cross section (MACS) from this work ($10.7\pm 5$~mb at 30 keV) as well as the minimum (1 mb at 30 keV) and maximum (19 mb at 30 keV) rates predicted by a global unconstrained TALYS calculation (i.e. maximum and minimum reaction rates as prescribed in Ref. \cite{McKay2020}). Additionally, the rate of $^{65}\mathrm{Ni}(n,\gamma)$ was also varied because $^{65}\mathrm{Ni}$ has a half-life of $\sim 2.5$ hr (55 hr for $^{66}\mathrm{Ni}$), which is fast enough to partially bypass the neutron capture during the i-process. For $^{65}\mathrm{Ni}(n,\gamma)$, we considered the minimum (10 mb at 30 keV) and maximum (70 mb at 30 keV) MACS predicted by TALYS for different NLD and $\gamma$SF models. 
We also considered the uncertainties related to the nominal $^{66}\mathrm{Ni}(n,\gamma)$ MACS evaluated in this study by varying it by $\pm 5$~mb at 30 keV (cases 1B, 1C, 2B and 2C). The different cases considered are reported in Table~\ref{tab:cases}.

\begin{table}
    \centering
    \caption{The different cases considered for the rates of $^{65}$Ni($n,\gamma$) and $^{66}$Ni($n,\gamma$) for the one-zone and stellar model calculations.} 
    \begin{tabular}{l|l|l} \hline
         & $^{65}$Ni($n,\gamma$) & $^{66}$Ni($n,\gamma$) \\ \hline
        case 1 & TALYS min & nominal \\
        case 1B & TALYS min & nominal $-\,5$~mb \\
        case 1C & TALYS min & nominal $+\,5$~mb \\
        case 2 & TALYS max & nominal \\
        case 2B & TALYS max & nominal $-\,5$~mb\\
        case 2C & TALYS max & nominal $+\,5$~mb \\
        case 3 & TALYS max & TALYS min \\
        case 4 & TALYS max & TALYS max \\
       \hline
    \end{tabular}
    \label{tab:cases}
\end{table}

The initial abundances of the one-zone model correspond to the chemical composition in the thermal pulse of the AGB stellar model, just before the start of the proton ingestion event. The temperature and density were fixed to 220 MK and 2500 g~cm$^{-3}$, respectively, which are typical values in the pulse of low-metallicity AGB models. The nucleosynthesis was calculated over a total time of $2 \times 10^{5}$~s ($\sim 2.3$ days). To mimic the proton ingestion episode in the one-zone model, an initial abundance of protons was set to $2 \times 10^{-5}$ and $3 \times 10^{-4}$ leading to maximum neutron densities of $N_{\rm n, max} = 1.5 \times 10^{14}$~cm$^{-3}$ and $6.7 \times 10^{14}$~cm$^{-3}$, respectively. For a low initial proton abundance, a weak i-process develops and elements up to the first peak ($A \lesssim 90$) are synthesized (Fig.~\ref{fig:ipro}, black pattern). With more protons, a larger neutron density is reached and elements of the third peak (Fig.~\ref{fig:ipro}, red pattern) are produced. For this particular case, the  final chemical composition is found to be very similar to that of the  full multi-zone AGB model where densities up to $N_n = 2.1 \times 10^{15}$~cm$^{-3}$ are reached.

As shown in Fig.~\ref{fig:onezone}, increasing the rate of $^{65}\mathrm{Ni}(n,\gamma)$ (case 2) leads to a smaller production of $^{65}\mathrm{Cu}$ (as a result of $^{65}\mathrm{Ni}$ decay) by a factor of $\sim 3$ in both weak and strong i-process cases.
Lowering $^{66}\mathrm{Ni}(n,\gamma)$ (case 3) has the strongest impact since $^{66}\mathrm{Ni}$ now acts as a bottleneck. In the weak i-process case (Fig.~\ref{fig:onezone}, top panel), it leads to an overproduction of $^{66}\mathrm{Zn}$ by a factor of $\sim 10$ to the detriment of $70<A<90$ isotopes ($30<Z<40$), which are underproduced by a factor of $\sim 3$ at maximum. For a higher neutron density (Fig.~\ref{fig:onezone}, bottom panel), the production of $66<A<135$ isotopes ($28<Z<54$) is enhanced by a factor of typically $2-4$. % to the detriment of heavier elements. 
The uncertainties on the nominal $^{66}\mathrm{Ni}(n,\gamma)$ MACS obtained in this work ($\pm 5$~mb at 30 keV) induce a variation of about a factor of $3$ at $A=66$ (dashed lines in Fig.~\ref{fig:onezone}, top panel).

Ref. \cite{McKay2020} investigated the case of a weak i-process using one-zone models with an initial metallicity of [Fe/H]~$=-1.6$ ([Fe/H]~$=-2.5$ here) and different reaction rates (most of ($n,\gamma$) rates coming from the JINA REACLIB library  \cite{cyburt_jina_2010}). Although the physical inputs are different, in the case of a weak i-process the results are qualitatively similar to those of Ref. \cite{McKay2020}: an underproduction of the elements with $32<Z<42$ when adopting a low $^{66}\mathrm{Ni}(n,\gamma)$ rate.

However, when considering a full multi-zone AGB calculation, the impact of  $^{66}\mathrm{Ni}(n,\gamma)$ is strongly dampened (Fig.~\ref{fig:agb}). The overproduction of $^{66}\mathrm{Zn}$ in case 3 is clearly present but the enrichment is a factor of $\sim 3$ lower. In the stellar model, multiple zones with different temperatures, densities, chemical compositions and irradiations are mixed together by convection leading to a dilute production of the elements, an effect that cannot be captured by one-zone calculations. %In particular, the i-process products, which are synthesized in deep layers, are eventually diluted in the large AGB convective envelope.

The impact of the $^{66}\mathrm{Ni}(n,\gamma)$ reaction appears to be marginal in low-metallicity AGB stars experiencing i-process nucleosynthesis. It shows that one-zone calculations are inherently approximate and cannot guarantee to reproduce the accurate physics picture for a system. %The conclusions based on one-zone modeling may not be valid (or at least can be different) when going to stellar modeling. 
We should also emphasize that only one i-process site was investigated and we cannot exclude a different impact of $^{66}\mathrm{Ni}(n,\gamma)$ in other sites such as e.g. rapidly accreting white dwarfs \cite{denissenkov17}.

\section{Summary}
In this paper we have presented experimental NLD and $\gamma$SF of the unstable $^{67}\mathrm{Ni}$ nucleus. It is the first time the Oslo method has been applied to an inverse kinematics experiment with a radioactive beam. Due to the lack of reliable nuclear data for $^{67}\mathrm{Ni}$ the NLD and $\gamma$SF are normalized to model predictions resulting in relatively large uncertainties. Hauser-Feshbach calculations were performed with the extracted NLD and $\gamma$SF to find the neutron capture cross section of $^{66}\mathrm{Ni}$. The result of these calculations showed a rather high capture rate compared to the theoretical range, and can help improve our understanding of the i-process nucleosynthesis. 
In agreement with Ref. \cite{McKay2020}, we find that the reaction $^{66}\mathrm{Ni}(n,\gamma)$ acts as a bottleneck when using one-zone models. Nevertheless, the impact is strongly dampened in multi-zone low-metallicity AGB stellar models experiencing i-process nucleosynthesis. This reaction may however have a different impact in other i-process sites.

\begin{acknowledgments}
The authors would like to thank the accelerator team at CERN ISOLDE for providing excellent running conditions. This study has been funded by the Research Council of Norway through its grants to the Norwegian Nuclear Research Centre (Project No. 341985) and the Norwegian Centre for CERN-related Research (Project No. 310713). Additional research project grants were provided by the Research Council of Norway (Grants No. 263030, 325714 and 245882), the National Research Foundation of South Africa (Grant No. 118846), the EU Horizon 2020 research and innovation program through ENSAR2 (Grant No. 654002), and the U.S. Department of Energy, Office of Science, Office of Nuclear Physics under Contract No. DE-AC02-05CH11231 (LBNL) and Contract No. DE-AC52-07NA27344 (LLNL). Funding was also provided by the German BMBF under contracts 05P18PKCIA, 05P21PKCI1, 05P15RDCIA, and ‘Verbundprojekt’ 05P18PKCI1, 05P18PKCI1. This work was supported by the Fonds de la Recherche Scientifique-FNRS under Grant No IISN 4.4502.19. L.S. and S.G. are senior FRS-F.N.R.S. research associates and A.C. is a post-doctorate FRS-F.N.R.S. fellow. G.R. acknowledges the support by the Bulgarian Ministry of Education and Science within the National Roadmap for Research Infrastructures (object CERN). Additional support was provided by the Research Foundation Flanders (FWO, Belgium) under GOA/2015/010 (BOF KU Leuven) and the FWO and F.R.S.-FNRS under the Excellence of Science (EOS) programme (No. 40007501).
\end{acknowledgments}

\appendix

\section{Prior and posterior parameters}\label{app:appendix}

\begin{table}
    \centering
    \caption{Numerical values of the centroids of the normalization parameter priors. The standard deviation was $10$ times the centroids for $A$ and $B$, $1$ 1/MeV for $\alpha$.}
    \begin{tabular}{c|c|c|c}
       Models & A (MeV$^{-1}$) & B (MeV$^{-3}$) & $\alpha$ (MeV$^{-1}$) \\ \hline
       CT+SMLO+M1 & 1.863 & 0.538 & -0.037 \\
       CT+SMLO+PDR+M1 & 1.745 & 0.484 & -0.00453 \\
       CT+QRPA+M1 & 2.378 & 0.677 & -0.141 \\
       CT+QRPA+PDR+M1 & 2.152 & 0.574 &  -0.111 \\
       BSFG+SMLO+M1 & 2.033 & 0.670 & -0.0781 \\
       BSFG+SMLO+PDR+M1 & 1.826 & 0.520 & -0.0177 \\
       BSFG+QRPA+M1 & 2.652 & 0.882 & -0.120 \\
       BSFG+QRPA+PDR+M1 & 2.034 & 0.541 & -0.0709 \\
       HFB+SMLO+M1 & 1.916 & 0.620 & -0.0446 \\
       HFB+SMLO+PDR+M1 & 1.880 & 0.580 & -0.0327 \\
       HFB+QRPA+M1 & 2.270 & 0.528 & -0.122 \\
       HFB+QRPA+PDR+M1 & 2.169 & 0.557 & -0.102 \\
       \hline
    \end{tabular}
    \label{tab:prior_normalization_parameters}
\end{table}

% \begin{table}
%     \centering
%     \caption{Posterior median values of the parameters for CT NLD model and SMLO+M1 $\gamma$SF model.}
%     \begin{tabular}{c|c}
%     Parameter & Value \\ \hline
%     $A$ & $1.94(21)$ MeV$^{-1}$ \\
%     $B$ & $0.59(15)$ MeV$^{-3}$ \\
%     $\alpha$ & $-0.050(50)$ MeV$^{-1}$ \\ \hline
%     $T$ & $0.834(95)$ MeV \\
%     $\delta$ & $-0.23(44)$ MeV \\
%     $\sigma_D$ & $2.50(23)$ \\
%     $\sigma(Sn)$ & $3.79(35)$ \\ \hline
%     $E_\text{GDR}$ & $17.42(13)$ MeV \\
%     $\Gamma_\text{GDR}$ & $7.30(59)$ MeV \\
%     $\sigma_\text{GDR}$ & $1118(44)$ mb \\
%     $E_\text{sf}$ & $9.92(71)$ MeV \\
%     $\Gamma_\text{sf}$ & $3.88(37)$ MeV \\
%     $\sigma_\text{sf}$ & $0.995(91)$ mb \\
%     $C$ & $7.3(14)\cdot10^{-8}$ MeV$^{-3}$ \\
%     $\eta$ & $0.936(59)$ MeV$^{-1}$ \\ \hline
%     \end{tabular}
%     \label{tab:CT_SMLO_M1}
% \end{table}

\begin{table}
    \centering
    \caption{Posterior mean value of the parameters for CT NLD model and SMLO+M1 $\gamma$SF model.}
    \begin{tabular}{c|c}
    Parameter & Value \\ \hline
    $A$ & $1.95(20)$ MeV$^{-1}$ \\
    $B$ & $0.60(15)$ MeV$^{-3}$ \\
    $\alpha$ & $-0.048(48)$ MeV$^{-1}$ \\ \hline
    $T$ & $0.843(97)$ MeV \\
    $\delta$ & $-0.28(46)$ MeV \\
    $\sigma_D$ & $2.51(23)$ \\
    $\sigma(Sn)$ & $3.79(36)$ \\ \hline
    $E_\text{GDR}$ & $17.42(12)$ MeV \\
    $\Gamma_\text{GDR}$ & $7.32(60)$ MeV \\
    $\sigma_\text{GDR}$ & $1121(44)$ mb \\
    $E_\text{sf}$ & $9.93(70)$ MeV \\
    $\Gamma_\text{sf}$ & $3.88(37)$ MeV \\
    $\sigma_\text{sf}$ & $0.993(93)$ mb \\
    $C$ & $7.36(142)\cdot10^{-8}$ MeV$^{-3}$ \\
    $\eta$ & $0.935(59)$ MeV$^{-1}$ \\ \hline
    \end{tabular}
    \label{tab:CT_SMLO_M1}
\end{table}

% \begin{table}
%     \centering
%      \caption{Posterior median values of the parameters for CT NLD model and SMLO+M1 $\gamma$SF model.}
%     \begin{tabular}{c|c}
%     Parameter & Value \\ \hline
%     $A$ & $1.92(22)$ MeV$^{-1}$ \\
%     $B$ & $0.61(16)$ MeV$^{-3}$ \\
%     $\alpha$ & $0.045(52)$ MeV$^{-1}$ \\ \hline
%     $T$ & $0.835(94)$ MeV \\
%     $\delta$ & $-0.24(44)$ MeV \\
%     $\sigma_D$ & $2.50(23)$ \\
%     $\sigma(Sn)$ & $3.79(35)$ \\ \hline
%     $E_\text{GDR}$ & $17.42(13)$ MeV \\
%     $\Gamma_\text{GDR}$ & $6.58(61)$ MeV \\
%     $\sigma_\text{GDR}$ & $1082(48)$ mb \\
%     $E_\text{PDR}$ & $9.71(13)$ MeV \\
%     $\Gamma_\text{PDR}$ & $0.94(18)$ MeV \\
%     $\sigma_\text{PDR}$ & $23.7(43)$ mb \\
%     $E_\text{sf}$ & $9.76(69)$ MeV \\
%     $\Gamma_\text{sf}$ & $3.88(36)$ MeV \\
%     $\sigma_\text{sf}$ & $0.990(91)$ mb \\
%     $C$ & $7.5(15)\cdot10^{-8}$ MeV$^{-3}$ \\
%     $\eta$ & $0.934(56)$ MeV$^{-1}$ \\ \hline
%     \end{tabular}
%     \label{tab:CT_SMLO_PDR_M1}
% \end{table}

\begin{table}
    \centering
    \caption{Posterior mean values of the parameters for CT NLD model and SMLO+M1 $\gamma$SF model.}
    \begin{tabular}{c|c}
    Parameter & Value \\ \hline
    $A$ & $1.94(21)$ MeV$^{-1}$ \\
    $B$ & $0.63(16)$ MeV$^{-3}$ \\
    $\alpha$ & $-0.045(50)$ MeV$^{-1}$ \\ \hline
    $T$ & $0.844(97)$ MeV \\
    $\delta$ & $-0.28(46)$ MeV \\
    $\sigma_D$ & $2.51(24)$ \\
    $\sigma(Sn)$ & $3.79(35)$ \\ \hline
    $E_\text{GDR}$ & $17.42(13)$ MeV \\
    $\Gamma_\text{GDR}$ & $6.61(61)$ MeV \\
    $\sigma_\text{GDR}$ & $1084(47)$ mb \\
    $E_\text{PDR}$ & $9.69(13)$ MeV \\
    $\Gamma_\text{PDR}$ & $0.93(18)$ MeV \\
    $\sigma_\text{PDR}$ & $24.1(44)$ mb \\
    $E_\text{sf}$ & $9.79(69)$ MeV \\
    $\Gamma_\text{sf}$ & $3.87(36)$ MeV \\
    $\sigma_\text{sf}$ & $0.988(92)$ mb \\
    $C$ & $7.6(15)\cdot10^{-8}$ MeV$^{-3}$ \\
    $\eta$ & $0.934(56)$ MeV$^{-1}$ \\ \hline
    \end{tabular}
    \label{tab:CT_SMLO_PDR_M1}
\end{table}

% \begin{table}
%     \centering
%     \caption{Posterior median values of the parameters for CT NLD model and QRPA+M1 $\gamma$SF model.}
%     \begin{tabular}{c|c}
%     Parameter & Value \\ \hline
%     $A$ & $2.02(22)$MeV$^{-1}$ \\
%     $B$ & $0.45(12)$ MeV$^{-3}$ \\
%     $\alpha$ & $-0.068(51)$ MeV$^{-1}$ \\ \hline
%     $T$ & $1.03(19)$ MeV \\
%     $\delta$ & $-1.13(93)$ MeV \\
%     $\sigma_D$ & $2.50(21)$ \\
%     $\sigma(Sn)$ & $3.77(34)$ \\ \hline
%     $c_\text{E1}$ & $0.819(41)$ \\
%     $\delta_\text{E1}$ & $-0.49(15)$ MeV \\
%     $f_0$ & $1.2(11)\cdot10^{-10}$ MeV$^{-4}$ \\
%     $\epsilon_0$ & $3.53(93)$ MeV \\
%     $U$ & $4.41(56)$ MeV \\
%     $E_\text{sf}$ & $10.21(62)$ MeV \\
%     $\Gamma_\text{sf}$ & $3.86(34)$ MeV \\
%     $\sigma_\text{sf}$ & $0.993(87)$ mb \\
%     $C$ & $5.7(11)\cdot10^{-8}$ MeV$^{-3}$ \\
%     $\eta$ & $0.924(49)$ MeV$^{-1}$ \\ \hline
%     \end{tabular}
%     \label{tab:CT_QRPA_M1}
% \end{table}

\begin{table}
    \centering
    \caption{Posterior mean values of the parameters for CT NLD model and QRPA+M1 $\gamma$SF model.}
    \begin{tabular}{c|c}
    Parameter & Value \\ \hline
    $A$ & $2.04(22)$ MeV$^{-1}$ \\
    $B$ & $0.47(13)$ MeV$^{-3}$ \\
    $\alpha$ & $-0.072(50)$ MeV$^{-1}$ \\ \hline
    $T$ & $1.06(21)$ MeV \\
    $\delta$ & $-1.3(11)$ MeV \\
    $\sigma_D$ & $2.50(23)$ \\
    $\sigma(Sn)$ & $3.77(34)$ \\ \hline
    $c_\text{E1}$ & $0.819(41)$ \\
    $\delta_\text{E1}$ & $-0.48(15)$ MeV \\
    $f_0$ & $1.4(11)\cdot10^{-10}$ MeV$^{-4}$ \\
    $\epsilon_0$ & $3.55(93)$ MeV \\
    $U$ & $4.44(52)$ MeV \\
    $E_\text{sf}$ & $10.22(63)$ MeV \\
    $\Gamma_\text{sf}$ & $3.85(35)$ MeV \\
    $\sigma_\text{sf}$ & $0.993(89)$ mb \\
    $C$ & $5.8(11)\cdot10^{-8}$ MeV$^{-3}$ \\
    $\eta$ & $0.923(48)$ MeV$^{-1}$ \\ \hline
    \end{tabular}
    \label{tab:CT_QRPA_M1}
\end{table}

% \begin{table}
%     \centering
%     \caption{Posterior median values of the parameters for CT NLD model and QRPA+PDR+M1 $\gamma$SF model.}
%     \begin{tabular}{c|c}
%     Parameter & Value \\ \hline
%     $A$ & $1.99(23)$ MeV$^{-1}$ \\
%     $B$ & $0.49(14)$ MeV$^{-3}$ \\
%     $\alpha$ & $-0.060(53)$ MeV$^{-1}$ \\ \hline
%     $T$ & $1.01(18)$ MeV \\
%     $\delta$ & $-1.07(89)$ MeV \\
%     $\sigma_D$ & $2.51(21)$ \\
%     $\sigma(Sn)$ & $3.80(33)$ \\ \hline
%     $c_\text{E1}$ & $0.797(43)$ \\
%     $\delta_\text{E1}$ & $-0.55(15)$ MeV \\
%     $f_0$ & $1.2(11)\cdot10^{-10}$ MeV$^{-4}$ \\
%     $\epsilon_0$ & $3.50(88)$ MeV \\
%     $U$ & $4.39(57)$ MeV \\
%     $E_\text{PDR}$ & $9.68(13)$ MeV \\
%     $\Gamma_\text{PDR}$ & $0.88(16)$ MeV \\
%     $\sigma_\text{PDR}$ & $23.3(40)$ mb \\
%     $E_\text{sf}$ & $9.90(66)$ MeV \\
%     $\Gamma_\text{sf}$ & $3.85(34)$ MeV \\
%     $\sigma_\text{sf}$ & $0.977(85)$ mb \\
%     $C$ & $6.2(12)\cdot10^{-8}$ MeV$^{-3}$ \\
%     $\eta$ & $0.922(51)$ MeV$^{-1}$ \\ \hline
%     \end{tabular}
%     \label{tab:CT_QRPA_PDR_M1}
% \end{table}

\begin{table}
    \centering
    \caption{Posterior mean values of the parameters for CT NLD model and QRPA+PDR+M1 $\gamma$SF model.}
    \begin{tabular}{c|c}
    Parameter & Value \\ \hline
    $A$ & $2.01(23)$ MeV$^{-1}$ \\
    $B$ & $0.52(15)$ MeV$^{-3}$ \\
    $\alpha$ & $-0.063(52)$ MeV$^{-1}$ \\ \hline
    $T$ & $1.05(22)$ MeV \\
    $\delta$ & $-1.3(11)$ MeV \\
    $\sigma_D$ & $2.51(21)$ \\
    $\sigma(Sn)$ & $3.79(34)$ \\ \hline
    $c_\text{E1}$ & $0.796(42)$ \\
    $\delta_\text{E1}$ & $-0.55(15)$ MeV \\
    $f_0$ & $1.5(12)\cdot10^{-10}$ MeV$^{-4}$ \\
    $\epsilon_0$ & $3.52(89)$ MeV \\
    $U$ & $4.42(52)$ MeV \\
    $E_\text{PDR}$ & $9.67(13)$ MeV \\
    $\Gamma_\text{PDR}$ & $0.87(16)$ MeV \\
    $\sigma_\text{PDR}$ & $23.5(40)$ mb \\
    $E_\text{sf}$ & $9.94(66)$ MeV \\
    $\Gamma_\text{sf}$ & $3.84(36)$ MeV \\
    $\sigma_\text{sf}$ & $0.979(86)$ mb \\
    $C$ & $6.3(12)\cdot10^{-8}$ MeV$^{-3}$ \\
    $\eta$ & $0.922(51)$ MeV$^{-1}$ \\ \hline
    \end{tabular}
    \label{tab:CT_QRPA_PDR_M1}
\end{table}

% \begin{table}
%     \centering
%     \caption{Posterior median values of the parameters for BSFG NLD model and SMLO+M1 $\gamma$SF model.}
%     \begin{tabular}{c|c}
%     Parameter & Value \\ \hline
%     $A$ & $1.92(21)$ MeV$^{-1}$ \\
%     $B$ & $0.56(14)$ MeV$^{-3}$ \\
%     $\alpha$ & $-0.045(49)$ MeV$^{-1}$ \\ \hline
%     $a$ & $8.72(84)$ MeV \\
%     $\delta$ & $0.27(32)$ MeV \\
%     $\sigma_D$ & $2.48(22)$ \\
%     $\sigma(Sn)$ & $3.85(34)$ \\ \hline
%     $E_\text{GDR}$ & $17.44(12)$ MeV \\
%     $\Gamma_\text{GDR}$ & $7.48(56)$ MeV \\
%     $\sigma_\text{GDR}$ & $1120(46)$ mb \\
%     $E_\text{sf}$ & $9.95(70)$ MeV \\
%     $\Gamma_\text{sf}$ & $3.89(36)$ MeV \\
%     $\sigma_\text{sf}$ & $0.994(89)$ mb \\
%     $C$ & $7.1(14)\cdot10^{-8}$ MeV$^{-3}$ \\
%     $\eta$ & $0.933(56)$ MeV$^{-1}$ \\ \hline
%     \end{tabular}
%     \label{tab:BSFG_SMLO_M1}
% \end{table}

\begin{table}
    \centering
    \caption{Posterior mean values of the parameters for BSFG NLD model and SMLO+M1 $\gamma$SF model.}
    \begin{tabular}{c|c}
    Parameter & Value \\ \hline
    $A$ & $1.94(21)$ MeV$^{-1}$ \\
    $B$ & $0.58(14)$ MeV$^{-3}$ \\
    $\alpha$ & $-0.046(48)$ MeV$^{-1}$ \\ \hline
    $a$ & $8.72(82)$ MeV \\
    $\delta$ & $0.24(32)$ MeV \\
    $\sigma_D$ & $2.48(23)$ \\
    $\sigma(Sn)$ & $3.85(35)$ \\ \hline
    $E_\text{GDR}$ & $17.44(12)$ MeV \\
    $\Gamma_\text{GDR}$ & $7.50(57)$ MeV \\
    $\sigma_\text{GDR}$ & $1122(46)$ mb \\
    $E_\text{sf}$ & $9.96(69)$ MeV \\
    $\Gamma_\text{sf}$ & $3.88(37)$ MeV \\
    $\sigma_\text{sf}$ & $0.996(92)$ mb \\
    $C$ & $7.2(14)\cdot10^{-8}$ MeV$^{-3}$ \\
    $\eta$ & $0.931(55)$ MeV$^{-1}$ \\ \hline
    \end{tabular}
    \label{tab:BSFG_SMLO_M1}
\end{table}

% \begin{table}
%     \centering
%     \caption{Posterior median values of the parameters for BSFG NLD model and SMLO+PDR+M1 $\gamma$SF model.}
%     \begin{tabular}{c|c}
%     Parameter & Value \\ \hline
%     $A$ & $1.92(21)$ MeV$^{-1}$ \\
%     $B$ & $0.59(14)$ MeV$^{-3}$ \\
%     $\alpha$ & $-0.044(50)$ MeV$^{-1}$ \\ \hline
%     $a$ & $8.65(83)$ MeV \\ 
%     $\delta$ & $0.24(33)$ MeV \\
%     $\sigma_D$ & $2.48(22)$ \\
%     $\sigma(Sn)$ & $3.84(34)$ \\ \hline
%     $E_\text{GDR}$ & $17.42(13)$ MeV \\
%     $\Gamma_\text{GDR}$ & $6.71(62)$ MeV \\
%     $\sigma_\text{GDR}$ & $1083(45)$ mb \\
%     $E_\text{PDR}$ & $9.71(12)$ MeV \\
%     $\Gamma_\text{PDR}$ & $0.94(18)$ MeV \\
%     $\sigma_\text{PDR}$ & $23.6(42)$ mb \\
%     $E_\text{sf}$ & $9.79(71)$ MeV \\
%     $\Gamma_\text{sf}$ & $3.87(35)$ MeV \\
%     $\sigma_\text{sf}$ & $0.986(89)$ mb \\
%     $C$ & $7.38(15)\cdot10^{-8}$ MeV$^{-3}$ \\
%     $\eta$ & $0.935(58)$ MeV$^{-1}$ \\ \hline
%     \end{tabular}
%     \label{tab:BSFG_SMLO_PDR_M1}
% \end{table}

\begin{table}
    \centering
    \caption{Posterior mean values of the parameters for BSFG NLD model and SMLO+PDR+M1 $\gamma$SF model.}
    \begin{tabular}{c|c}
    Parameter & Value \\ \hline
    $A$ & $1.93(20)$ MeV$^{-1}$ \\
    $B$ & $0.61(15)$ MeV$^{-3}$ \\
    $\alpha$ & $-0.044(48)$ MeV$^{-1}$ \\ \hline
    $a$ & $8.64(83)$ MeV \\ 
    $\delta$ & $0.21(34)$ MeV \\
    $\sigma_D$ & $2.48(23)$ \\
    $\sigma(Sn)$ & $3.84(35)$ \\ \hline
    $E_\text{GDR}$ & $17.41(13)$ MeV \\
    $\Gamma_\text{GDR}$ & $6.74(62)$ MeV \\
    $\sigma_\text{GDR}$ & $1084(46)$ mb \\
    $E_\text{PDR}$ & $9.70(13)$ MeV \\
    $\Gamma_\text{PDR}$ & $0.93(18)$ MeV \\
    $\sigma_\text{PDR}$ & $23.9(43)$ mb \\
    $E_\text{sf}$ & $9.81(70)$ MeV \\
    $\Gamma_\text{sf}$ & $3.86(36)$ MeV \\
    $\sigma_\text{sf}$ & $0.986(93)$ mb \\
    $C$ & $7.5(14)\cdot10^{-8}$ MeV$^{-3}$ \\
    $\eta$ & $0.934(57)$ MeV$^{-1}$ \\ \hline
    \end{tabular}
    \label{tab:BSFG_SMLO_PDR_M1}
\end{table}

% \begin{table}
%     \centering
%     \caption{Posterior mean values of the parameters for BSFG NLD model and QRPA+M1 $\gamma$SF model.}
%     \begin{tabular}{c|c}
%     Parameter & Value \\ \hline
%     $A$ & $1.95(24)$ MeV$^{-1}$ \\
%     $B$ & $0.40(12)$ MeV$^{-3}$ \\
%     $\alpha$ & $-0.050(57)$ MeV$^{-1}$ \\ \hline
%     $a$ & $7.89(95)$ MeV \\
%     $\delta$ & $-0.08(45)$ MeV \\
%     $\sigma_D$ & $2.50(22)$ \\
%     $\sigma(Sn)$ & $3.79(34)$ \\ \hline
%     $c_\text{E1}$ & $0.823(45)$ \\
%     $\delta_\text{E1}$ & $-0.50(16)$ MeV \\
%     $f_0$ & $1.21(11)\cdot10^{-10}$ MeV$^{-4}$ \\
%     $\epsilon_0$ & $3.43(96)$ MeV \\
%     $U$ & $4.48(61)$ MeV \\
%     $E_\text{sf}$ & $10.21(67)$ MeV \\
%     $\Gamma_\text{sf}$ & $3.87(35)$ MeV \\
%     $\sigma_\text{sf}$ & $1.000(90)$ mb \\
%     $C$ & $5.31(11)\cdot10^{-8}$ MeV$^{-3}$ \\
%     $\eta$ & $0.927(57)$ MeV$^{-1}$ \\ \hline
%     \end{tabular}
%     \label{tab:BSFG_QRPA_M1}
% \end{table}

\begin{table}
    \centering
    \caption{Posterior mean values of the parameters for BSFG NLD model and QRPA+M1 $\gamma$SF model.}
    \begin{tabular}{c|c}
    Parameter & Value \\ \hline
    $A$ & $1.97(24)$ MeV$^{-1}$ \\
    $B$ & $0.43(13)$ MeV$^{-3}$ \\
    $\alpha$ & $-0.052(56)$ MeV$^{-1}$ \\ \hline
    $a$ & $7.88(93)$ MeV \\
    $\delta$ & $-0.14(47)$ MeV \\
    $\sigma_D$ & $2.50(23)$ \\
    $\sigma(Sn)$ & $3.79(35)$ \\ \hline
    $c_\text{E1}$ & $0.822(45)$ \\
    $\delta_\text{E1}$ & $-0.49(16)$ MeV \\
    $f_0$ & $1.5(12)\cdot10^{-10}$ MeV$^{-4}$ \\
    $\epsilon_0$ & $3.47(98)$ MeV \\
    $U$ & $4.49(54)$ MeV \\
    $E_\text{sf}$ & $10.20(66)$ MeV \\
    $\Gamma_\text{sf}$ & $3.86(36)$ MeV \\
    $\sigma_\text{sf}$ & $0.998(93)$ mb \\
    $C$ & $5.4(12)\cdot10^{-8}$ MeV$^{-3}$ \\
    $\eta$ & $0.924(56)$ MeV$^{-1}$ \\ \hline
    \end{tabular}
    \label{tab:BSFG_QRPA_M1}
\end{table}

% \begin{table}
%     \centering
%     \caption{Posterior median values of the parameters for BSFG NLD model and QRPA+PDR+M1 $\gamma$SF model.}
%     \begin{tabular}{c|c}
%     Parameter & Value \\ \hline
%     $A$ & $1.93(22)$ MeV$^{-1}$ \\
%     $B$ & $0.46(12)$ MeV$^{-3}$ \\
%     $\alpha$ & $-0.047(53)$ MeV$^{-1}$ \\ \hline
%     $a$ & $7.86(94)$ MeV \\
%     $\delta$ & $-0.10(45)$ MeV \\
%     $\sigma_D$ & $2.50(22)$ \\
%     $\sigma(Sn)$ & $3.79(34)$ \\ \hline
%     $c_\text{E1}$ & $0.799(44)$ \\
%     $\delta_\text{E1}$ & $-0.56(16)$ MeV \\
%     $f_0$ & $1.2(11)\cdot10^{-10}$ MeV$^{-4}$ \\
%     $\epsilon_0$ & $3.45(94)$ MeV \\
%     $U$ & $4.41(59)$ MeV \\
%     $E_\text{PDR}$ & $9.68(14)$ MeV \\
%     $\Gamma_\text{PDR}$ & $0.88(16)$ MeV \\
%     $\sigma_\text{PDR}$ & $23.1(38)$ mb \\
%     $E_\text{sf}$ & $9.90(68)$ MeV \\
%     $\Gamma_\text{sf}$ & $3.85(35)$ MeV \\
%     $\sigma_\text{sf}$ & $0.982(89)$ mb \\
%     $C$ & $5.95(120)\cdot10^{-8}$ MeV$^{-3}$ \\
%     $\eta$ & $0.924(55)$ MeV$^{-1}$ \\ \hline
%     \end{tabular}
%     \label{tab:BSFG_QRPA_PDR_M1}
% \end{table}

\begin{table}
    \centering
    \caption{Posterior mean values of the parameters for BSFG NLD model and QRPA+PDR+M1 $\gamma$SF model.}
    \begin{tabular}{c|c}
    Parameter & Value \\ \hline
    $A$ & $1.95(22)$ MeV$^{-1}$ \\
    $B$ & $0.48(13)$ MeV$^{-3}$ \\
    $\alpha$ & $-0.047(52)$ MeV$^{-1}$ \\ \hline
    $a$ & $7.85(91)$ MeV \\
    $\delta$ & $-0.16(47)$ MeV \\
    $\sigma_D$ & $2.50(23)$ \\
    $\sigma(Sn)$ & $3.79(35)$ \\ \hline
    $c_\text{E1}$ & $0.799(43)$ \\
    $\delta_\text{E1}$ & $-0.55(16)$ MeV \\
    $f_0$ & $1.4(12)\cdot10^{-10}$ MeV$^{-4}$ \\
    $\epsilon_0$ & $3.47(95)$ MeV \\
    $U$ & $4.43(53)$ MeV \\
    $E_\text{PDR}$ & $9.66(14)$ MeV \\
    $\Gamma_\text{PDR}$ & $0.87(16)$ MeV \\
    $\sigma_\text{PDR}$ & $23.3(40)$ mb \\
    $E_\text{sf}$ & $9.92(68)$ MeV \\
    $\Gamma_\text{sf}$ & $3.84(36)$ MeV \\
    $\sigma_\text{sf}$ & $0.982(92)$ mb \\
    $C$ & $6.0(12)\cdot10^{-8}$ MeV$^{-3}$ \\
    $\eta$ & $0.922(54)$ MeV$^{-1}$ \\ \hline
    \end{tabular}
    \label{tab:BSFG_QRPA_PDR_M1}
\end{table}

% \begin{table}
%     \centering
%     \caption{Posterior median values of the parameters for HFB NLD model and SMLO+M1 $\gamma$SF model.}
%     \begin{tabular}{c|c}
%     Parameter & Value \\ \hline
%     $A$ & $1.93(20)$ MeV$^{-1}$ \\
%     $B$ & $0.60(15)$ MeV$^{-3}$ \\
%     $\alpha$ & $-0.048(49)$ MeV$^{-1}$ \\ \hline
%     $c$ & $0.42(28)$ MeV$^{-1/2}$ \\
%     $\delta$ & $0.26(37)$ MeV \\ \hline
%     $E_\text{GDR}$ & $17.43(12)$ MeV \\
%     $\Gamma_\text{GDR}$ & $7.23(55)$ MeV \\
%     $\sigma_\text{GDR}$ & $1118(45)$ mb \\
%     $E_\text{sf}$ & $9.90(69)$ MeV \\
%     $\Gamma_\text{sf}$ & $3.88(34)$ MeV \\
%     $\sigma_\text{sf}$ & $0.988(88)$ mb \\
%     $C$ & $7.4(14)\cdot10^{-8}$ MeV$^{-3}$ \\
%     $\eta$ & $0.937(57)$ MeV$^{-1}$ \\ \hline
%     \end{tabular}
%     \label{tab:HFB_SMLO_M1}
% \end{table}

\begin{table}
    \centering
    \caption{Posterior mean values of the parameters for HFB NLD model and SMLO+M1 $\gamma$SF model.}
    \begin{tabular}{c|c}
    Parameter & Value \\ \hline
    $A$ & $1.94(20)$ MeV$^{-1}$ \\
    $B$ & $0.62(15)$ MeV$^{-3}$ \\
    $\alpha$ & $-0.047(48)$ MeV$^{-1}$ \\ \hline
    $c$ & $0.41(28)$ MeV$^{-1/2}$ \\
    $\delta$ & $0.23(37)$ MeV \\ \hline
    $E_\text{GDR}$ & $17.43(12)$ MeV \\
    $\Gamma_\text{GDR}$ & $7.25(56)$ MeV \\
    $\sigma_\text{GDR}$ & $1121(44)$ mb \\
    $E_\text{sf}$ & $9.92(68)$ MeV \\
    $\Gamma_\text{sf}$ & $3.89(35)$ MeV \\
    $\sigma_\text{sf}$ & $0.991(92)$ mb \\
    $C$ & $7.4(14)\cdot10^{-8}$ MeV$^{-3}$ \\
    $\eta$ & $0.935(57)$ MeV$^{-1}$ \\ \hline
    \end{tabular}
    \label{tab:HFB_SMLO_M1}
\end{table}

% \begin{table}
%     \centering
%     \caption{Posterior median values of the parameters for HFB NLD model and SMLO+PDR+M1 $\gamma$SF model.}
%     \begin{tabular}{c|c}
%     Parameter & Value \\ \hline
%     $A$ & $1.92(19)$ MeV$^{-1}$ \\
%     $B$ & $0.62(14)$ MeV$^{-3}$ \\
%     $\alpha$ & $-0.045(46)$ MeV$^{-1}$ \\ \hline
%     $c$ & $0.38(27)$ MeV$^{-1/2}$ \\
%     $\delta$ & $0.21(37)$ MeV \\ \hline
%     $E_\text{GDR}$ & $17.42(12)$ MeV \\
%     $\Gamma_\text{GDR}$ & $6.50(57)$ MeV \\
%     $\sigma_\text{GDR}$ & $1080(45)$ mb \\
%     $E_\text{PDR}$ & $9.71(12)$ MeV \\
%     $\Gamma_\text{PDR}$ & $0.94(17)$ MeV \\
%     $\sigma_\text{PDR}$ & $23.8(38)$ mb \\
%     $E_\text{sf}$ & $9.73(68)$ MeV \\
%     $\Gamma_\text{sf}$ & $3.87(34)$ MeV \\
%     $\sigma_\text{sf}$ & $0.989(85)$ mb \\
%     $C$ & $7.51(138)\cdot10^{-8}$ MeV$^{-3}$ \\
%     $\eta$ & $0.931(56)$ MeV$^{-1}$ \\ \hline
%     \end{tabular}
%     \label{tab:HFB_SMLO_PDR_M1}
% \end{table}

\begin{table}
    \centering
    \caption{Posterior mean values of the parameters for HFB NLD model and SMLO+PDR+M1 $\gamma$SF model.}
    \begin{tabular}{c|c}
    Parameter & Value \\ \hline
    $A$ & $1.93(19)$ MeV$^{-1}$ \\
    $B$ & $0.64(15)$ MeV$^{-3}$ \\
    $\alpha$ & $-0.045(45)$ MeV$^{-1}$ \\ \hline
    $c$ & $0.38(27)$ MeV$^{-1/2}$ \\
    $\delta$ & $0.18(38)$ MeV \\ \hline
    $E_\text{GDR}$ & $17.42(12)$ MeV \\
    $\Gamma_\text{GDR}$ & $6.52(58)$ MeV \\
    $\sigma_\text{GDR}$ & $1082(46)$ mb \\
    $E_\text{PDR}$ & $9.69(13)$ MeV \\
    $\Gamma_\text{PDR}$ & $0.93(18)$ MeV \\
    $\sigma_\text{PDR}$ & $24.0(39)$ mb \\
    $E_\text{sf}$ & $9.77(69)$ MeV \\
    $\Gamma_\text{sf}$ & $3.88(36)$ MeV \\
    $\sigma_\text{sf}$ & $0.988(89)$ mb \\
    $C$ & $7.6(14)\cdot10^{-8}$ MeV$^{-3}$ \\
    $\eta$ & $0.930(56)$ MeV$^{-1}$ \\ \hline
    \end{tabular}
    \label{tab:HFB_SMLO_PDR_M1}
\end{table}

% \begin{table}
%     \centering
%     \caption{Posterior median values of the parameters for HFB NLD model and QRPA+M1 $\gamma$SF model.}
%     \begin{tabular}{c|c}
%     Parameter & Value \\ \hline
%     $A$ & $1.94(23)$ MeV$^{-1}$ \\
%     $B$ & $0.40(11)$ MeV$^{-3}$ \\
%     $\alpha$ & $-0.049(54)$ MeV$^{-1}$ \\ \hline
%     $c$ & $0.10(33)$ MeV$^{-1/2}$ \\
%     $\delta$ & $-0.19(51)$ MeV \\ \hline
%     $c_\text{E1}$ & $0.820(45)$ \\
%     $\delta_\text{E1}$ & $-0.49(16)$ MeV \\
%     $f_0$ & $1.2(11)\cdot10^{-10}$ MeV$^{-4}$ \\
%     $\epsilon_0$ & $3.48(94)$ MeV \\
%     $U$ & $4.39(59)$ MeV \\
%     $E_\text{sf}$ & $10.18(65)$ MeV \\
%     $\Gamma_\text{sf}$ & $3.85(35)$ MeV \\
%     $\sigma_\text{sf}$ & $0.993(87)$ mb \\
%     $C$ & $5.3(11)\cdot10^{-8}$ MeV$^{-3}$ \\
%     $\eta$ & $0.922(52)$ MeV$^{-1}$ \\ \hline
%     \end{tabular}
%     \label{tab:HFB_QRPA_M1}
% \end{table}

\begin{table}
    \centering
    \caption{Posterior mean values of the parameters for HFB NLD model and QRPA+M1 $\gamma$SF model.}
    \begin{tabular}{c|c}
    Parameter & Value \\ \hline
    $A$ & $1.96(23)$ MeV$^{-1}$ \\
    $B$ & $0.42(12)$ MeV$^{-3}$ \\
    $\alpha$ & $-0.050(52)$ MeV$^{-1}$ \\ \hline
    $c$ & $0.10(32)$ MeV$^{-1/2}$ \\
    $\delta$ & $-0.23(50)$ MeV \\ \hline
    $c_\text{E1}$ & $0.820(46)$ \\
    $\delta_\text{E1}$ & $-0.49(16)$ MeV \\
    $f_0$ & $1.4(11)\cdot10^{-10}$ MeV$^{-4}$ \\
    $\epsilon_0$ & $3.50(94)$ MeV \\
    $U$ & $4.43(53)$ MeV \\
    $E_\text{sf}$ & $10.17(65)$ MeV \\
    $\Gamma_\text{sf}$ & $3.85(37)$ MeV \\
    $\sigma_\text{sf}$ & $0.993(92)$ mb \\
    $C$ & $5.4(11)\cdot10^{-8}$ MeV$^{-3}$ \\
    $\eta$ & $0.921(52)$ MeV$^{-1}$ \\ \hline
    \end{tabular}
    \label{tab:HFB_QRPA_M1}
\end{table}

% \begin{table}
%     \centering
%     \caption{Posterior median values of the parameters for HFB NLD model and QRPA+PDR+M1 $\gamma$SF model.}
%     \begin{tabular}{c|c}
%     Parameter & Value \\ \hline
%     $A$ & $1.93(19)$ MeV$^{-1}$ \\
%     $B$ & $0.45(11)$ MeV$^{-3}$ \\
%     $\alpha$ & $-0.045(48)$ MeV$^{-1}$ \\ \hline
%     $c$ & $0.09(31)$ MeV$^{-1/2}$ \\
%     $\delta$ & $-0.21(49)$ MeV \\ \hline
%     $c_\text{E1}$ & $0.797(45)$ \\
%     $\delta_\text{E1}$ & $-0.56(16)$ MeV \\
%     $f_0$ & $1.2(11)\cdot10^{-10}$ MeV$^{-4}$ \\
%     $\epsilon_0$ & $3.51(89)$ MeV \\
%     $U$ & $4.45(59)$ MeV \\
%     $E_\text{PDR}$ & $9.69(12)$ MeV \\
%     $\Gamma_\text{PDR}$ & $0.87(16)$ MeV \\
%     $\sigma_\text{PDR}$ & $23.5(41)$ mb \\
%     $E_\text{sf}$ & $9.89(67)$ MeV \\
%     $\Gamma_\text{sf}$ & $3.85(34)$ MeV \\
%     $\sigma_\text{sf}$ & $0.979(90)$ mb \\
%     $C$ & $5.8(11)\cdot10^{-8}$ MeV$^{-3}$ \\
%     $\eta$ & $0.920(54)$ MeV$^{-1}$ \\ \hline
%     \end{tabular}
%     \label{tab:HFB_QRPA_PDR_M1}
% \end{table}

\begin{table}
    \centering
    \caption{Posterior median values of the parameters for HFB NLD model and QRPA+PDR+M1 $\gamma$SF model.}
    \begin{tabular}{c|c}
    Parameter & Value \\ \hline
    $A$ & $1.94(20)$ MeV$^{-1}$ \\
    $B$ & $0.48(12)$ MeV$^{-3}$ \\
    $\alpha$ & $-0.047(47)$ MeV$^{-1}$ \\ \hline
    $c$ & $0.10(30)$ MeV$^{-1/2}$ \\
    $\delta$ & $-0.24(48)$ MeV \\ \hline
    $c_\text{E1}$ & $0.795(45)$ \\
    $\delta_\text{E1}$ & $-0.55(16)$ MeV \\
    $f_0$ & $1.4(12)\cdot10^{-10}$ MeV$^{-4}$ \\
    $\epsilon_0$ & $3.51(93)$ MeV \\
    $U$ & $4.47(53)$ MeV \\
    $E_\text{PDR}$ & $9.68(13)$ MeV \\
    $\Gamma_\text{PDR}$ & $0.87(16)$ MeV \\
    $\sigma_\text{PDR}$ & $23.6(41)$ mb \\
    $E_\text{sf}$ & $9.92(67)$ MeV \\
    $\Gamma_\text{sf}$ & $3.85(36)$ MeV \\
    $\sigma_\text{sf}$ & $0.980(92)$ mb \\
    $C$ & $5.9(11)\cdot10^{-8}$ MeV$^{-3}$ \\
    $\eta$ & $0.919(55)$ MeV$^{-1}$ \\ \hline
    \end{tabular}
    \label{tab:HFB_QRPA_PDR_M1}
\end{table}

\clearpage
\bibliography{ref.bib}

\begin{thebibliography}{95}%
\makeatletter
\providecommand \@ifxundefined [1]{%
 \@ifx{#1\undefined}
}%
\providecommand \@ifnum [1]{%
 \ifnum #1\expandafter \@firstoftwo
 \else \expandafter \@secondoftwo
 \fi
}%
\providecommand \@ifx [1]{%
 \ifx #1\expandafter \@firstoftwo
 \else \expandafter \@secondoftwo
 \fi
}%
\providecommand \natexlab [1]{#1}%
\providecommand \enquote  [1]{``#1''}%
\providecommand \bibnamefont  [1]{#1}%
\providecommand \bibfnamefont [1]{#1}%
\providecommand \citenamefont [1]{#1}%
\providecommand \href@noop [0]{\@secondoftwo}%
\providecommand \href [0]{\begingroup \@sanitize@url \@href}%
\providecommand \@href[1]{\@@startlink{#1}\@@href}%
\providecommand \@@href[1]{\endgroup#1\@@endlink}%
\providecommand \@sanitize@url [0]{\catcode `\\12\catcode `\$12\catcode
  `\&12\catcode `\#12\catcode `\^12\catcode `\_12\catcode `\%12\relax}%
\providecommand \@@startlink[1]{}%
\providecommand \@@endlink[0]{}%
\providecommand \url  [0]{\begingroup\@sanitize@url \@url }%
\providecommand \@url [1]{\endgroup\@href {#1}{\urlprefix }}%
\providecommand \urlprefix  [0]{URL }%
\providecommand \Eprint [0]{\href }%
\providecommand \doibase [0]{https://doi.org/}%
\providecommand \selectlanguage [0]{\@gobble}%
\providecommand \bibinfo  [0]{\@secondoftwo}%
\providecommand \bibfield  [0]{\@secondoftwo}%
\providecommand \translation [1]{[#1]}%
\providecommand \BibitemOpen [0]{}%
\providecommand \bibitemStop [0]{}%
\providecommand \bibitemNoStop [0]{.\EOS\space}%
\providecommand \EOS [0]{\spacefactor3000\relax}%
\providecommand \BibitemShut  [1]{\csname bibitem#1\endcsname}%
\let\auto@bib@innerbib\@empty
%</preamble>
\bibitem [{\citenamefont {Merrill}(1952)}]{Merrill1952}%
  \BibitemOpen
  \bibfield  {author} {\bibinfo {author} {\bibfnamefont {P.}~\bibnamefont
  {Merrill}},\ }\bibfield  {title} {\bibinfo {title} {{Technetium in the
  stars}},\ }\href@noop {} {\bibfield  {journal} {\bibinfo  {journal}
  {Science}\ }\textbf {\bibinfo {volume} {115}},\ \bibinfo {pages} {479}
  (\bibinfo {year} {1952})}\BibitemShut {NoStop}%
\bibitem [{\citenamefont {Burbidge}\ \emph {et~al.}(1957)\citenamefont
  {Burbidge}, \citenamefont {Burbidge}, \citenamefont {Fowler},\ and\
  \citenamefont {Hoyle}}]{RevModPhys.29.547}%
  \BibitemOpen
  \bibfield  {author} {\bibinfo {author} {\bibfnamefont {E.~M.}\ \bibnamefont
  {Burbidge}}, \bibinfo {author} {\bibfnamefont {G.~R.}\ \bibnamefont
  {Burbidge}}, \bibinfo {author} {\bibfnamefont {W.~A.}\ \bibnamefont
  {Fowler}},\ and\ \bibinfo {author} {\bibfnamefont {F.}~\bibnamefont
  {Hoyle}},\ }\bibfield  {title} {\bibinfo {title} {{Synthesis of the elements
  in stars}},\ }\href {https://doi.org/10.1103/RevModPhys.29.547} {\bibfield
  {journal} {\bibinfo  {journal} {Reviews of Modern Physics}\ }\textbf
  {\bibinfo {volume} {29}},\ \bibinfo {pages} {547} (\bibinfo {year}
  {1957})}\BibitemShut {NoStop}%
\bibitem [{\citenamefont {Cowan}\ and\ \citenamefont {Rose}(1977)}]{Cowan1977}%
  \BibitemOpen
  \bibfield  {author} {\bibinfo {author} {\bibfnamefont {J.~J.}\ \bibnamefont
  {Cowan}}\ and\ \bibinfo {author} {\bibfnamefont {W.~K.}\ \bibnamefont
  {Rose}},\ }\bibfield  {title} {\bibinfo {title} {{Production of C-14 and
  neutrons in red giants}},\ }\href {https://doi.org/10.1086/155030} {\bibfield
   {journal} {\bibinfo  {journal} {The Astrophysical Journal}\ }\textbf
  {\bibinfo {volume} {212}},\ \bibinfo {pages} {149} (\bibinfo {year}
  {1977})}\BibitemShut {NoStop}%
\bibitem [{\citenamefont {Roederer}\ \emph {et~al.}(2016)\citenamefont
  {Roederer}, \citenamefont {Karakas}, \citenamefont {Pignatari},\ and\
  \citenamefont {Herwig}}]{Roederer2016}%
  \BibitemOpen
  \bibfield  {author} {\bibinfo {author} {\bibfnamefont {I.~U.}\ \bibnamefont
  {Roederer}}, \bibinfo {author} {\bibfnamefont {A.~I.}\ \bibnamefont
  {Karakas}}, \bibinfo {author} {\bibfnamefont {M.}~\bibnamefont {Pignatari}},\
  and\ \bibinfo {author} {\bibfnamefont {F.}~\bibnamefont {Herwig}},\
  }\bibfield  {title} {\bibinfo {title} {{ The Diverse Origines of
  Neutron-capture Elements in The Metal-poor Star HD 94028: Possible Detection
  of Products of i-process Nucleosynthesis}},\ }\href
  {https://doi.org/10.3847/0004-637x/821/1/37} {\bibfield  {journal} {\bibinfo
  {journal} {The Astrophysical Journal}\ }\textbf {\bibinfo {volume} {821}},\
  \bibinfo {pages} {37} (\bibinfo {year} {2016})}\BibitemShut {NoStop}%
\bibitem [{\citenamefont {Goriely}\ \emph {et~al.}(2019)\citenamefont
  {Goriely}, \citenamefont {Dimitriou}, \citenamefont {Wiedeking},
  \citenamefont {Belgya}, \citenamefont {Firestone}, \citenamefont {Kopecky},
  \citenamefont {Krti{\v{c}}ka}, \citenamefont {Plujko}, \citenamefont
  {Schwengner}, \citenamefont {Siem}, \citenamefont {Utsunomiya}, \citenamefont
  {Hilaire}, \citenamefont {P{\'{e}}ru}, \citenamefont {Cho}, \citenamefont
  {Filipescu}, \citenamefont {Iwamoto}, \citenamefont {Kawano}, \citenamefont
  {Varlamov},\ and\ \citenamefont {Xu}}]{GorielyCRP2019}%
  \BibitemOpen
  \bibfield  {author} {\bibinfo {author} {\bibfnamefont {S.}~\bibnamefont
  {Goriely}}, \bibinfo {author} {\bibfnamefont {P.}~\bibnamefont {Dimitriou}},
  \bibinfo {author} {\bibfnamefont {M.}~\bibnamefont {Wiedeking}}, \bibinfo
  {author} {\bibfnamefont {T.}~\bibnamefont {Belgya}}, \bibinfo {author}
  {\bibfnamefont {R.}~\bibnamefont {Firestone}}, \bibinfo {author}
  {\bibfnamefont {J.}~\bibnamefont {Kopecky}}, \bibinfo {author} {\bibfnamefont
  {M.}~\bibnamefont {Krti{\v{c}}ka}}, \bibinfo {author} {\bibfnamefont
  {V.}~\bibnamefont {Plujko}}, \bibinfo {author} {\bibfnamefont
  {R.}~\bibnamefont {Schwengner}}, \bibinfo {author} {\bibfnamefont
  {S.}~\bibnamefont {Siem}}, \bibinfo {author} {\bibfnamefont {H.}~\bibnamefont
  {Utsunomiya}}, \bibinfo {author} {\bibfnamefont {S.}~\bibnamefont {Hilaire}},
  \bibinfo {author} {\bibfnamefont {S.}~\bibnamefont {P{\'{e}}ru}}, \bibinfo
  {author} {\bibfnamefont {Y.~S.}\ \bibnamefont {Cho}}, \bibinfo {author}
  {\bibfnamefont {D.~M.}\ \bibnamefont {Filipescu}}, \bibinfo {author}
  {\bibfnamefont {N.}~\bibnamefont {Iwamoto}}, \bibinfo {author} {\bibfnamefont
  {T.}~\bibnamefont {Kawano}}, \bibinfo {author} {\bibfnamefont
  {V.}~\bibnamefont {Varlamov}},\ and\ \bibinfo {author} {\bibfnamefont
  {R.}~\bibnamefont {Xu}},\ }\bibfield  {title} {\bibinfo {title} {{Reference
  database for photon strength functions}},\ }\href
  {https://doi.org/10.1140/epja/i2019-12840-1} {\bibfield  {journal} {\bibinfo
  {journal} {European Physical Journal A}\ }\textbf {\bibinfo {volume} {55}},\
  \bibinfo {pages} {172} (\bibinfo {year} {2019})}\BibitemShut {NoStop}%
\bibitem [{\citenamefont {Larsen}\ \emph {et~al.}(2019)\citenamefont {Larsen},
  \citenamefont {Spyrou}, \citenamefont {Liddick},\ and\ \citenamefont
  {Guttormsen}}]{Larsen2019}%
  \BibitemOpen
  \bibfield  {author} {\bibinfo {author} {\bibfnamefont {A.~C.}\ \bibnamefont
  {Larsen}}, \bibinfo {author} {\bibfnamefont {A.}~\bibnamefont {Spyrou}},
  \bibinfo {author} {\bibfnamefont {S.~N.}\ \bibnamefont {Liddick}},\ and\
  \bibinfo {author} {\bibfnamefont {M.}~\bibnamefont {Guttormsen}},\ }\href
  {https://doi.org/10.1016/j.ppnp.2019.04.002} {\bibinfo {title} {{Novel
  techniques for constraining neutron-capture rates relevant for r-process
  heavy-element nucleosynthesis}}} (\bibinfo {year} {2019})\BibitemShut
  {NoStop}%
\bibitem [{\citenamefont {Wiedeking}\ and\ \citenamefont
  {Goriely}(2024)}]{Wiedeking24}%
  \BibitemOpen
  \bibfield  {author} {\bibinfo {author} {\bibfnamefont {M.}~\bibnamefont
  {Wiedeking}}\ and\ \bibinfo {author} {\bibfnamefont {S.}~\bibnamefont
  {Goriely}},\ }\href@noop {} {\bibfield  {journal} {\bibinfo  {journal} {Phil.
  Trans. R. Soc. A.}\ }\textbf {\bibinfo {volume} {382}},\ \bibinfo {pages}
  {20230125} (\bibinfo {year} {2024})}\BibitemShut {NoStop}%
\bibitem [{\citenamefont {Schiller}\ \emph {et~al.}(2000)\citenamefont
  {Schiller}, \citenamefont {Bergholt}, \citenamefont {Guttormsen},
  \citenamefont {Melby}, \citenamefont {Rekstad},\ and\ \citenamefont
  {Siem}}]{OsloMethodNIM}%
  \BibitemOpen
  \bibfield  {author} {\bibinfo {author} {\bibfnamefont {A.}~\bibnamefont
  {Schiller}}, \bibinfo {author} {\bibfnamefont {L.}~\bibnamefont {Bergholt}},
  \bibinfo {author} {\bibfnamefont {M.}~\bibnamefont {Guttormsen}}, \bibinfo
  {author} {\bibfnamefont {E.}~\bibnamefont {Melby}}, \bibinfo {author}
  {\bibfnamefont {J.}~\bibnamefont {Rekstad}},\ and\ \bibinfo {author}
  {\bibfnamefont {S.}~\bibnamefont {Siem}},\ }\bibfield  {title} {\bibinfo
  {title} {{Extraction of level density and {$\gamma$} strength function from
  primary {$\gamma$} spectra}},\ }\href
  {https://doi.org/10.1016/S0168-9002(99)01187-0} {\bibfield  {journal}
  {\bibinfo  {journal} {Nuclear Instruments and Methods in Physics Research,
  Section A: Accelerators, Spectrometers, Detectors and Associated Equipment}\
  }\textbf {\bibinfo {volume} {447}},\ \bibinfo {pages} {498} (\bibinfo {year}
  {2000})}\BibitemShut {NoStop}%
\bibitem [{\citenamefont {Midtb{\o}}\ \emph {et~al.}(2021)\citenamefont
  {Midtb{\o}}, \citenamefont {Zeiser}, \citenamefont {Lima}, \citenamefont
  {Larsen}, \citenamefont {Tveten}, \citenamefont {Guttormsen}, \citenamefont
  {Bello~Garrote}, \citenamefont {Kvellestad},\ and\ \citenamefont
  {Renstr{\o}m}}]{Midtb2019}%
  \BibitemOpen
  \bibfield  {author} {\bibinfo {author} {\bibfnamefont {J.~E.}\ \bibnamefont
  {Midtb{\o}}}, \bibinfo {author} {\bibfnamefont {F.}~\bibnamefont {Zeiser}},
  \bibinfo {author} {\bibfnamefont {E.}~\bibnamefont {Lima}}, \bibinfo {author}
  {\bibfnamefont {A.~C.}\ \bibnamefont {Larsen}}, \bibinfo {author}
  {\bibfnamefont {G.~M.}\ \bibnamefont {Tveten}}, \bibinfo {author}
  {\bibfnamefont {M.}~\bibnamefont {Guttormsen}}, \bibinfo {author}
  {\bibfnamefont {F.~L.}\ \bibnamefont {Bello~Garrote}}, \bibinfo {author}
  {\bibfnamefont {A.}~\bibnamefont {Kvellestad}},\ and\ \bibinfo {author}
  {\bibfnamefont {T.}~\bibnamefont {Renstr{\o}m}},\ }\bibfield  {title}
  {\bibinfo {title} {{A new software implementation of the Oslo method with
  rigorous statistical uncertainty propagation}},\ }\bibfield  {journal}
  {\bibinfo  {journal} {Computer Physics Communications}\ }\textbf {\bibinfo
  {volume} {262}},\ \href {https://doi.org/10.1016/j.cpc.2020.107795}
  {10.1016/j.cpc.2020.107795} (\bibinfo {year} {2021})\BibitemShut {NoStop}%
\bibitem [{\citenamefont {Hauser}\ and\ \citenamefont
  {Feshbach}(1952)}]{PhysRev.87.366}%
  \BibitemOpen
  \bibfield  {author} {\bibinfo {author} {\bibfnamefont {W.}~\bibnamefont
  {Hauser}}\ and\ \bibinfo {author} {\bibfnamefont {H.}~\bibnamefont
  {Feshbach}},\ }\bibfield  {title} {\bibinfo {title} {{The Inelastic
  Scattering of Neutrons}},\ }\href {https://doi.org/10.1103/PhysRev.87.366}
  {\bibfield  {journal} {\bibinfo  {journal} {Phys. Rev.}\ }\textbf {\bibinfo
  {volume} {87}},\ \bibinfo {pages} {366} (\bibinfo {year} {1952})}\BibitemShut
  {NoStop}%
\bibitem [{\citenamefont {Spyrou}\ \emph {et~al.}(2014)\citenamefont {Spyrou},
  \citenamefont {Liddick}, \citenamefont {Larsen}, \citenamefont {Guttormsen},
  \citenamefont {Cooper}, \citenamefont {Dombos}, \citenamefont {Morrissey},
  \citenamefont {Naqvi}, \citenamefont {Perdikakis}, \citenamefont {Quinn},
  \citenamefont {Renstr\o{}m}, \citenamefont {Rodriguez}, \citenamefont
  {Simon}, \citenamefont {Sumithrarachchi},\ and\ \citenamefont
  {Zegers}}]{PhysRevLett.113.232502}%
  \BibitemOpen
  \bibfield  {author} {\bibinfo {author} {\bibfnamefont {A.}~\bibnamefont
  {Spyrou}}, \bibinfo {author} {\bibfnamefont {S.~N.}\ \bibnamefont {Liddick}},
  \bibinfo {author} {\bibfnamefont {A.~C.}\ \bibnamefont {Larsen}}, \bibinfo
  {author} {\bibfnamefont {M.}~\bibnamefont {Guttormsen}}, \bibinfo {author}
  {\bibfnamefont {K.}~\bibnamefont {Cooper}}, \bibinfo {author} {\bibfnamefont
  {A.~C.}\ \bibnamefont {Dombos}}, \bibinfo {author} {\bibfnamefont {D.~J.}\
  \bibnamefont {Morrissey}}, \bibinfo {author} {\bibfnamefont {F.}~\bibnamefont
  {Naqvi}}, \bibinfo {author} {\bibfnamefont {G.}~\bibnamefont {Perdikakis}},
  \bibinfo {author} {\bibfnamefont {S.~J.}\ \bibnamefont {Quinn}}, \bibinfo
  {author} {\bibfnamefont {T.}~\bibnamefont {Renstr\o{}m}}, \bibinfo {author}
  {\bibfnamefont {J.~A.}\ \bibnamefont {Rodriguez}}, \bibinfo {author}
  {\bibfnamefont {A.}~\bibnamefont {Simon}}, \bibinfo {author} {\bibfnamefont
  {C.~S.}\ \bibnamefont {Sumithrarachchi}},\ and\ \bibinfo {author}
  {\bibfnamefont {R.~G.~T.}\ \bibnamefont {Zegers}},\ }\bibfield  {title}
  {\bibinfo {title} {{Novel technique for constraining r -process (n,
  {$\gamma$}) reaction rates}},\ }\href
  {https://doi.org/10.1103/PhysRevLett.113.232502} {\bibfield  {journal}
  {\bibinfo  {journal} {Physical Review Letters}\ }\textbf {\bibinfo {volume}
  {113}},\ \bibinfo {pages} {232502} (\bibinfo {year} {2014})}\BibitemShut
  {NoStop}%
\bibitem [{\citenamefont {Larsen}\ \emph {et~al.}(2018)\citenamefont {Larsen},
  \citenamefont {Midtb{\o}}, \citenamefont {Guttormsen}, \citenamefont
  {Renstr{\o}m}, \citenamefont {Liddick}, \citenamefont {Spyrou}, \citenamefont
  {Karampagia}, \citenamefont {Brown}, \citenamefont {Achakovskiy},
  \citenamefont {Kamerdzhiev}, \citenamefont {Bleuel}, \citenamefont {Couture},
  \citenamefont {Campo}, \citenamefont {Crider}, \citenamefont {Dombos},
  \citenamefont {Lewis}, \citenamefont {Mosby}, \citenamefont {Naqvi},
  \citenamefont {Perdikakis}, \citenamefont {Prokop}, \citenamefont {Quinn},\
  and\ \citenamefont {Siem}}]{PhysRevC.97.054329}%
  \BibitemOpen
  \bibfield  {author} {\bibinfo {author} {\bibfnamefont {A.~C.}\ \bibnamefont
  {Larsen}}, \bibinfo {author} {\bibfnamefont {J.~E.}\ \bibnamefont
  {Midtb{\o}}}, \bibinfo {author} {\bibfnamefont {M.}~\bibnamefont
  {Guttormsen}}, \bibinfo {author} {\bibfnamefont {T.}~\bibnamefont
  {Renstr{\o}m}}, \bibinfo {author} {\bibfnamefont {S.~N.}\ \bibnamefont
  {Liddick}}, \bibinfo {author} {\bibfnamefont {A.}~\bibnamefont {Spyrou}},
  \bibinfo {author} {\bibfnamefont {S.}~\bibnamefont {Karampagia}}, \bibinfo
  {author} {\bibfnamefont {B.~A.}\ \bibnamefont {Brown}}, \bibinfo {author}
  {\bibfnamefont {O.}~\bibnamefont {Achakovskiy}}, \bibinfo {author}
  {\bibfnamefont {S.}~\bibnamefont {Kamerdzhiev}}, \bibinfo {author}
  {\bibfnamefont {D.~L.}\ \bibnamefont {Bleuel}}, \bibinfo {author}
  {\bibfnamefont {A.}~\bibnamefont {Couture}}, \bibinfo {author} {\bibfnamefont
  {L.~C.}\ \bibnamefont {Campo}}, \bibinfo {author} {\bibfnamefont {B.~P.}\
  \bibnamefont {Crider}}, \bibinfo {author} {\bibfnamefont {A.~C.}\
  \bibnamefont {Dombos}}, \bibinfo {author} {\bibfnamefont {R.}~\bibnamefont
  {Lewis}}, \bibinfo {author} {\bibfnamefont {S.}~\bibnamefont {Mosby}},
  \bibinfo {author} {\bibfnamefont {F.}~\bibnamefont {Naqvi}}, \bibinfo
  {author} {\bibfnamefont {G.}~\bibnamefont {Perdikakis}}, \bibinfo {author}
  {\bibfnamefont {C.~J.}\ \bibnamefont {Prokop}}, \bibinfo {author}
  {\bibfnamefont {S.~J.}\ \bibnamefont {Quinn}},\ and\ \bibinfo {author}
  {\bibfnamefont {S.}~\bibnamefont {Siem}},\ }\bibfield  {title} {\bibinfo
  {title} {{Enhanced low-energy {$\gamma$}-decay strength of $^{70}$Ni and its
  robustness within the shell model}},\ }\href
  {https://doi.org/10.1103/PhysRevC.97.054329} {\bibfield  {journal} {\bibinfo
  {journal} {Physical Review C}\ }\textbf {\bibinfo {volume} {97}},\ \bibinfo
  {pages} {054329} (\bibinfo {year} {2018})}\BibitemShut {NoStop}%
\bibitem [{\citenamefont {Lewis}\ \emph {et~al.}(2019)\citenamefont {Lewis},
  \citenamefont {Liddick}, \citenamefont {Larsen}, \citenamefont {Spyrou},
  \citenamefont {Bleuel}, \citenamefont {Couture}, \citenamefont {Campo},
  \citenamefont {Crider}, \citenamefont {Dombos}, \citenamefont {Guttormsen},
  \citenamefont {Mosby}, \citenamefont {Naqvi}, \citenamefont {Perdikakis},
  \citenamefont {Prokop}, \citenamefont {Quinn}, \citenamefont {Renstr{\o}m},\
  and\ \citenamefont {Siem}}]{lewis19}%
  \BibitemOpen
  \bibfield  {author} {\bibinfo {author} {\bibfnamefont {R.}~\bibnamefont
  {Lewis}}, \bibinfo {author} {\bibfnamefont {S.~N.}\ \bibnamefont {Liddick}},
  \bibinfo {author} {\bibfnamefont {A.~C.}\ \bibnamefont {Larsen}}, \bibinfo
  {author} {\bibfnamefont {A.}~\bibnamefont {Spyrou}}, \bibinfo {author}
  {\bibfnamefont {D.~L.}\ \bibnamefont {Bleuel}}, \bibinfo {author}
  {\bibfnamefont {A.}~\bibnamefont {Couture}}, \bibinfo {author} {\bibfnamefont
  {L.~C.}\ \bibnamefont {Campo}}, \bibinfo {author} {\bibfnamefont {B.~P.}\
  \bibnamefont {Crider}}, \bibinfo {author} {\bibfnamefont {A.~C.}\
  \bibnamefont {Dombos}}, \bibinfo {author} {\bibfnamefont {M.}~\bibnamefont
  {Guttormsen}}, \bibinfo {author} {\bibfnamefont {S.}~\bibnamefont {Mosby}},
  \bibinfo {author} {\bibfnamefont {F.}~\bibnamefont {Naqvi}}, \bibinfo
  {author} {\bibfnamefont {G.}~\bibnamefont {Perdikakis}}, \bibinfo {author}
  {\bibfnamefont {C.~J.}\ \bibnamefont {Prokop}}, \bibinfo {author}
  {\bibfnamefont {S.~J.}\ \bibnamefont {Quinn}}, \bibinfo {author}
  {\bibfnamefont {T.}~\bibnamefont {Renstr{\o}m}},\ and\ \bibinfo {author}
  {\bibfnamefont {S.}~\bibnamefont {Siem}},\ }\bibfield  {title} {\bibinfo
  {title} {{Experimental constraints on the
  $^{73}\mathrm{Zn}(n,\gamma)^{74}\mathrm{Zn}$ reaction rate}},\ }\href
  {https://doi.org/10.1103/PhysRevC.99.034601} {\bibfield  {journal} {\bibinfo
  {journal} {Physical Review C}\ }\textbf {\bibinfo {volume} {99}},\ \bibinfo
  {pages} {034601} (\bibinfo {year} {2019})}\BibitemShut {NoStop}%
\bibitem [{\citenamefont {Liddick}\ \emph {et~al.}(2016)\citenamefont
  {Liddick}, \citenamefont {Spyrou}, \citenamefont {Crider}, \citenamefont
  {Naqvi}, \citenamefont {Larsen}, \citenamefont {Guttormsen}, \citenamefont
  {Mumpower}, \citenamefont {Surman}, \citenamefont {Perdikakis}, \citenamefont
  {Bleuel}, \citenamefont {Couture}, \citenamefont {Crespo~Campo},
  \citenamefont {Dombos}, \citenamefont {Lewis}, \citenamefont {Mosby},
  \citenamefont {Nikas}, \citenamefont {Prokop}, \citenamefont {Renstrom},
  \citenamefont {Rubio}, \citenamefont {Siem},\ and\ \citenamefont
  {Quinn}}]{PhysRevLett.116.242502}%
  \BibitemOpen
  \bibfield  {author} {\bibinfo {author} {\bibfnamefont {S.~N.}\ \bibnamefont
  {Liddick}}, \bibinfo {author} {\bibfnamefont {A.}~\bibnamefont {Spyrou}},
  \bibinfo {author} {\bibfnamefont {B.~P.}\ \bibnamefont {Crider}}, \bibinfo
  {author} {\bibfnamefont {F.}~\bibnamefont {Naqvi}}, \bibinfo {author}
  {\bibfnamefont {A.~C.}\ \bibnamefont {Larsen}}, \bibinfo {author}
  {\bibfnamefont {M.}~\bibnamefont {Guttormsen}}, \bibinfo {author}
  {\bibfnamefont {M.}~\bibnamefont {Mumpower}}, \bibinfo {author}
  {\bibfnamefont {R.}~\bibnamefont {Surman}}, \bibinfo {author} {\bibfnamefont
  {G.}~\bibnamefont {Perdikakis}}, \bibinfo {author} {\bibfnamefont {D.~L.}\
  \bibnamefont {Bleuel}}, \bibinfo {author} {\bibfnamefont {A.}~\bibnamefont
  {Couture}}, \bibinfo {author} {\bibfnamefont {L.}~\bibnamefont
  {Crespo~Campo}}, \bibinfo {author} {\bibfnamefont {A.~C.}\ \bibnamefont
  {Dombos}}, \bibinfo {author} {\bibfnamefont {R.}~\bibnamefont {Lewis}},
  \bibinfo {author} {\bibfnamefont {S.}~\bibnamefont {Mosby}}, \bibinfo
  {author} {\bibfnamefont {S.}~\bibnamefont {Nikas}}, \bibinfo {author}
  {\bibfnamefont {C.~J.}\ \bibnamefont {Prokop}}, \bibinfo {author}
  {\bibfnamefont {T.}~\bibnamefont {Renstrom}}, \bibinfo {author}
  {\bibfnamefont {B.}~\bibnamefont {Rubio}}, \bibinfo {author} {\bibfnamefont
  {S.}~\bibnamefont {Siem}},\ and\ \bibinfo {author} {\bibfnamefont {S.~J.}\
  \bibnamefont {Quinn}},\ }\bibfield  {title} {\bibinfo {title} {{Experimental
  Neutron Capture Rate Constraint Far from Stability}},\ }\href
  {https://doi.org/10.1103/PhysRevLett.116.242502} {\bibfield  {journal}
  {\bibinfo  {journal} {Physical Review Letters}\ }\textbf {\bibinfo {volume}
  {116}},\ \bibinfo {pages} {242502} (\bibinfo {year} {2016})}\BibitemShut
  {NoStop}%
\bibitem [{\citenamefont {Liddick}\ \emph {et~al.}(2019)\citenamefont
  {Liddick}, \citenamefont {Larsen}, \citenamefont {Guttormsen}, \citenamefont
  {Spyrou}, \citenamefont {Crider}, \citenamefont {Naqvi}, \citenamefont
  {Midtb\o{}}, \citenamefont {Bello~Garrote}, \citenamefont {Bleuel},
  \citenamefont {Crespo~Campo}, \citenamefont {Couture}, \citenamefont
  {Dombos}, \citenamefont {Giacoppo}, \citenamefont {G\"orgen}, \citenamefont
  {Hadynska-Klek}, \citenamefont {Hagen}, \citenamefont {Ingeberg},
  \citenamefont {Kheswa}, \citenamefont {Lewis}, \citenamefont {Mosby},
  \citenamefont {Perdikakis}, \citenamefont {Prokop}, \citenamefont {Quinn},
  \citenamefont {Renstr\o{}m}, \citenamefont {Rose}, \citenamefont {Sahin},
  \citenamefont {Siem}, \citenamefont {Tveten}, \citenamefont {Wiedeking},\
  and\ \citenamefont {Zeiser}}]{Liddick2019BenchmarkingMethod}%
  \BibitemOpen
  \bibfield  {author} {\bibinfo {author} {\bibfnamefont {S.~N.}\ \bibnamefont
  {Liddick}}, \bibinfo {author} {\bibfnamefont {A.~C.}\ \bibnamefont {Larsen}},
  \bibinfo {author} {\bibfnamefont {M.}~\bibnamefont {Guttormsen}}, \bibinfo
  {author} {\bibfnamefont {A.}~\bibnamefont {Spyrou}}, \bibinfo {author}
  {\bibfnamefont {B.~P.}\ \bibnamefont {Crider}}, \bibinfo {author}
  {\bibfnamefont {F.}~\bibnamefont {Naqvi}}, \bibinfo {author} {\bibfnamefont
  {J.~E.}\ \bibnamefont {Midtb\o{}}}, \bibinfo {author} {\bibfnamefont {F.~L.}\
  \bibnamefont {Bello~Garrote}}, \bibinfo {author} {\bibfnamefont {D.~L.}\
  \bibnamefont {Bleuel}}, \bibinfo {author} {\bibfnamefont {L.}~\bibnamefont
  {Crespo~Campo}}, \bibinfo {author} {\bibfnamefont {A.}~\bibnamefont
  {Couture}}, \bibinfo {author} {\bibfnamefont {A.~C.}\ \bibnamefont {Dombos}},
  \bibinfo {author} {\bibfnamefont {F.}~\bibnamefont {Giacoppo}}, \bibinfo
  {author} {\bibfnamefont {A.}~\bibnamefont {G\"orgen}}, \bibinfo {author}
  {\bibfnamefont {K.}~\bibnamefont {Hadynska-Klek}}, \bibinfo {author}
  {\bibfnamefont {T.~W.}\ \bibnamefont {Hagen}}, \bibinfo {author}
  {\bibfnamefont {V.~W.}\ \bibnamefont {Ingeberg}}, \bibinfo {author}
  {\bibfnamefont {B.~V.}\ \bibnamefont {Kheswa}}, \bibinfo {author}
  {\bibfnamefont {R.}~\bibnamefont {Lewis}}, \bibinfo {author} {\bibfnamefont
  {S.}~\bibnamefont {Mosby}}, \bibinfo {author} {\bibfnamefont
  {G.}~\bibnamefont {Perdikakis}}, \bibinfo {author} {\bibfnamefont {C.~J.}\
  \bibnamefont {Prokop}}, \bibinfo {author} {\bibfnamefont {S.~J.}\
  \bibnamefont {Quinn}}, \bibinfo {author} {\bibfnamefont {T.}~\bibnamefont
  {Renstr\o{}m}}, \bibinfo {author} {\bibfnamefont {S.~J.}\ \bibnamefont
  {Rose}}, \bibinfo {author} {\bibfnamefont {E.}~\bibnamefont {Sahin}},
  \bibinfo {author} {\bibfnamefont {S.}~\bibnamefont {Siem}}, \bibinfo {author}
  {\bibfnamefont {G.~M.}\ \bibnamefont {Tveten}}, \bibinfo {author}
  {\bibfnamefont {M.}~\bibnamefont {Wiedeking}},\ and\ \bibinfo {author}
  {\bibfnamefont {F.}~\bibnamefont {Zeiser}},\ }\bibfield  {title} {\bibinfo
  {title} {Benchmarking the extraction of statistical neutron capture cross
  sections on short-lived nuclei for applications using the
  $\ensuremath{\beta}$-oslo method},\ }\href
  {https://doi.org/10.1103/PhysRevC.100.024624} {\bibfield  {journal} {\bibinfo
   {journal} {Phys. Rev. C}\ }\textbf {\bibinfo {volume} {100}},\ \bibinfo
  {pages} {024624} (\bibinfo {year} {2019})}\BibitemShut {NoStop}%
\bibitem [{\citenamefont {Ingeberg}\ \emph {et~al.}(2020)\citenamefont
  {Ingeberg}, \citenamefont {Siem}, \citenamefont {Wiedeking}, \citenamefont
  {Sieja}, \citenamefont {Bleuel}, \citenamefont {Brits}, \citenamefont
  {Bucher}, \citenamefont {Dinoko}, \citenamefont {Easton}, \citenamefont
  {G{\"{o}}rgen}, \citenamefont {Guttormsen}, \citenamefont {Jones},
  \citenamefont {Kheswa}, \citenamefont {Khumalo}, \citenamefont {Larsen},
  \citenamefont {Lawrie}, \citenamefont {Lawrie}, \citenamefont {Majola},
  \citenamefont {Malatji}, \citenamefont {Makhathini}, \citenamefont
  {Maqabuka}, \citenamefont {Negi}, \citenamefont {Noncolela}, \citenamefont
  {Papka}, \citenamefont {Sahin}, \citenamefont {Schwengner}, \citenamefont
  {Tveten}, \citenamefont {Zeiser},\ and\ \citenamefont
  {Zikhali}}]{Ingeberg2020}%
  \BibitemOpen
  \bibfield  {author} {\bibinfo {author} {\bibfnamefont {V.~W.}\ \bibnamefont
  {Ingeberg}}, \bibinfo {author} {\bibfnamefont {S.}~\bibnamefont {Siem}},
  \bibinfo {author} {\bibfnamefont {M.}~\bibnamefont {Wiedeking}}, \bibinfo
  {author} {\bibfnamefont {K.}~\bibnamefont {Sieja}}, \bibinfo {author}
  {\bibfnamefont {D.~L.}\ \bibnamefont {Bleuel}}, \bibinfo {author}
  {\bibfnamefont {C.~P.}\ \bibnamefont {Brits}}, \bibinfo {author}
  {\bibfnamefont {T.~D.}\ \bibnamefont {Bucher}}, \bibinfo {author}
  {\bibfnamefont {T.~S.}\ \bibnamefont {Dinoko}}, \bibinfo {author}
  {\bibfnamefont {J.~L.}\ \bibnamefont {Easton}}, \bibinfo {author}
  {\bibfnamefont {A.}~\bibnamefont {G{\"{o}}rgen}}, \bibinfo {author}
  {\bibfnamefont {M.}~\bibnamefont {Guttormsen}}, \bibinfo {author}
  {\bibfnamefont {P.}~\bibnamefont {Jones}}, \bibinfo {author} {\bibfnamefont
  {B.~V.}\ \bibnamefont {Kheswa}}, \bibinfo {author} {\bibfnamefont {N.~A.}\
  \bibnamefont {Khumalo}}, \bibinfo {author} {\bibfnamefont {A.~C.}\
  \bibnamefont {Larsen}}, \bibinfo {author} {\bibfnamefont {E.~A.}\
  \bibnamefont {Lawrie}}, \bibinfo {author} {\bibfnamefont {J.~J.}\
  \bibnamefont {Lawrie}}, \bibinfo {author} {\bibfnamefont {S.~N.}\
  \bibnamefont {Majola}}, \bibinfo {author} {\bibfnamefont {K.~L.}\
  \bibnamefont {Malatji}}, \bibinfo {author} {\bibfnamefont {L.}~\bibnamefont
  {Makhathini}}, \bibinfo {author} {\bibfnamefont {B.}~\bibnamefont
  {Maqabuka}}, \bibinfo {author} {\bibfnamefont {D.}~\bibnamefont {Negi}},
  \bibinfo {author} {\bibfnamefont {S.~P.}\ \bibnamefont {Noncolela}}, \bibinfo
  {author} {\bibfnamefont {P.}~\bibnamefont {Papka}}, \bibinfo {author}
  {\bibfnamefont {E.}~\bibnamefont {Sahin}}, \bibinfo {author} {\bibfnamefont
  {R.}~\bibnamefont {Schwengner}}, \bibinfo {author} {\bibfnamefont {G.~M.}\
  \bibnamefont {Tveten}}, \bibinfo {author} {\bibfnamefont {F.}~\bibnamefont
  {Zeiser}},\ and\ \bibinfo {author} {\bibfnamefont {B.~R.}\ \bibnamefont
  {Zikhali}},\ }\bibfield  {title} {\bibinfo {title} {{First application of the
  Oslo method in inverse kinematics}},\ }\href
  {https://doi.org/10.1140/epja/s10050-020-00070-7} {\bibfield  {journal}
  {\bibinfo  {journal} {European Physical Journal A}\ }\textbf {\bibinfo
  {volume} {56}},\ \bibinfo {pages} {68} (\bibinfo {year} {2020})}\BibitemShut
  {NoStop}%
\bibitem [{\citenamefont {McKay}\ \emph {et~al.}(2020)\citenamefont {McKay},
  \citenamefont {Denissenkov}, \citenamefont {Herwig}, \citenamefont
  {Perdikakis},\ and\ \citenamefont {Schatz}}]{McKay2020}%
  \BibitemOpen
  \bibfield  {author} {\bibinfo {author} {\bibfnamefont {J.~E.}\ \bibnamefont
  {McKay}}, \bibinfo {author} {\bibfnamefont {P.~A.}\ \bibnamefont
  {Denissenkov}}, \bibinfo {author} {\bibfnamefont {F.}~\bibnamefont {Herwig}},
  \bibinfo {author} {\bibfnamefont {G.}~\bibnamefont {Perdikakis}},\ and\
  \bibinfo {author} {\bibfnamefont {H.}~\bibnamefont {Schatz}},\ }\bibfield
  {title} {\bibinfo {title} {{The impact of (n,$\gamma$) reaction rate
  uncertainties on the predicted abundances of i-process elements with 32
  $\leq$ Z $\leq$ 48 in the metal-poor star HD94028}},\ }\href
  {https://doi.org/10.1093/mnras/stz3322} {\bibfield  {journal} {\bibinfo
  {journal} {Monthly Notices of the Royal Astronomical Society}\ }\textbf
  {\bibinfo {volume} {491}},\ \bibinfo {pages} {5179} (\bibinfo {year}
  {2020})}\BibitemShut {NoStop}%
\bibitem [{\citenamefont {Kadi}\ \emph {et~al.}(2018)\citenamefont {Kadi},
  \citenamefont {Fraser},\ and\ \citenamefont
  {Papageorgiou-Koufidou}}]{Kadi18}%
  \BibitemOpen
  \bibfield  {author} {\bibinfo {author} {\bibfnamefont {Y.}~\bibnamefont
  {Kadi}}, \bibinfo {author} {\bibfnamefont {M.~A.}\ \bibnamefont {Fraser}},\
  and\ \bibinfo {author} {\bibfnamefont {A.}~\bibnamefont
  {Papageorgiou-Koufidou}},\ }\href {https://doi.org/10.23731/CYRM-2018-001}
  {\emph {\bibinfo {title} {{HIE-ISOLDE: technical design report for the energy
  upgrade}}}},\ edited by\ \bibinfo {editor} {\bibfnamefont {Y.}~\bibnamefont
  {Kadi}}\ (\bibinfo  {publisher} {CERN},\ \bibinfo {address} {Geneva},\
  \bibinfo {year} {2018})\BibitemShut {NoStop}%
\bibitem [{\citenamefont {Fedoseyev}\ \emph {et~al.}(2000)\citenamefont
  {Fedoseyev}, \citenamefont {Huber}, \citenamefont {K{\"{o}}ster},
  \citenamefont {Lettry}, \citenamefont {Mishin}, \citenamefont {Ravn},\ and\
  \citenamefont {Sebastian}}]{Fedoseyev2000}%
  \BibitemOpen
  \bibfield  {author} {\bibinfo {author} {\bibfnamefont {V.~N.}\ \bibnamefont
  {Fedoseyev}}, \bibinfo {author} {\bibfnamefont {G.}~\bibnamefont {Huber}},
  \bibinfo {author} {\bibfnamefont {U.}~\bibnamefont {K{\"{o}}ster}}, \bibinfo
  {author} {\bibfnamefont {J.}~\bibnamefont {Lettry}}, \bibinfo {author}
  {\bibfnamefont {V.~I.}\ \bibnamefont {Mishin}}, \bibinfo {author}
  {\bibfnamefont {H.}~\bibnamefont {Ravn}},\ and\ \bibinfo {author}
  {\bibfnamefont {V.}~\bibnamefont {Sebastian}},\ }\bibfield  {title} {\bibinfo
  {title} {{The ISOLDE laser ion source for exotic nuclei}},\ }\href
  {https://doi.org/10.1023/A:1012609515865} {\bibfield  {journal} {\bibinfo
  {journal} {Hyperfine Interactions}\ }\textbf {\bibinfo {volume} {127}},\
  \bibinfo {pages} {409} (\bibinfo {year} {2000})}\BibitemShut {NoStop}%
\bibitem [{\citenamefont {Ames}\ \emph {et~al.}(2005)\citenamefont {Ames},
  \citenamefont {Bollen}, \citenamefont {Delahaye}, \citenamefont {Forstner},
  \citenamefont {Huber}, \citenamefont {Kester}, \citenamefont {Reisinger},\
  and\ \citenamefont {Schmidt}}]{AMES200517}%
  \BibitemOpen
  \bibfield  {author} {\bibinfo {author} {\bibfnamefont {F.}~\bibnamefont
  {Ames}}, \bibinfo {author} {\bibfnamefont {G.}~\bibnamefont {Bollen}},
  \bibinfo {author} {\bibfnamefont {P.}~\bibnamefont {Delahaye}}, \bibinfo
  {author} {\bibfnamefont {O.}~\bibnamefont {Forstner}}, \bibinfo {author}
  {\bibfnamefont {G.}~\bibnamefont {Huber}}, \bibinfo {author} {\bibfnamefont
  {O.}~\bibnamefont {Kester}}, \bibinfo {author} {\bibfnamefont
  {K.}~\bibnamefont {Reisinger}},\ and\ \bibinfo {author} {\bibfnamefont
  {P.}~\bibnamefont {Schmidt}},\ }\bibfield  {title} {\bibinfo {title}
  {{Cooling of radioactive ions with the Penning trap REXTRAP}},\ }\href
  {https://doi.org/10.1016/j.nima.2004.08.119} {\bibfield  {journal} {\bibinfo
  {journal} {Nuclear Instruments and Methods in Physics Research, Section A:
  Accelerators, Spectrometers, Detectors and Associated Equipment}\ }\textbf
  {\bibinfo {volume} {538}},\ \bibinfo {pages} {17} (\bibinfo {year}
  {2005})}\BibitemShut {NoStop}%
\bibitem [{\citenamefont {Wenander}(2002)}]{WENANDER2002528}%
  \BibitemOpen
  \bibfield  {author} {\bibinfo {author} {\bibfnamefont {F.}~\bibnamefont
  {Wenander}},\ }\bibfield  {title} {\bibinfo {title} {{EBIS as charge breeder
  for radioactive ion beam accelerators}},\ }\href
  {https://doi.org/10.1016/S0375-9474(01)01640-2} {\bibfield  {journal}
  {\bibinfo  {journal} {Nuclear Physics A}\ }\textbf {\bibinfo {volume}
  {701}},\ \bibinfo {pages} {528} (\bibinfo {year} {2002})}\BibitemShut
  {NoStop}%
\bibitem [{\citenamefont {Wenander}(2010)}]{Wenander_2010}%
  \BibitemOpen
  \bibfield  {author} {\bibinfo {author} {\bibfnamefont {F.}~\bibnamefont
  {Wenander}},\ }\bibfield  {title} {\bibinfo {title} {{Charge breeding of
  radioactive ions with EBIS and EBIT}},\ }\href
  {https://doi.org/10.1088/1748-0221/5/10/C10004} {\bibfield  {journal}
  {\bibinfo  {journal} {Journal of Instrumentation}\ }\textbf {\bibinfo
  {volume} {5}}\bibinfo  {number} { (10)},\ \bibinfo {pages}
  {C10004}}\BibitemShut {NoStop}%
\bibitem [{\citenamefont {Kheswa}\ \emph {et~al.}(2020)\citenamefont {Kheswa},
  \citenamefont {Dinoko},\ and\ \citenamefont
  {Wiedeking}}]{kheswa_production_2020}%
  \BibitemOpen
\bibfield  {number} {  }\bibfield  {author} {\bibinfo {author} {\bibfnamefont
  {N.}~\bibnamefont {Kheswa}}, \bibinfo {author} {\bibfnamefont
  {T.}~\bibnamefont {Dinoko}},\ and\ \bibinfo {author} {\bibfnamefont
  {M.}~\bibnamefont {Wiedeking}},\ }\bibfield  {title} {\bibinfo {title} {The
  production of deuterated polyethylene targets},\ }\href
  {https://doi.org/10.1051/epjconf/202022903003} {\bibfield  {journal}
  {\bibinfo  {journal} {EPJ Web of Conferences}\ }\textbf {\bibinfo {volume}
  {229}},\ \bibinfo {pages} {03003} (\bibinfo {year} {2020})}\BibitemShut
  {NoStop}%
\bibitem [{\citenamefont {Diriken}\ \emph {et~al.}(2014)\citenamefont
  {Diriken}, \citenamefont {Patronis}, \citenamefont {Andreyev}, \citenamefont
  {Antalic}, \citenamefont {Bildstein}, \citenamefont {Blazhev}, \citenamefont
  {Darby}, \citenamefont {De~Witte}, \citenamefont {Eberth}, \citenamefont
  {Elseviers}, \citenamefont {Fedosseev}, \citenamefont {Flavigny},
  \citenamefont {Fransen}, \citenamefont {Georgiev}, \citenamefont
  {Gernhauser}, \citenamefont {Hess}, \citenamefont {Huyse}, \citenamefont
  {Jolie}, \citenamefont {Kröll}, \citenamefont {Krücken}, \citenamefont
  {Lutter}, \citenamefont {Marsh}, \citenamefont {Mertzimekis}, \citenamefont
  {Muecher}, \citenamefont {Nowacki}, \citenamefont {Orlandi}, \citenamefont
  {Pakou}, \citenamefont {Raabe}, \citenamefont {Randisi}, \citenamefont
  {Reiter}, \citenamefont {Roger}, \citenamefont {Seidlitz}, \citenamefont
  {Seliverstov}, \citenamefont {Sieja}, \citenamefont {Sotty}, \citenamefont
  {Tornqvist}, \citenamefont {Van De~Walle}, \citenamefont {Van~Duppen},
  \citenamefont {Voulot}, \citenamefont {Warr}, \citenamefont {Wenander},\ and\
  \citenamefont {Wimmer}}]{diriken_study_2014}%
  \BibitemOpen
  \bibfield  {author} {\bibinfo {author} {\bibfnamefont {J.}~\bibnamefont
  {Diriken}}, \bibinfo {author} {\bibfnamefont {N.}~\bibnamefont {Patronis}},
  \bibinfo {author} {\bibfnamefont {A.}~\bibnamefont {Andreyev}}, \bibinfo
  {author} {\bibfnamefont {S.}~\bibnamefont {Antalic}}, \bibinfo {author}
  {\bibfnamefont {V.}~\bibnamefont {Bildstein}}, \bibinfo {author}
  {\bibfnamefont {A.}~\bibnamefont {Blazhev}}, \bibinfo {author} {\bibfnamefont
  {I.}~\bibnamefont {Darby}}, \bibinfo {author} {\bibfnamefont
  {H.}~\bibnamefont {De~Witte}}, \bibinfo {author} {\bibfnamefont
  {J.}~\bibnamefont {Eberth}}, \bibinfo {author} {\bibfnamefont
  {J.}~\bibnamefont {Elseviers}}, \bibinfo {author} {\bibfnamefont
  {V.}~\bibnamefont {Fedosseev}}, \bibinfo {author} {\bibfnamefont
  {F.}~\bibnamefont {Flavigny}}, \bibinfo {author} {\bibfnamefont
  {C.}~\bibnamefont {Fransen}}, \bibinfo {author} {\bibfnamefont
  {G.}~\bibnamefont {Georgiev}}, \bibinfo {author} {\bibfnamefont
  {R.}~\bibnamefont {Gernhauser}}, \bibinfo {author} {\bibfnamefont
  {H.}~\bibnamefont {Hess}}, \bibinfo {author} {\bibfnamefont {M.}~\bibnamefont
  {Huyse}}, \bibinfo {author} {\bibfnamefont {J.}~\bibnamefont {Jolie}},
  \bibinfo {author} {\bibfnamefont {T.}~\bibnamefont {Kröll}}, \bibinfo
  {author} {\bibfnamefont {R.}~\bibnamefont {Krücken}}, \bibinfo {author}
  {\bibfnamefont {R.}~\bibnamefont {Lutter}}, \bibinfo {author} {\bibfnamefont
  {B.}~\bibnamefont {Marsh}}, \bibinfo {author} {\bibfnamefont
  {T.}~\bibnamefont {Mertzimekis}}, \bibinfo {author} {\bibfnamefont
  {D.}~\bibnamefont {Muecher}}, \bibinfo {author} {\bibfnamefont
  {F.}~\bibnamefont {Nowacki}}, \bibinfo {author} {\bibfnamefont
  {R.}~\bibnamefont {Orlandi}}, \bibinfo {author} {\bibfnamefont
  {A.}~\bibnamefont {Pakou}}, \bibinfo {author} {\bibfnamefont
  {R.}~\bibnamefont {Raabe}}, \bibinfo {author} {\bibfnamefont
  {G.}~\bibnamefont {Randisi}}, \bibinfo {author} {\bibfnamefont
  {P.}~\bibnamefont {Reiter}}, \bibinfo {author} {\bibfnamefont
  {T.}~\bibnamefont {Roger}}, \bibinfo {author} {\bibfnamefont
  {M.}~\bibnamefont {Seidlitz}}, \bibinfo {author} {\bibfnamefont
  {M.}~\bibnamefont {Seliverstov}}, \bibinfo {author} {\bibfnamefont
  {K.}~\bibnamefont {Sieja}}, \bibinfo {author} {\bibfnamefont
  {C.}~\bibnamefont {Sotty}}, \bibinfo {author} {\bibfnamefont
  {H.}~\bibnamefont {Tornqvist}}, \bibinfo {author} {\bibfnamefont
  {J.}~\bibnamefont {Van De~Walle}}, \bibinfo {author} {\bibfnamefont
  {P.}~\bibnamefont {Van~Duppen}}, \bibinfo {author} {\bibfnamefont
  {D.}~\bibnamefont {Voulot}}, \bibinfo {author} {\bibfnamefont
  {N.}~\bibnamefont {Warr}}, \bibinfo {author} {\bibfnamefont {F.}~\bibnamefont
  {Wenander}},\ and\ \bibinfo {author} {\bibfnamefont {K.}~\bibnamefont
  {Wimmer}},\ }\bibfield  {title} {\bibinfo {title} {Study of the
  deformation-driving $\nu$d$_{5/2}$ orbital in $^{67}_{28}\mathrm{Ni}_{39}$
  using one-neutron transfer reactions},\ }\href
  {https://doi.org/10.1016/j.physletb.2014.08.004} {\bibfield  {journal}
  {\bibinfo  {journal} {Physics Letters B}\ }\textbf {\bibinfo {volume}
  {736}},\ \bibinfo {pages} {533} (\bibinfo {year} {2014})}\BibitemShut
  {NoStop}%
\bibitem [{\citenamefont {Hellgartner}\ \emph {et~al.}(2023)\citenamefont
  {Hellgartner}, \citenamefont {Mücher}, \citenamefont {Wimmer}, \citenamefont
  {Bildstein}, \citenamefont {Egido}, \citenamefont {Gernhäuser},
  \citenamefont {Krücken}, \citenamefont {Nowak}, \citenamefont {Zielińska},
  \citenamefont {Bauer}, \citenamefont {Benito}, \citenamefont {Bottoni},
  \citenamefont {De~Witte}, \citenamefont {Elseviers}, \citenamefont {Fedorov},
  \citenamefont {Flavigny}, \citenamefont {Illana}, \citenamefont
  {Klintefjord}, \citenamefont {Kröll}, \citenamefont {Lutter}, \citenamefont
  {Marsh}, \citenamefont {Orlandi}, \citenamefont {Pakarinen}, \citenamefont
  {Raabe}, \citenamefont {Rapisarda}, \citenamefont {Reichert}, \citenamefont
  {Reiter}, \citenamefont {Scheck}, \citenamefont {Seidlitz}, \citenamefont
  {Siebeck}, \citenamefont {Siesling}, \citenamefont {Steinbach}, \citenamefont
  {Stora}, \citenamefont {Vermeulen}, \citenamefont {Voulot}, \citenamefont
  {Warr},\ and\ \citenamefont {Wenander}}]{hellgartner_axial_2023}%
  \BibitemOpen
  \bibfield  {author} {\bibinfo {author} {\bibfnamefont {S.}~\bibnamefont
  {Hellgartner}}, \bibinfo {author} {\bibfnamefont {D.}~\bibnamefont
  {Mücher}}, \bibinfo {author} {\bibfnamefont {K.}~\bibnamefont {Wimmer}},
  \bibinfo {author} {\bibfnamefont {V.}~\bibnamefont {Bildstein}}, \bibinfo
  {author} {\bibfnamefont {J.}~\bibnamefont {Egido}}, \bibinfo {author}
  {\bibfnamefont {R.}~\bibnamefont {Gernhäuser}}, \bibinfo {author}
  {\bibfnamefont {R.}~\bibnamefont {Krücken}}, \bibinfo {author}
  {\bibfnamefont {A.}~\bibnamefont {Nowak}}, \bibinfo {author} {\bibfnamefont
  {M.}~\bibnamefont {Zielińska}}, \bibinfo {author} {\bibfnamefont
  {C.}~\bibnamefont {Bauer}}, \bibinfo {author} {\bibfnamefont
  {M.}~\bibnamefont {Benito}}, \bibinfo {author} {\bibfnamefont
  {S.}~\bibnamefont {Bottoni}}, \bibinfo {author} {\bibfnamefont
  {H.}~\bibnamefont {De~Witte}}, \bibinfo {author} {\bibfnamefont
  {J.}~\bibnamefont {Elseviers}}, \bibinfo {author} {\bibfnamefont
  {D.}~\bibnamefont {Fedorov}}, \bibinfo {author} {\bibfnamefont
  {F.}~\bibnamefont {Flavigny}}, \bibinfo {author} {\bibfnamefont
  {A.}~\bibnamefont {Illana}}, \bibinfo {author} {\bibfnamefont
  {M.}~\bibnamefont {Klintefjord}}, \bibinfo {author} {\bibfnamefont
  {T.}~\bibnamefont {Kröll}}, \bibinfo {author} {\bibfnamefont
  {R.}~\bibnamefont {Lutter}}, \bibinfo {author} {\bibfnamefont
  {B.}~\bibnamefont {Marsh}}, \bibinfo {author} {\bibfnamefont
  {R.}~\bibnamefont {Orlandi}}, \bibinfo {author} {\bibfnamefont
  {J.}~\bibnamefont {Pakarinen}}, \bibinfo {author} {\bibfnamefont
  {R.}~\bibnamefont {Raabe}}, \bibinfo {author} {\bibfnamefont
  {E.}~\bibnamefont {Rapisarda}}, \bibinfo {author} {\bibfnamefont
  {S.}~\bibnamefont {Reichert}}, \bibinfo {author} {\bibfnamefont
  {P.}~\bibnamefont {Reiter}}, \bibinfo {author} {\bibfnamefont
  {M.}~\bibnamefont {Scheck}}, \bibinfo {author} {\bibfnamefont
  {M.}~\bibnamefont {Seidlitz}}, \bibinfo {author} {\bibfnamefont
  {B.}~\bibnamefont {Siebeck}}, \bibinfo {author} {\bibfnamefont
  {E.}~\bibnamefont {Siesling}}, \bibinfo {author} {\bibfnamefont
  {T.}~\bibnamefont {Steinbach}}, \bibinfo {author} {\bibfnamefont
  {T.}~\bibnamefont {Stora}}, \bibinfo {author} {\bibfnamefont
  {M.}~\bibnamefont {Vermeulen}}, \bibinfo {author} {\bibfnamefont
  {D.}~\bibnamefont {Voulot}}, \bibinfo {author} {\bibfnamefont
  {N.}~\bibnamefont {Warr}},\ and\ \bibinfo {author} {\bibfnamefont
  {F.}~\bibnamefont {Wenander}},\ }\bibfield  {title} {\bibinfo {title} {{Axial
  and triaxial degrees of freedom in $^{72}$Zn}},\ }\href
  {https://doi.org/10.1016/j.physletb.2023.137933} {\bibfield  {journal}
  {\bibinfo  {journal} {Physics Letters B}\ }\textbf {\bibinfo {volume}
  {841}},\ \bibinfo {pages} {137933} (\bibinfo {year} {2023})}\BibitemShut
  {NoStop}%
\bibitem [{\citenamefont {Hellgartner}(2015)}]{Hellgartner15}%
  \BibitemOpen
  \bibfield  {author} {\bibinfo {author} {\bibfnamefont {S.~C.}\ \bibnamefont
  {Hellgartner}},\ }\emph {\bibinfo {title} {Probing Nuclear Shell Structure
  beyond the N= 40 Subshell using Multiple Coulomb Excitation and Transfer
  Experiments}},\ \href {https://mediatum .ub .tum .de /node ?id =1277804}
  {Ph.D. thesis},\ \bibinfo  {school} {Technische Universität München,
  Lehrstuhl E12 für Experimentalphysik} (\bibinfo {year} {2015})\BibitemShut
  {NoStop}%
\bibitem [{\citenamefont {Eberth}\ \emph {et~al.}(2001)\citenamefont {Eberth},
  \citenamefont {Pascovici}, \citenamefont {Thomas}, \citenamefont {Warr},
  \citenamefont {Weisshaar}, \citenamefont {Habs}, \citenamefont {Reiter},
  \citenamefont {Thirolf}, \citenamefont {Schwalm}, \citenamefont {Gund},
  \citenamefont {Scheit}, \citenamefont {Lauer}, \citenamefont {Van~Duppen},
  \citenamefont {Franchoo}, \citenamefont {Huyse}, \citenamefont {Lieder},
  \citenamefont {Gast}, \citenamefont {Gerl},\ and\ \citenamefont
  {Lieb}}]{EBERTH2001389}%
  \BibitemOpen
  \bibfield  {author} {\bibinfo {author} {\bibfnamefont {J.}~\bibnamefont
  {Eberth}}, \bibinfo {author} {\bibfnamefont {G.}~\bibnamefont {Pascovici}},
  \bibinfo {author} {\bibfnamefont {H.~G.}\ \bibnamefont {Thomas}}, \bibinfo
  {author} {\bibfnamefont {N.}~\bibnamefont {Warr}}, \bibinfo {author}
  {\bibfnamefont {D.}~\bibnamefont {Weisshaar}}, \bibinfo {author}
  {\bibfnamefont {D.}~\bibnamefont {Habs}}, \bibinfo {author} {\bibfnamefont
  {P.}~\bibnamefont {Reiter}}, \bibinfo {author} {\bibfnamefont
  {P.}~\bibnamefont {Thirolf}}, \bibinfo {author} {\bibfnamefont
  {D.}~\bibnamefont {Schwalm}}, \bibinfo {author} {\bibfnamefont
  {C.}~\bibnamefont {Gund}}, \bibinfo {author} {\bibfnamefont {H.}~\bibnamefont
  {Scheit}}, \bibinfo {author} {\bibfnamefont {M.}~\bibnamefont {Lauer}},
  \bibinfo {author} {\bibfnamefont {P.}~\bibnamefont {Van~Duppen}}, \bibinfo
  {author} {\bibfnamefont {S.}~\bibnamefont {Franchoo}}, \bibinfo {author}
  {\bibfnamefont {M.}~\bibnamefont {Huyse}}, \bibinfo {author} {\bibfnamefont
  {R.~M.}\ \bibnamefont {Lieder}}, \bibinfo {author} {\bibfnamefont
  {W.}~\bibnamefont {Gast}}, \bibinfo {author} {\bibfnamefont {J.}~\bibnamefont
  {Gerl}},\ and\ \bibinfo {author} {\bibfnamefont {K.~P.}\ \bibnamefont
  {Lieb}},\ }\bibfield  {title} {\bibinfo {title} {{MINIBALL A Ge detector
  array for radioactive ion beam facilities}},\ }\href
  {https://doi.org/https://doi.org/10.1016/S0146-6410(01)00145-4} {\bibfield
  {journal} {\bibinfo  {journal} {Progress in Particle and Nuclear Physics}\
  }\textbf {\bibinfo {volume} {46}},\ \bibinfo {pages} {389} (\bibinfo {year}
  {2001})}\BibitemShut {NoStop}%
\bibitem [{\citenamefont {Warr}\ \emph {et~al.}(2013)\citenamefont {Warr},
  \citenamefont {Van~de Walle}, \citenamefont {Albers}, \citenamefont {Ames},
  \citenamefont {Bastin}, \citenamefont {Bauer}, \citenamefont {Bildstein},
  \citenamefont {Blazhev}, \citenamefont {B{\"{o}}nig}, \citenamefont {Bree},
  \citenamefont {Bruyneel}, \citenamefont {Butler}, \citenamefont
  {Cederk{\"{a}}ll}, \citenamefont {Cl{\'{e}}ment}, \citenamefont {Cocolios},
  \citenamefont {Davinson}, \citenamefont {De~Witte}, \citenamefont {Delahaye},
  \citenamefont {DiJulio}, \citenamefont {Diriken}, \citenamefont {Eberth},
  \citenamefont {Ekstr{\"{o}}m}, \citenamefont {Elseviers}, \citenamefont
  {Emhofer}, \citenamefont {Fedorov}, \citenamefont {Fedosseev}, \citenamefont
  {Franchoo}, \citenamefont {Fransen}, \citenamefont {Gaffney}, \citenamefont
  {Gerl}, \citenamefont {Georgiev}, \citenamefont {Gernh{\"{a}}user},
  \citenamefont {Grahn}, \citenamefont {Habs}, \citenamefont {Hess},
  \citenamefont {Hurst}, \citenamefont {Huyse}, \citenamefont {Ivanov},
  \citenamefont {Iwanicki}, \citenamefont {Jenkins}, \citenamefont {Jolie},
  \citenamefont {Kesteloot}, \citenamefont {Kester}, \citenamefont
  {K{\"{o}}ster}, \citenamefont {Krauth}, \citenamefont {Kr{\"{o}}ll},
  \citenamefont {Kr{\"{u}}cken}, \citenamefont {Lauer}, \citenamefont {Leske},
  \citenamefont {Lieb}, \citenamefont {Lutter}, \citenamefont {Maier},
  \citenamefont {Marsh}, \citenamefont {M{\"{u}}cher}, \citenamefont
  {M{\"{u}}nch}, \citenamefont {Niedermaier}, \citenamefont {Pakarinen},
  \citenamefont {Pantea}, \citenamefont {Pascovici}, \citenamefont {Patronis},
  \citenamefont {Pauwels}, \citenamefont {Petts}, \citenamefont {Pietralla},
  \citenamefont {Raabe}, \citenamefont {Rapisarda}, \citenamefont {Reiter},
  \citenamefont {Richter}, \citenamefont {Schaile}, \citenamefont {Scheck},
  \citenamefont {Scheit}, \citenamefont {Schrieder}, \citenamefont {Schwalm},
  \citenamefont {Seidlitz}, \citenamefont {Seliverstov}, \citenamefont
  {Sieber}, \citenamefont {Simon}, \citenamefont {Speidel}, \citenamefont
  {Stahl}, \citenamefont {Stefanescu}, \citenamefont {Thirolf}, \citenamefont
  {Thomas}, \citenamefont {Th{\"{u}}rauf}, \citenamefont {Van~Duppen},
  \citenamefont {Voulot}, \citenamefont {Wadsworth}, \citenamefont {Walter},
  \citenamefont {Wei{\ss}haar}, \citenamefont {Wenander}, \citenamefont
  {Wiens}, \citenamefont {Wimmer}, \citenamefont {Wolf}, \citenamefont {Woods},
  \citenamefont {Wrzosek-Lipska},\ and\ \citenamefont {Zell}}]{Warr2013}%
  \BibitemOpen
  \bibfield  {author} {\bibinfo {author} {\bibfnamefont {N.}~\bibnamefont
  {Warr}}, \bibinfo {author} {\bibfnamefont {J.}~\bibnamefont {Van~de Walle}},
  \bibinfo {author} {\bibfnamefont {M.}~\bibnamefont {Albers}}, \bibinfo
  {author} {\bibfnamefont {F.}~\bibnamefont {Ames}}, \bibinfo {author}
  {\bibfnamefont {B.}~\bibnamefont {Bastin}}, \bibinfo {author} {\bibfnamefont
  {C.}~\bibnamefont {Bauer}}, \bibinfo {author} {\bibfnamefont
  {V.}~\bibnamefont {Bildstein}}, \bibinfo {author} {\bibfnamefont
  {A.}~\bibnamefont {Blazhev}}, \bibinfo {author} {\bibfnamefont
  {S.}~\bibnamefont {B{\"{o}}nig}}, \bibinfo {author} {\bibfnamefont
  {N.}~\bibnamefont {Bree}}, \bibinfo {author} {\bibfnamefont {B.}~\bibnamefont
  {Bruyneel}}, \bibinfo {author} {\bibfnamefont {P.~A.}\ \bibnamefont
  {Butler}}, \bibinfo {author} {\bibfnamefont {J.}~\bibnamefont
  {Cederk{\"{a}}ll}}, \bibinfo {author} {\bibfnamefont {E.}~\bibnamefont
  {Cl{\'{e}}ment}}, \bibinfo {author} {\bibfnamefont {T.~E.}\ \bibnamefont
  {Cocolios}}, \bibinfo {author} {\bibfnamefont {T.}~\bibnamefont {Davinson}},
  \bibinfo {author} {\bibfnamefont {H.}~\bibnamefont {De~Witte}}, \bibinfo
  {author} {\bibfnamefont {P.}~\bibnamefont {Delahaye}}, \bibinfo {author}
  {\bibfnamefont {D.~D.}\ \bibnamefont {DiJulio}}, \bibinfo {author}
  {\bibfnamefont {J.}~\bibnamefont {Diriken}}, \bibinfo {author} {\bibfnamefont
  {J.}~\bibnamefont {Eberth}}, \bibinfo {author} {\bibfnamefont
  {A.}~\bibnamefont {Ekstr{\"{o}}m}}, \bibinfo {author} {\bibfnamefont
  {J.}~\bibnamefont {Elseviers}}, \bibinfo {author} {\bibfnamefont
  {S.}~\bibnamefont {Emhofer}}, \bibinfo {author} {\bibfnamefont {D.~V.}\
  \bibnamefont {Fedorov}}, \bibinfo {author} {\bibfnamefont {V.~N.}\
  \bibnamefont {Fedosseev}}, \bibinfo {author} {\bibfnamefont {S.}~\bibnamefont
  {Franchoo}}, \bibinfo {author} {\bibfnamefont {C.}~\bibnamefont {Fransen}},
  \bibinfo {author} {\bibfnamefont {L.~P.}\ \bibnamefont {Gaffney}}, \bibinfo
  {author} {\bibfnamefont {J.}~\bibnamefont {Gerl}}, \bibinfo {author}
  {\bibfnamefont {G.}~\bibnamefont {Georgiev}}, \bibinfo {author}
  {\bibfnamefont {R.}~\bibnamefont {Gernh{\"{a}}user}}, \bibinfo {author}
  {\bibfnamefont {T.}~\bibnamefont {Grahn}}, \bibinfo {author} {\bibfnamefont
  {D.}~\bibnamefont {Habs}}, \bibinfo {author} {\bibfnamefont {H.}~\bibnamefont
  {Hess}}, \bibinfo {author} {\bibfnamefont {A.~M.}\ \bibnamefont {Hurst}},
  \bibinfo {author} {\bibfnamefont {M.}~\bibnamefont {Huyse}}, \bibinfo
  {author} {\bibfnamefont {O.}~\bibnamefont {Ivanov}}, \bibinfo {author}
  {\bibfnamefont {J.}~\bibnamefont {Iwanicki}}, \bibinfo {author}
  {\bibfnamefont {D.~G.}\ \bibnamefont {Jenkins}}, \bibinfo {author}
  {\bibfnamefont {J.}~\bibnamefont {Jolie}}, \bibinfo {author} {\bibfnamefont
  {N.}~\bibnamefont {Kesteloot}}, \bibinfo {author} {\bibfnamefont
  {O.}~\bibnamefont {Kester}}, \bibinfo {author} {\bibfnamefont
  {U.}~\bibnamefont {K{\"{o}}ster}}, \bibinfo {author} {\bibfnamefont
  {M.}~\bibnamefont {Krauth}}, \bibinfo {author} {\bibfnamefont
  {T.}~\bibnamefont {Kr{\"{o}}ll}}, \bibinfo {author} {\bibfnamefont
  {R.}~\bibnamefont {Kr{\"{u}}cken}}, \bibinfo {author} {\bibfnamefont
  {M.}~\bibnamefont {Lauer}}, \bibinfo {author} {\bibfnamefont
  {J.}~\bibnamefont {Leske}}, \bibinfo {author} {\bibfnamefont {K.~P.}\
  \bibnamefont {Lieb}}, \bibinfo {author} {\bibfnamefont {R.}~\bibnamefont
  {Lutter}}, \bibinfo {author} {\bibfnamefont {L.}~\bibnamefont {Maier}},
  \bibinfo {author} {\bibfnamefont {B.~A.}\ \bibnamefont {Marsh}}, \bibinfo
  {author} {\bibfnamefont {D.}~\bibnamefont {M{\"{u}}cher}}, \bibinfo {author}
  {\bibfnamefont {M.}~\bibnamefont {M{\"{u}}nch}}, \bibinfo {author}
  {\bibfnamefont {O.}~\bibnamefont {Niedermaier}}, \bibinfo {author}
  {\bibfnamefont {J.}~\bibnamefont {Pakarinen}}, \bibinfo {author}
  {\bibfnamefont {M.}~\bibnamefont {Pantea}}, \bibinfo {author} {\bibfnamefont
  {G.}~\bibnamefont {Pascovici}}, \bibinfo {author} {\bibfnamefont
  {N.}~\bibnamefont {Patronis}}, \bibinfo {author} {\bibfnamefont
  {D.}~\bibnamefont {Pauwels}}, \bibinfo {author} {\bibfnamefont
  {A.}~\bibnamefont {Petts}}, \bibinfo {author} {\bibfnamefont
  {N.}~\bibnamefont {Pietralla}}, \bibinfo {author} {\bibfnamefont
  {R.}~\bibnamefont {Raabe}}, \bibinfo {author} {\bibfnamefont
  {E.}~\bibnamefont {Rapisarda}}, \bibinfo {author} {\bibfnamefont
  {P.}~\bibnamefont {Reiter}}, \bibinfo {author} {\bibfnamefont
  {A.}~\bibnamefont {Richter}}, \bibinfo {author} {\bibfnamefont
  {O.}~\bibnamefont {Schaile}}, \bibinfo {author} {\bibfnamefont
  {M.}~\bibnamefont {Scheck}}, \bibinfo {author} {\bibfnamefont
  {H.}~\bibnamefont {Scheit}}, \bibinfo {author} {\bibfnamefont
  {G.}~\bibnamefont {Schrieder}}, \bibinfo {author} {\bibfnamefont
  {D.}~\bibnamefont {Schwalm}}, \bibinfo {author} {\bibfnamefont
  {M.}~\bibnamefont {Seidlitz}}, \bibinfo {author} {\bibfnamefont
  {M.}~\bibnamefont {Seliverstov}}, \bibinfo {author} {\bibfnamefont
  {T.}~\bibnamefont {Sieber}}, \bibinfo {author} {\bibfnamefont
  {H.}~\bibnamefont {Simon}}, \bibinfo {author} {\bibfnamefont {K.~H.}\
  \bibnamefont {Speidel}}, \bibinfo {author} {\bibfnamefont {C.}~\bibnamefont
  {Stahl}}, \bibinfo {author} {\bibfnamefont {I.}~\bibnamefont {Stefanescu}},
  \bibinfo {author} {\bibfnamefont {P.~G.}\ \bibnamefont {Thirolf}}, \bibinfo
  {author} {\bibfnamefont {H.~G.}\ \bibnamefont {Thomas}}, \bibinfo {author}
  {\bibfnamefont {M.}~\bibnamefont {Th{\"{u}}rauf}}, \bibinfo {author}
  {\bibfnamefont {P.}~\bibnamefont {Van~Duppen}}, \bibinfo {author}
  {\bibfnamefont {D.}~\bibnamefont {Voulot}}, \bibinfo {author} {\bibfnamefont
  {R.}~\bibnamefont {Wadsworth}}, \bibinfo {author} {\bibfnamefont
  {G.}~\bibnamefont {Walter}}, \bibinfo {author} {\bibfnamefont
  {D.}~\bibnamefont {Wei{\ss}haar}}, \bibinfo {author} {\bibfnamefont
  {F.}~\bibnamefont {Wenander}}, \bibinfo {author} {\bibfnamefont
  {A.}~\bibnamefont {Wiens}}, \bibinfo {author} {\bibfnamefont
  {K.}~\bibnamefont {Wimmer}}, \bibinfo {author} {\bibfnamefont {B.~H.}\
  \bibnamefont {Wolf}}, \bibinfo {author} {\bibfnamefont {P.~J.}\ \bibnamefont
  {Woods}}, \bibinfo {author} {\bibfnamefont {K.}~\bibnamefont
  {Wrzosek-Lipska}},\ and\ \bibinfo {author} {\bibfnamefont {K.~O.}\
  \bibnamefont {Zell}},\ }\bibfield  {title} {\bibinfo {title} {{The Miniball
  spectrometer}},\ }\href {https://doi.org/10.1140/epja/i2013-13040-9}
  {\bibfield  {journal} {\bibinfo  {journal} {The European Physical Journal A}\
  }\textbf {\bibinfo {volume} {49}},\ \bibinfo {pages} {40} (\bibinfo {year}
  {2013})}\BibitemShut {NoStop}%
\bibitem [{\citenamefont {Guttormsen}\ \emph {et~al.}(1996)\citenamefont
  {Guttormsen}, \citenamefont {Tveter}, \citenamefont {Bergholt}, \citenamefont
  {Ingebretsen},\ and\ \citenamefont {Rekstad}}]{UnfoldingNIM}%
  \BibitemOpen
  \bibfield  {author} {\bibinfo {author} {\bibfnamefont {M.}~\bibnamefont
  {Guttormsen}}, \bibinfo {author} {\bibfnamefont {T.~S.}\ \bibnamefont
  {Tveter}}, \bibinfo {author} {\bibfnamefont {L.}~\bibnamefont {Bergholt}},
  \bibinfo {author} {\bibfnamefont {F.}~\bibnamefont {Ingebretsen}},\ and\
  \bibinfo {author} {\bibfnamefont {J.}~\bibnamefont {Rekstad}},\ }\bibfield
  {title} {\bibinfo {title} {{The unfolding of continuum {$\gamma$}-ray
  spectra}},\ }\href {https://doi.org/10.1016/0168-9002(96)00197-0} {\bibfield
  {journal} {\bibinfo  {journal} {Nuclear Instruments and Methods in Physics
  Research, Section A: Accelerators, Spectrometers, Detectors and Associated
  Equipment}\ }\textbf {\bibinfo {volume} {374}},\ \bibinfo {pages} {371}
  (\bibinfo {year} {1996})}\BibitemShut {NoStop}%
\bibitem [{\citenamefont {Agostinelli}\ \emph {et~al.}(2003)\citenamefont
  {Agostinelli}, \citenamefont {Allison}, \citenamefont {Amako}, \citenamefont
  {Apostolakis}, \citenamefont {Araujo}, \citenamefont {Arce}, \citenamefont
  {Asai}, \citenamefont {Axen}, \citenamefont {Banerjee}, \citenamefont
  {Barrand}, \citenamefont {Behner}, \citenamefont {Bellagamba}, \citenamefont
  {Boudreau}, \citenamefont {Broglia}, \citenamefont {Brunengo}, \citenamefont
  {Burkhardt}, \citenamefont {Chauvie}, \citenamefont {Chuma}, \citenamefont
  {Chytracek}, \citenamefont {Cooperman}, \citenamefont {Cosmo}, \citenamefont
  {Degtyarenko}, \citenamefont {Dell'Acqua}, \citenamefont {Depaola},
  \citenamefont {Dietrich}, \citenamefont {Enami}, \citenamefont {Feliciello},
  \citenamefont {Ferguson}, \citenamefont {Fesefeldt}, \citenamefont {Folger},
  \citenamefont {Foppiano}, \citenamefont {Forti}, \citenamefont {Garelli},
  \citenamefont {Giani}, \citenamefont {Giannitrapani}, \citenamefont {Gibin},
  \citenamefont {Cadenas}, \citenamefont {Gonz{\'{a}}lez}, \citenamefont
  {Abril}, \citenamefont {Greeniaus}, \citenamefont {Greiner}, \citenamefont
  {Grichine}, \citenamefont {Grossheim}, \citenamefont {Guatelli},
  \citenamefont {Gumplinger}, \citenamefont {Hamatsu}, \citenamefont
  {Hashimoto}, \citenamefont {Hasui}, \citenamefont {Heikkinen}, \citenamefont
  {Howard}, \citenamefont {Ivanchenko}, \citenamefont {Johnson}, \citenamefont
  {Jones}, \citenamefont {Kallenbach}, \citenamefont {Kanaya}, \citenamefont
  {Kawabata}, \citenamefont {Kawabata}, \citenamefont {Kawaguti}, \citenamefont
  {Kelner}, \citenamefont {Kent}, \citenamefont {Kimura}, \citenamefont
  {Kodama}, \citenamefont {Kokoulin}, \citenamefont {Kossov}, \citenamefont
  {Kurashige}, \citenamefont {Lamanna}, \citenamefont {Lamp{\'{e}}n},
  \citenamefont {Lara}, \citenamefont {Lefebure}, \citenamefont {Lei},
  \citenamefont {Liendl}, \citenamefont {Lockman}, \citenamefont {Longo},
  \citenamefont {Magni}, \citenamefont {Maire}, \citenamefont {Medernach},
  \citenamefont {Minamimoto}, \citenamefont {de~Freitas}, \citenamefont
  {Morita}, \citenamefont {Murakami}, \citenamefont {Nagamatu}, \citenamefont
  {Nartallo}, \citenamefont {Nieminen}, \citenamefont {Nishimura},
  \citenamefont {Ohtsubo}, \citenamefont {Okamura}, \citenamefont {O'Neale},
  \citenamefont {Oohata}, \citenamefont {Paech}, \citenamefont {Perl},
  \citenamefont {Pfeiffer}, \citenamefont {Pia}, \citenamefont {Ranjard},
  \citenamefont {Rybin}, \citenamefont {Sadilov}, \citenamefont {Salvo},
  \citenamefont {Santin}, \citenamefont {Sasaki}, \citenamefont {Savvas},
  \citenamefont {Sawada}, \citenamefont {Scherer}, \citenamefont {Sei},
  \citenamefont {Sirotenko}, \citenamefont {Smith}, \citenamefont {Starkov},
  \citenamefont {Stoecker}, \citenamefont {Sulkimo}, \citenamefont {Takahata},
  \citenamefont {Tanaka}, \citenamefont {Tcherniaev}, \citenamefont {Tehrani},
  \citenamefont {Tropeano}, \citenamefont {Truscott}, \citenamefont {Uno},
  \citenamefont {Urban}, \citenamefont {Urban}, \citenamefont {Verderi},
  \citenamefont {Walkden}, \citenamefont {Wander}, \citenamefont {Weber},
  \citenamefont {Wellisch}, \citenamefont {Wenaus}, \citenamefont {Williams},
  \citenamefont {Wright}, \citenamefont {Yamada}, \citenamefont {Yoshida},\
  and\ \citenamefont {Zschiesche}}]{AGOSTINELLI2003250}%
  \BibitemOpen
  \bibfield  {author} {\bibinfo {author} {\bibfnamefont {S.}~\bibnamefont
  {Agostinelli}}, \bibinfo {author} {\bibfnamefont {J.}~\bibnamefont
  {Allison}}, \bibinfo {author} {\bibfnamefont {K.}~\bibnamefont {Amako}},
  \bibinfo {author} {\bibfnamefont {J.}~\bibnamefont {Apostolakis}}, \bibinfo
  {author} {\bibfnamefont {H.}~\bibnamefont {Araujo}}, \bibinfo {author}
  {\bibfnamefont {P.}~\bibnamefont {Arce}}, \bibinfo {author} {\bibfnamefont
  {M.}~\bibnamefont {Asai}}, \bibinfo {author} {\bibfnamefont {D.}~\bibnamefont
  {Axen}}, \bibinfo {author} {\bibfnamefont {S.}~\bibnamefont {Banerjee}},
  \bibinfo {author} {\bibfnamefont {G.}~\bibnamefont {Barrand}}, \bibinfo
  {author} {\bibfnamefont {F.}~\bibnamefont {Behner}}, \bibinfo {author}
  {\bibfnamefont {L.}~\bibnamefont {Bellagamba}}, \bibinfo {author}
  {\bibfnamefont {J.}~\bibnamefont {Boudreau}}, \bibinfo {author}
  {\bibfnamefont {L.}~\bibnamefont {Broglia}}, \bibinfo {author} {\bibfnamefont
  {A.}~\bibnamefont {Brunengo}}, \bibinfo {author} {\bibfnamefont
  {H.}~\bibnamefont {Burkhardt}}, \bibinfo {author} {\bibfnamefont
  {S.}~\bibnamefont {Chauvie}}, \bibinfo {author} {\bibfnamefont
  {J.}~\bibnamefont {Chuma}}, \bibinfo {author} {\bibfnamefont
  {R.}~\bibnamefont {Chytracek}}, \bibinfo {author} {\bibfnamefont
  {G.}~\bibnamefont {Cooperman}}, \bibinfo {author} {\bibfnamefont
  {G.}~\bibnamefont {Cosmo}}, \bibinfo {author} {\bibfnamefont
  {P.}~\bibnamefont {Degtyarenko}}, \bibinfo {author} {\bibfnamefont
  {A.}~\bibnamefont {Dell'Acqua}}, \bibinfo {author} {\bibfnamefont
  {G.}~\bibnamefont {Depaola}}, \bibinfo {author} {\bibfnamefont
  {D.}~\bibnamefont {Dietrich}}, \bibinfo {author} {\bibfnamefont
  {R.}~\bibnamefont {Enami}}, \bibinfo {author} {\bibfnamefont
  {A.}~\bibnamefont {Feliciello}}, \bibinfo {author} {\bibfnamefont
  {C.}~\bibnamefont {Ferguson}}, \bibinfo {author} {\bibfnamefont
  {H.}~\bibnamefont {Fesefeldt}}, \bibinfo {author} {\bibfnamefont
  {G.}~\bibnamefont {Folger}}, \bibinfo {author} {\bibfnamefont
  {F.}~\bibnamefont {Foppiano}}, \bibinfo {author} {\bibfnamefont
  {A.}~\bibnamefont {Forti}}, \bibinfo {author} {\bibfnamefont
  {S.}~\bibnamefont {Garelli}}, \bibinfo {author} {\bibfnamefont
  {S.}~\bibnamefont {Giani}}, \bibinfo {author} {\bibfnamefont
  {R.}~\bibnamefont {Giannitrapani}}, \bibinfo {author} {\bibfnamefont
  {D.}~\bibnamefont {Gibin}}, \bibinfo {author} {\bibfnamefont {J.~J.~G.}\
  \bibnamefont {Cadenas}}, \bibinfo {author} {\bibfnamefont {I.}~\bibnamefont
  {Gonz{\'{a}}lez}}, \bibinfo {author} {\bibfnamefont {G.~G.}\ \bibnamefont
  {Abril}}, \bibinfo {author} {\bibfnamefont {G.}~\bibnamefont {Greeniaus}},
  \bibinfo {author} {\bibfnamefont {W.}~\bibnamefont {Greiner}}, \bibinfo
  {author} {\bibfnamefont {V.}~\bibnamefont {Grichine}}, \bibinfo {author}
  {\bibfnamefont {A.}~\bibnamefont {Grossheim}}, \bibinfo {author}
  {\bibfnamefont {S.}~\bibnamefont {Guatelli}}, \bibinfo {author}
  {\bibfnamefont {P.}~\bibnamefont {Gumplinger}}, \bibinfo {author}
  {\bibfnamefont {R.}~\bibnamefont {Hamatsu}}, \bibinfo {author} {\bibfnamefont
  {K.}~\bibnamefont {Hashimoto}}, \bibinfo {author} {\bibfnamefont
  {H.}~\bibnamefont {Hasui}}, \bibinfo {author} {\bibfnamefont
  {A.}~\bibnamefont {Heikkinen}}, \bibinfo {author} {\bibfnamefont
  {A.}~\bibnamefont {Howard}}, \bibinfo {author} {\bibfnamefont
  {V.}~\bibnamefont {Ivanchenko}}, \bibinfo {author} {\bibfnamefont
  {A.}~\bibnamefont {Johnson}}, \bibinfo {author} {\bibfnamefont {F.~W.}\
  \bibnamefont {Jones}}, \bibinfo {author} {\bibfnamefont {J.}~\bibnamefont
  {Kallenbach}}, \bibinfo {author} {\bibfnamefont {N.}~\bibnamefont {Kanaya}},
  \bibinfo {author} {\bibfnamefont {M.}~\bibnamefont {Kawabata}}, \bibinfo
  {author} {\bibfnamefont {Y.}~\bibnamefont {Kawabata}}, \bibinfo {author}
  {\bibfnamefont {M.}~\bibnamefont {Kawaguti}}, \bibinfo {author}
  {\bibfnamefont {S.}~\bibnamefont {Kelner}}, \bibinfo {author} {\bibfnamefont
  {P.}~\bibnamefont {Kent}}, \bibinfo {author} {\bibfnamefont {A.}~\bibnamefont
  {Kimura}}, \bibinfo {author} {\bibfnamefont {T.}~\bibnamefont {Kodama}},
  \bibinfo {author} {\bibfnamefont {R.}~\bibnamefont {Kokoulin}}, \bibinfo
  {author} {\bibfnamefont {M.}~\bibnamefont {Kossov}}, \bibinfo {author}
  {\bibfnamefont {H.}~\bibnamefont {Kurashige}}, \bibinfo {author}
  {\bibfnamefont {E.}~\bibnamefont {Lamanna}}, \bibinfo {author} {\bibfnamefont
  {T.}~\bibnamefont {Lamp{\'{e}}n}}, \bibinfo {author} {\bibfnamefont
  {V.}~\bibnamefont {Lara}}, \bibinfo {author} {\bibfnamefont {V.}~\bibnamefont
  {Lefebure}}, \bibinfo {author} {\bibfnamefont {F.}~\bibnamefont {Lei}},
  \bibinfo {author} {\bibfnamefont {M.}~\bibnamefont {Liendl}}, \bibinfo
  {author} {\bibfnamefont {W.}~\bibnamefont {Lockman}}, \bibinfo {author}
  {\bibfnamefont {F.}~\bibnamefont {Longo}}, \bibinfo {author} {\bibfnamefont
  {S.}~\bibnamefont {Magni}}, \bibinfo {author} {\bibfnamefont
  {M.}~\bibnamefont {Maire}}, \bibinfo {author} {\bibfnamefont
  {E.}~\bibnamefont {Medernach}}, \bibinfo {author} {\bibfnamefont
  {K.}~\bibnamefont {Minamimoto}}, \bibinfo {author} {\bibfnamefont {P.~M.}\
  \bibnamefont {de~Freitas}}, \bibinfo {author} {\bibfnamefont
  {Y.}~\bibnamefont {Morita}}, \bibinfo {author} {\bibfnamefont
  {K.}~\bibnamefont {Murakami}}, \bibinfo {author} {\bibfnamefont
  {M.}~\bibnamefont {Nagamatu}}, \bibinfo {author} {\bibfnamefont
  {R.}~\bibnamefont {Nartallo}}, \bibinfo {author} {\bibfnamefont
  {P.}~\bibnamefont {Nieminen}}, \bibinfo {author} {\bibfnamefont
  {T.}~\bibnamefont {Nishimura}}, \bibinfo {author} {\bibfnamefont
  {K.}~\bibnamefont {Ohtsubo}}, \bibinfo {author} {\bibfnamefont
  {M.}~\bibnamefont {Okamura}}, \bibinfo {author} {\bibfnamefont
  {S.}~\bibnamefont {O'Neale}}, \bibinfo {author} {\bibfnamefont
  {Y.}~\bibnamefont {Oohata}}, \bibinfo {author} {\bibfnamefont
  {K.}~\bibnamefont {Paech}}, \bibinfo {author} {\bibfnamefont
  {J.}~\bibnamefont {Perl}}, \bibinfo {author} {\bibfnamefont {A.}~\bibnamefont
  {Pfeiffer}}, \bibinfo {author} {\bibfnamefont {M.~G.}\ \bibnamefont {Pia}},
  \bibinfo {author} {\bibfnamefont {F.}~\bibnamefont {Ranjard}}, \bibinfo
  {author} {\bibfnamefont {A.}~\bibnamefont {Rybin}}, \bibinfo {author}
  {\bibfnamefont {S.}~\bibnamefont {Sadilov}}, \bibinfo {author} {\bibfnamefont
  {E.~D.}\ \bibnamefont {Salvo}}, \bibinfo {author} {\bibfnamefont
  {G.}~\bibnamefont {Santin}}, \bibinfo {author} {\bibfnamefont
  {T.}~\bibnamefont {Sasaki}}, \bibinfo {author} {\bibfnamefont
  {N.}~\bibnamefont {Savvas}}, \bibinfo {author} {\bibfnamefont
  {Y.}~\bibnamefont {Sawada}}, \bibinfo {author} {\bibfnamefont
  {S.}~\bibnamefont {Scherer}}, \bibinfo {author} {\bibfnamefont
  {S.}~\bibnamefont {Sei}}, \bibinfo {author} {\bibfnamefont {V.}~\bibnamefont
  {Sirotenko}}, \bibinfo {author} {\bibfnamefont {D.}~\bibnamefont {Smith}},
  \bibinfo {author} {\bibfnamefont {N.}~\bibnamefont {Starkov}}, \bibinfo
  {author} {\bibfnamefont {H.}~\bibnamefont {Stoecker}}, \bibinfo {author}
  {\bibfnamefont {J.}~\bibnamefont {Sulkimo}}, \bibinfo {author} {\bibfnamefont
  {M.}~\bibnamefont {Takahata}}, \bibinfo {author} {\bibfnamefont
  {S.}~\bibnamefont {Tanaka}}, \bibinfo {author} {\bibfnamefont
  {E.}~\bibnamefont {Tcherniaev}}, \bibinfo {author} {\bibfnamefont {E.~S.}\
  \bibnamefont {Tehrani}}, \bibinfo {author} {\bibfnamefont {M.}~\bibnamefont
  {Tropeano}}, \bibinfo {author} {\bibfnamefont {P.}~\bibnamefont {Truscott}},
  \bibinfo {author} {\bibfnamefont {H.}~\bibnamefont {Uno}}, \bibinfo {author}
  {\bibfnamefont {L.}~\bibnamefont {Urban}}, \bibinfo {author} {\bibfnamefont
  {P.}~\bibnamefont {Urban}}, \bibinfo {author} {\bibfnamefont
  {M.}~\bibnamefont {Verderi}}, \bibinfo {author} {\bibfnamefont
  {A.}~\bibnamefont {Walkden}}, \bibinfo {author} {\bibfnamefont
  {W.}~\bibnamefont {Wander}}, \bibinfo {author} {\bibfnamefont
  {H.}~\bibnamefont {Weber}}, \bibinfo {author} {\bibfnamefont {J.~P.}\
  \bibnamefont {Wellisch}}, \bibinfo {author} {\bibfnamefont {T.}~\bibnamefont
  {Wenaus}}, \bibinfo {author} {\bibfnamefont {D.~C.}\ \bibnamefont
  {Williams}}, \bibinfo {author} {\bibfnamefont {D.}~\bibnamefont {Wright}},
  \bibinfo {author} {\bibfnamefont {T.}~\bibnamefont {Yamada}}, \bibinfo
  {author} {\bibfnamefont {H.}~\bibnamefont {Yoshida}},\ and\ \bibinfo {author}
  {\bibfnamefont {D.}~\bibnamefont {Zschiesche}},\ }\bibfield  {title}
  {\bibinfo {title} {{Geant4—a simulation toolkit}},\ }\href
  {https://doi.org/10.1016/S0168-9002(03)01368-8} {\bibfield  {journal}
  {\bibinfo  {journal} {Nuclear Instruments and Methods in Physics Research
  Section A: Accelerators, Spectrometers, Detectors and Associated Equipment}\
  }\textbf {\bibinfo {volume} {506}},\ \bibinfo {pages} {250} (\bibinfo {year}
  {2003})}\BibitemShut {NoStop}%
\bibitem [{\citenamefont {Ingeberg}(2022)}]{Ingeberg2022d}%
  \BibitemOpen
  \bibfield  {author} {\bibinfo {author} {\bibfnamefont {V.~W.}\ \bibnamefont
  {Ingeberg}},\ }\href {https://doi.org/10.5281/ZENODO.6075853} {\bibinfo
  {title} {{vetlewi/MiniballREX: Ni67,
  https://github.com/vetlewi/MiniballREX}}} (\bibinfo {year}
  {2022})\BibitemShut {NoStop}%
\bibitem [{\citenamefont {Guttormsen}\ \emph {et~al.}(1987)\citenamefont
  {Guttormsen}, \citenamefont {Rams{\o}y},\ and\ \citenamefont
  {Rekstad}}]{FirstGenerationNIM}%
  \BibitemOpen
  \bibfield  {author} {\bibinfo {author} {\bibfnamefont {M.}~\bibnamefont
  {Guttormsen}}, \bibinfo {author} {\bibfnamefont {T.}~\bibnamefont
  {Rams{\o}y}},\ and\ \bibinfo {author} {\bibfnamefont {J.}~\bibnamefont
  {Rekstad}},\ }\bibfield  {title} {\bibinfo {title} {{The first generation of
  {$\gamma$}-rays from hot nuclei}},\ }\href
  {https://doi.org/10.1016/0168-9002(87)91221-6} {\bibfield  {journal}
  {\bibinfo  {journal} {Nuclear Inst. and Methods in Physics Research, A}\
  }\textbf {\bibinfo {volume} {255}},\ \bibinfo {pages} {518} (\bibinfo {year}
  {1987})}\BibitemShut {NoStop}%
\bibitem [{\citenamefont {Larsen}\ \emph {et~al.}(2011)\citenamefont {Larsen},
  \citenamefont {Guttormsen}, \citenamefont {Krti\ifmmode~\check{c}\else
  \v{c}\fi{}ka}, \citenamefont {B\ifmmode~\check{e}\else \v{e}\fi{}t\'ak},
  \citenamefont {B\"urger}, \citenamefont {G\"orgen}, \citenamefont {Nyhus},
  \citenamefont {Rekstad}, \citenamefont {Schiller}, \citenamefont {Siem},
  \citenamefont {Toft}, \citenamefont {Tveten}, \citenamefont {Voinov},\ and\
  \citenamefont {Wikan}}]{Larsen2011}%
  \BibitemOpen
  \bibfield  {author} {\bibinfo {author} {\bibfnamefont {A.~C.}\ \bibnamefont
  {Larsen}}, \bibinfo {author} {\bibfnamefont {M.}~\bibnamefont {Guttormsen}},
  \bibinfo {author} {\bibfnamefont {M.}~\bibnamefont
  {Krti\ifmmode~\check{c}\else \v{c}\fi{}ka}}, \bibinfo {author} {\bibfnamefont
  {E.}~\bibnamefont {B\ifmmode~\check{e}\else \v{e}\fi{}t\'ak}}, \bibinfo
  {author} {\bibfnamefont {A.}~\bibnamefont {B\"urger}}, \bibinfo {author}
  {\bibfnamefont {A.}~\bibnamefont {G\"orgen}}, \bibinfo {author}
  {\bibfnamefont {H.~T.}\ \bibnamefont {Nyhus}}, \bibinfo {author}
  {\bibfnamefont {J.}~\bibnamefont {Rekstad}}, \bibinfo {author} {\bibfnamefont
  {A.}~\bibnamefont {Schiller}}, \bibinfo {author} {\bibfnamefont
  {S.}~\bibnamefont {Siem}}, \bibinfo {author} {\bibfnamefont {H.~K.}\
  \bibnamefont {Toft}}, \bibinfo {author} {\bibfnamefont {G.~M.}\ \bibnamefont
  {Tveten}}, \bibinfo {author} {\bibfnamefont {A.~V.}\ \bibnamefont {Voinov}},\
  and\ \bibinfo {author} {\bibfnamefont {K.}~\bibnamefont {Wikan}},\ }\bibfield
   {title} {\bibinfo {title} {Analysis of possible systematic errors in the
  oslo method},\ }\href {https://doi.org/10.1103/PhysRevC.83.034315} {\bibfield
   {journal} {\bibinfo  {journal} {Physical Review C}\ }\textbf {\bibinfo
  {volume} {83}},\ \bibinfo {pages} {034315} (\bibinfo {year}
  {2011})}\BibitemShut {NoStop}%
\bibitem [{\citenamefont {Ingeberg}\ \emph {et~al.}(2022)\citenamefont
  {Ingeberg}, \citenamefont {Jones}, \citenamefont {Msebi}, \citenamefont
  {Siem}, \citenamefont {Wiedeking}, \citenamefont {Avaa}, \citenamefont
  {Chisapi}, \citenamefont {Lawrie}, \citenamefont {Malatji}, \citenamefont
  {Makhathini}, \citenamefont {Noncolela},\ and\ \citenamefont
  {Shirinda}}]{Ingeberg2022b}%
  \BibitemOpen
  \bibfield  {author} {\bibinfo {author} {\bibfnamefont {V.~W.}\ \bibnamefont
  {Ingeberg}}, \bibinfo {author} {\bibfnamefont {P.}~\bibnamefont {Jones}},
  \bibinfo {author} {\bibfnamefont {L.}~\bibnamefont {Msebi}}, \bibinfo
  {author} {\bibfnamefont {S.}~\bibnamefont {Siem}}, \bibinfo {author}
  {\bibfnamefont {M.}~\bibnamefont {Wiedeking}}, \bibinfo {author}
  {\bibfnamefont {A.~A.}\ \bibnamefont {Avaa}}, \bibinfo {author}
  {\bibfnamefont {M.~V.}\ \bibnamefont {Chisapi}}, \bibinfo {author}
  {\bibfnamefont {E.~A.}\ \bibnamefont {Lawrie}}, \bibinfo {author}
  {\bibfnamefont {K.~L.}\ \bibnamefont {Malatji}}, \bibinfo {author}
  {\bibfnamefont {L.}~\bibnamefont {Makhathini}}, \bibinfo {author}
  {\bibfnamefont {S.~P.}\ \bibnamefont {Noncolela}},\ and\ \bibinfo {author}
  {\bibfnamefont {O.}~\bibnamefont {Shirinda}},\ }\bibfield  {title} {\bibinfo
  {title} {Nuclear level density and $\ensuremath{\gamma}$-ray strength
  function of $^{63}\mathrm{Ni}$},\ }\href
  {https://doi.org/10.1103/PhysRevC.106.054315} {\bibfield  {journal} {\bibinfo
   {journal} {Physical Review C}\ }\textbf {\bibinfo {volume} {106}},\ \bibinfo
  {pages} {054315} (\bibinfo {year} {2022})}\BibitemShut {NoStop}%
\bibitem [{\citenamefont {Larsen}\ \emph {et~al.}(2016)\citenamefont {Larsen},
  \citenamefont {Guttormsen}, \citenamefont {Schwengner}, \citenamefont
  {Bleuel}, \citenamefont {Goriely}, \citenamefont {Harissopulos},
  \citenamefont {Bello~Garrote}, \citenamefont {Byun}, \citenamefont {Eriksen},
  \citenamefont {Giacoppo}, \citenamefont {G\"orgen}, \citenamefont {Hagen},
  \citenamefont {Klintefjord}, \citenamefont {Renstr\o{}m}, \citenamefont
  {Rose}, \citenamefont {Sahin}, \citenamefont {Siem}, \citenamefont {Tornyi},
  \citenamefont {Tveten}, \citenamefont {Voinov},\ and\ \citenamefont
  {Wiedeking}}]{Larsen2016}%
  \BibitemOpen
  \bibfield  {author} {\bibinfo {author} {\bibfnamefont {A.~C.}\ \bibnamefont
  {Larsen}}, \bibinfo {author} {\bibfnamefont {M.}~\bibnamefont {Guttormsen}},
  \bibinfo {author} {\bibfnamefont {R.}~\bibnamefont {Schwengner}}, \bibinfo
  {author} {\bibfnamefont {D.~L.}\ \bibnamefont {Bleuel}}, \bibinfo {author}
  {\bibfnamefont {S.}~\bibnamefont {Goriely}}, \bibinfo {author} {\bibfnamefont
  {S.}~\bibnamefont {Harissopulos}}, \bibinfo {author} {\bibfnamefont {F.~L.}\
  \bibnamefont {Bello~Garrote}}, \bibinfo {author} {\bibfnamefont
  {Y.}~\bibnamefont {Byun}}, \bibinfo {author} {\bibfnamefont {T.~K.}\
  \bibnamefont {Eriksen}}, \bibinfo {author} {\bibfnamefont {F.}~\bibnamefont
  {Giacoppo}}, \bibinfo {author} {\bibfnamefont {A.}~\bibnamefont {G\"orgen}},
  \bibinfo {author} {\bibfnamefont {T.~W.}\ \bibnamefont {Hagen}}, \bibinfo
  {author} {\bibfnamefont {M.}~\bibnamefont {Klintefjord}}, \bibinfo {author}
  {\bibfnamefont {T.}~\bibnamefont {Renstr\o{}m}}, \bibinfo {author}
  {\bibfnamefont {S.~J.}\ \bibnamefont {Rose}}, \bibinfo {author}
  {\bibfnamefont {E.}~\bibnamefont {Sahin}}, \bibinfo {author} {\bibfnamefont
  {S.}~\bibnamefont {Siem}}, \bibinfo {author} {\bibfnamefont {T.~G.}\
  \bibnamefont {Tornyi}}, \bibinfo {author} {\bibfnamefont {G.~M.}\
  \bibnamefont {Tveten}}, \bibinfo {author} {\bibfnamefont {A.~V.}\
  \bibnamefont {Voinov}},\ and\ \bibinfo {author} {\bibfnamefont
  {M.}~\bibnamefont {Wiedeking}},\ }\bibfield  {title} {\bibinfo {title}
  {Experimentally constrained $(p,\ensuremath{\gamma})^{89}\mathrm{Y}$ and
  $(n,\ensuremath{\gamma})^{89}\mathrm{Y}$ reaction rates relevant to
  $p$-process nucleosynthesis},\ }\href
  {https://doi.org/10.1103/PhysRevC.93.045810} {\bibfield  {journal} {\bibinfo
  {journal} {Physical Review C}\ }\textbf {\bibinfo {volume} {93}},\ \bibinfo
  {pages} {045810} (\bibinfo {year} {2016})}\BibitemShut {NoStop}%
\bibitem [{\citenamefont {Brits}\ \emph {et~al.}(2019)\citenamefont {Brits},
  \citenamefont {Malatji}, \citenamefont {Wiedeking}, \citenamefont {Kheswa},
  \citenamefont {Goriely}, \citenamefont {Garrote}, \citenamefont {Bleuel},
  \citenamefont {Giacoppo}, \citenamefont {G\"orgen}, \citenamefont
  {Guttormsen}, \citenamefont {Hadynska-Klek}, \citenamefont {Hagen},
  \citenamefont {Hilaire}, \citenamefont {Ingeberg}, \citenamefont {Jia},
  \citenamefont {Klintefjord}, \citenamefont {Larsen}, \citenamefont {Majola},
  \citenamefont {Papka}, \citenamefont {P\'eru}, \citenamefont {Qi},
  \citenamefont {Renstr\o{}m}, \citenamefont {Rose}, \citenamefont {Sahin},
  \citenamefont {Siem}, \citenamefont {Tveten},\ and\ \citenamefont
  {Zeiser}}]{Brits2019}%
  \BibitemOpen
  \bibfield  {author} {\bibinfo {author} {\bibfnamefont {C.~P.}\ \bibnamefont
  {Brits}}, \bibinfo {author} {\bibfnamefont {K.~L.}\ \bibnamefont {Malatji}},
  \bibinfo {author} {\bibfnamefont {M.}~\bibnamefont {Wiedeking}}, \bibinfo
  {author} {\bibfnamefont {B.~V.}\ \bibnamefont {Kheswa}}, \bibinfo {author}
  {\bibfnamefont {S.}~\bibnamefont {Goriely}}, \bibinfo {author} {\bibfnamefont
  {F.~L.~B.}\ \bibnamefont {Garrote}}, \bibinfo {author} {\bibfnamefont
  {D.~L.}\ \bibnamefont {Bleuel}}, \bibinfo {author} {\bibfnamefont
  {F.}~\bibnamefont {Giacoppo}}, \bibinfo {author} {\bibfnamefont
  {A.}~\bibnamefont {G\"orgen}}, \bibinfo {author} {\bibfnamefont
  {M.}~\bibnamefont {Guttormsen}}, \bibinfo {author} {\bibfnamefont
  {K.}~\bibnamefont {Hadynska-Klek}}, \bibinfo {author} {\bibfnamefont {T.~W.}\
  \bibnamefont {Hagen}}, \bibinfo {author} {\bibfnamefont {S.}~\bibnamefont
  {Hilaire}}, \bibinfo {author} {\bibfnamefont {V.~W.}\ \bibnamefont
  {Ingeberg}}, \bibinfo {author} {\bibfnamefont {H.}~\bibnamefont {Jia}},
  \bibinfo {author} {\bibfnamefont {M.}~\bibnamefont {Klintefjord}}, \bibinfo
  {author} {\bibfnamefont {A.~C.}\ \bibnamefont {Larsen}}, \bibinfo {author}
  {\bibfnamefont {S.~N.~T.}\ \bibnamefont {Majola}}, \bibinfo {author}
  {\bibfnamefont {P.}~\bibnamefont {Papka}}, \bibinfo {author} {\bibfnamefont
  {S.}~\bibnamefont {P\'eru}}, \bibinfo {author} {\bibfnamefont
  {B.}~\bibnamefont {Qi}}, \bibinfo {author} {\bibfnamefont {T.}~\bibnamefont
  {Renstr\o{}m}}, \bibinfo {author} {\bibfnamefont {S.~J.}\ \bibnamefont
  {Rose}}, \bibinfo {author} {\bibfnamefont {E.}~\bibnamefont {Sahin}},
  \bibinfo {author} {\bibfnamefont {S.}~\bibnamefont {Siem}}, \bibinfo {author}
  {\bibfnamefont {G.~M.}\ \bibnamefont {Tveten}},\ and\ \bibinfo {author}
  {\bibfnamefont {F.}~\bibnamefont {Zeiser}},\ }\bibfield  {title} {\bibinfo
  {title} {Nuclear level densities and $\ensuremath{\gamma}$-ray strength
  functions of $^{180,181,182}\mathrm{Ta}$},\ }\href
  {https://doi.org/10.1103/PhysRevC.99.054330} {\bibfield  {journal} {\bibinfo
  {journal} {Phys. Rev. C}\ }\textbf {\bibinfo {volume} {99}},\ \bibinfo
  {pages} {054330} (\bibinfo {year} {2019})}\BibitemShut {NoStop}%
\bibitem [{\citenamefont {Kheswa}\ \emph {et~al.}(2015)\citenamefont {Kheswa},
  \citenamefont {Wiedeking}, \citenamefont {Giacoppo}, \citenamefont {Goriely},
  \citenamefont {Guttormsen}, \citenamefont {Larsen}, \citenamefont
  {Bello~Garrote}, \citenamefont {Eriksen}, \citenamefont {G{\"{o}}rgen},
  \citenamefont {Hagen}, \citenamefont {Koehler}, \citenamefont {Klintefjord},
  \citenamefont {Nyhus}, \citenamefont {Papka}, \citenamefont {Renstr{\o}m},
  \citenamefont {Rose}, \citenamefont {Sahin}, \citenamefont {Siem},\ and\
  \citenamefont {Tornyi}}]{Kheswa2015GalacticProperties}%
  \BibitemOpen
  \bibfield  {author} {\bibinfo {author} {\bibfnamefont {B.~V.}\ \bibnamefont
  {Kheswa}}, \bibinfo {author} {\bibfnamefont {M.}~\bibnamefont {Wiedeking}},
  \bibinfo {author} {\bibfnamefont {F.}~\bibnamefont {Giacoppo}}, \bibinfo
  {author} {\bibfnamefont {S.}~\bibnamefont {Goriely}}, \bibinfo {author}
  {\bibfnamefont {M.}~\bibnamefont {Guttormsen}}, \bibinfo {author}
  {\bibfnamefont {A.~C.}\ \bibnamefont {Larsen}}, \bibinfo {author}
  {\bibfnamefont {F.~L.}\ \bibnamefont {Bello~Garrote}}, \bibinfo {author}
  {\bibfnamefont {T.~K.}\ \bibnamefont {Eriksen}}, \bibinfo {author}
  {\bibfnamefont {A.}~\bibnamefont {G{\"{o}}rgen}}, \bibinfo {author}
  {\bibfnamefont {T.~W.}\ \bibnamefont {Hagen}}, \bibinfo {author}
  {\bibfnamefont {P.~E.}\ \bibnamefont {Koehler}}, \bibinfo {author}
  {\bibfnamefont {M.}~\bibnamefont {Klintefjord}}, \bibinfo {author}
  {\bibfnamefont {H.~T.}\ \bibnamefont {Nyhus}}, \bibinfo {author}
  {\bibfnamefont {P.}~\bibnamefont {Papka}}, \bibinfo {author} {\bibfnamefont
  {T.}~\bibnamefont {Renstr{\o}m}}, \bibinfo {author} {\bibfnamefont
  {S.}~\bibnamefont {Rose}}, \bibinfo {author} {\bibfnamefont {E.}~\bibnamefont
  {Sahin}}, \bibinfo {author} {\bibfnamefont {S.}~\bibnamefont {Siem}},\ and\
  \bibinfo {author} {\bibfnamefont {T.}~\bibnamefont {Tornyi}},\ }\bibfield
  {title} {\bibinfo {title} {{Galactic production of138La: Impact of138,139La
  statistical properties}},\ }\href
  {https://doi.org/10.1016/j.physletb.2015.03.065} {\bibfield  {journal}
  {\bibinfo  {journal} {Physics Letters, Section B: Nuclear, Elementary
  Particle and High-Energy Physics}\ }\textbf {\bibinfo {volume} {744}},\
  \bibinfo {pages} {268} (\bibinfo {year} {2015})}\BibitemShut {NoStop}%
\bibitem [{\citenamefont {Sweet}\ \emph {et~al.}(2024)\citenamefont {Sweet},
  \citenamefont {Bleuel}, \citenamefont {Scielzo}, \citenamefont {Bernstein},
  \citenamefont {Dombos}, \citenamefont {Goldblum}, \citenamefont {Harris},
  \citenamefont {Laplace}, \citenamefont {Larsen}, \citenamefont {Lewis},
  \citenamefont {Liddick}, \citenamefont {Lyons}, \citenamefont {Naqvi},
  \citenamefont {Palmisano-Kyle}, \citenamefont {Richard}, \citenamefont
  {Smith}, \citenamefont {Spyrou}, \citenamefont {Vujic},\ and\ \citenamefont
  {Wiedeking}}]{PhysRevC.109.054305}%
  \BibitemOpen
  \bibfield  {author} {\bibinfo {author} {\bibfnamefont {A.}~\bibnamefont
  {Sweet}}, \bibinfo {author} {\bibfnamefont {D.~L.}\ \bibnamefont {Bleuel}},
  \bibinfo {author} {\bibfnamefont {N.~D.}\ \bibnamefont {Scielzo}}, \bibinfo
  {author} {\bibfnamefont {L.~A.}\ \bibnamefont {Bernstein}}, \bibinfo {author}
  {\bibfnamefont {A.~C.}\ \bibnamefont {Dombos}}, \bibinfo {author}
  {\bibfnamefont {B.~L.}\ \bibnamefont {Goldblum}}, \bibinfo {author}
  {\bibfnamefont {C.~M.}\ \bibnamefont {Harris}}, \bibinfo {author}
  {\bibfnamefont {T.~A.}\ \bibnamefont {Laplace}}, \bibinfo {author}
  {\bibfnamefont {A.~C.}\ \bibnamefont {Larsen}}, \bibinfo {author}
  {\bibfnamefont {R.}~\bibnamefont {Lewis}}, \bibinfo {author} {\bibfnamefont
  {S.~N.}\ \bibnamefont {Liddick}}, \bibinfo {author} {\bibfnamefont {S.~M.}\
  \bibnamefont {Lyons}}, \bibinfo {author} {\bibfnamefont {F.}~\bibnamefont
  {Naqvi}}, \bibinfo {author} {\bibfnamefont {A.}~\bibnamefont
  {Palmisano-Kyle}}, \bibinfo {author} {\bibfnamefont {A.~L.}\ \bibnamefont
  {Richard}}, \bibinfo {author} {\bibfnamefont {M.~K.}\ \bibnamefont {Smith}},
  \bibinfo {author} {\bibfnamefont {A.}~\bibnamefont {Spyrou}}, \bibinfo
  {author} {\bibfnamefont {J.}~\bibnamefont {Vujic}},\ and\ \bibinfo {author}
  {\bibfnamefont {M.}~\bibnamefont {Wiedeking}},\ }\bibfield  {title} {\bibinfo
  {title} {Nuclear level density and $\ensuremath{\gamma}$-decay strength of
  $^{93}\mathrm{Sr}$},\ }\href {https://doi.org/10.1103/PhysRevC.109.054305}
  {\bibfield  {journal} {\bibinfo  {journal} {Phys. Rev. C}\ }\textbf {\bibinfo
  {volume} {109}},\ \bibinfo {pages} {054305} (\bibinfo {year}
  {2024})}\BibitemShut {NoStop}%
\bibitem [{\citenamefont {Spyrou}\ \emph {et~al.}(2024)\citenamefont {Spyrou},
  \citenamefont {M\"ucher}, \citenamefont {Denissenkov}, \citenamefont
  {Herwig}, \citenamefont {Good}, \citenamefont {Balk}, \citenamefont {Berg},
  \citenamefont {Bleuel}, \citenamefont {Clark}, \citenamefont {Dembski},
  \citenamefont {DeYoung}, \citenamefont {Greaves}, \citenamefont {Guttormsen},
  \citenamefont {Harris}, \citenamefont {Larsen}, \citenamefont {Liddick},
  \citenamefont {Lyons}, \citenamefont {Markova}, \citenamefont {Mogannam},
  \citenamefont {Nikas}, \citenamefont {Owens-Fryar}, \citenamefont
  {Palmisano-Kyle}, \citenamefont {Perdikakis}, \citenamefont {Pogliano},
  \citenamefont {Quintieri}, \citenamefont {Richard}, \citenamefont
  {Santiago-Gonzalez}, \citenamefont {Savard}, \citenamefont {Smith},
  \citenamefont {Sweet}, \citenamefont {Tsantiri},\ and\ \citenamefont
  {Wiedeking}}]{PhysRevLett.132.202701}%
  \BibitemOpen
  \bibfield  {author} {\bibinfo {author} {\bibfnamefont {A.}~\bibnamefont
  {Spyrou}}, \bibinfo {author} {\bibfnamefont {D.}~\bibnamefont {M\"ucher}},
  \bibinfo {author} {\bibfnamefont {P.~A.}\ \bibnamefont {Denissenkov}},
  \bibinfo {author} {\bibfnamefont {F.}~\bibnamefont {Herwig}}, \bibinfo
  {author} {\bibfnamefont {E.~C.}\ \bibnamefont {Good}}, \bibinfo {author}
  {\bibfnamefont {G.}~\bibnamefont {Balk}}, \bibinfo {author} {\bibfnamefont
  {H.~C.}\ \bibnamefont {Berg}}, \bibinfo {author} {\bibfnamefont {D.~L.}\
  \bibnamefont {Bleuel}}, \bibinfo {author} {\bibfnamefont {J.~A.}\
  \bibnamefont {Clark}}, \bibinfo {author} {\bibfnamefont {C.}~\bibnamefont
  {Dembski}}, \bibinfo {author} {\bibfnamefont {P.~A.}\ \bibnamefont
  {DeYoung}}, \bibinfo {author} {\bibfnamefont {B.}~\bibnamefont {Greaves}},
  \bibinfo {author} {\bibfnamefont {M.}~\bibnamefont {Guttormsen}}, \bibinfo
  {author} {\bibfnamefont {C.}~\bibnamefont {Harris}}, \bibinfo {author}
  {\bibfnamefont {A.~C.}\ \bibnamefont {Larsen}}, \bibinfo {author}
  {\bibfnamefont {S.~N.}\ \bibnamefont {Liddick}}, \bibinfo {author}
  {\bibfnamefont {S.}~\bibnamefont {Lyons}}, \bibinfo {author} {\bibfnamefont
  {M.}~\bibnamefont {Markova}}, \bibinfo {author} {\bibfnamefont {M.~J.}\
  \bibnamefont {Mogannam}}, \bibinfo {author} {\bibfnamefont {S.}~\bibnamefont
  {Nikas}}, \bibinfo {author} {\bibfnamefont {J.}~\bibnamefont {Owens-Fryar}},
  \bibinfo {author} {\bibfnamefont {A.}~\bibnamefont {Palmisano-Kyle}},
  \bibinfo {author} {\bibfnamefont {G.}~\bibnamefont {Perdikakis}}, \bibinfo
  {author} {\bibfnamefont {F.}~\bibnamefont {Pogliano}}, \bibinfo {author}
  {\bibfnamefont {M.}~\bibnamefont {Quintieri}}, \bibinfo {author}
  {\bibfnamefont {A.~L.}\ \bibnamefont {Richard}}, \bibinfo {author}
  {\bibfnamefont {D.}~\bibnamefont {Santiago-Gonzalez}}, \bibinfo {author}
  {\bibfnamefont {G.}~\bibnamefont {Savard}}, \bibinfo {author} {\bibfnamefont
  {M.~K.}\ \bibnamefont {Smith}}, \bibinfo {author} {\bibfnamefont
  {A.}~\bibnamefont {Sweet}}, \bibinfo {author} {\bibfnamefont
  {A.}~\bibnamefont {Tsantiri}},\ and\ \bibinfo {author} {\bibfnamefont
  {M.}~\bibnamefont {Wiedeking}},\ }\bibfield  {title} {\bibinfo {title} {First
  study of the $^{139}\mathrm{Ba}(n,\ensuremath{\gamma})^{140}\mathrm{Ba}$
  reaction to constrain the conditions for the astrophysical $i$ process},\
  }\href {https://doi.org/10.1103/PhysRevLett.132.202701} {\bibfield  {journal}
  {\bibinfo  {journal} {Phys. Rev. Lett.}\ }\textbf {\bibinfo {volume} {132}},\
  \bibinfo {pages} {202701} (\bibinfo {year} {2024})}\BibitemShut {NoStop}%
\bibitem [{\citenamefont {M\"ucher}\ \emph {et~al.}(2023)\citenamefont
  {M\"ucher}, \citenamefont {Spyrou}, \citenamefont {Wiedeking}, \citenamefont
  {Guttormsen}, \citenamefont {Larsen}, \citenamefont {Zeiser}, \citenamefont
  {Harris}, \citenamefont {Richard}, \citenamefont {Smith}, \citenamefont
  {G\"orgen}, \citenamefont {Liddick}, \citenamefont {Siem}, \citenamefont
  {Berg}, \citenamefont {Clark}, \citenamefont {DeYoung}, \citenamefont
  {Dombos}, \citenamefont {Greaves}, \citenamefont {Hicks}, \citenamefont
  {Kelmar}, \citenamefont {Lyons}, \citenamefont {Owens-Fryar}, \citenamefont
  {Palmisano}, \citenamefont {Santiago-Gonzalez}, \citenamefont {Savard},\ and\
  \citenamefont {von Seeger}}]{PhysRevC.107.L011602}%
  \BibitemOpen
  \bibfield  {author} {\bibinfo {author} {\bibfnamefont {D.}~\bibnamefont
  {M\"ucher}}, \bibinfo {author} {\bibfnamefont {A.}~\bibnamefont {Spyrou}},
  \bibinfo {author} {\bibfnamefont {M.}~\bibnamefont {Wiedeking}}, \bibinfo
  {author} {\bibfnamefont {M.}~\bibnamefont {Guttormsen}}, \bibinfo {author}
  {\bibfnamefont {A.~C.}\ \bibnamefont {Larsen}}, \bibinfo {author}
  {\bibfnamefont {F.}~\bibnamefont {Zeiser}}, \bibinfo {author} {\bibfnamefont
  {C.}~\bibnamefont {Harris}}, \bibinfo {author} {\bibfnamefont {A.~L.}\
  \bibnamefont {Richard}}, \bibinfo {author} {\bibfnamefont {M.~K.}\
  \bibnamefont {Smith}}, \bibinfo {author} {\bibfnamefont {A.}~\bibnamefont
  {G\"orgen}}, \bibinfo {author} {\bibfnamefont {S.~N.}\ \bibnamefont
  {Liddick}}, \bibinfo {author} {\bibfnamefont {S.}~\bibnamefont {Siem}},
  \bibinfo {author} {\bibfnamefont {H.~C.}\ \bibnamefont {Berg}}, \bibinfo
  {author} {\bibfnamefont {J.~A.}\ \bibnamefont {Clark}}, \bibinfo {author}
  {\bibfnamefont {P.~A.}\ \bibnamefont {DeYoung}}, \bibinfo {author}
  {\bibfnamefont {A.~C.}\ \bibnamefont {Dombos}}, \bibinfo {author}
  {\bibfnamefont {B.}~\bibnamefont {Greaves}}, \bibinfo {author} {\bibfnamefont
  {L.}~\bibnamefont {Hicks}}, \bibinfo {author} {\bibfnamefont
  {R.}~\bibnamefont {Kelmar}}, \bibinfo {author} {\bibfnamefont
  {S.}~\bibnamefont {Lyons}}, \bibinfo {author} {\bibfnamefont
  {J.}~\bibnamefont {Owens-Fryar}}, \bibinfo {author} {\bibfnamefont
  {A.}~\bibnamefont {Palmisano}}, \bibinfo {author} {\bibfnamefont
  {D.}~\bibnamefont {Santiago-Gonzalez}}, \bibinfo {author} {\bibfnamefont
  {G.}~\bibnamefont {Savard}},\ and\ \bibinfo {author} {\bibfnamefont {W.~W.}\
  \bibnamefont {von Seeger}},\ }\bibfield  {title} {\bibinfo {title}
  {Extracting model-independent nuclear level densities away from stability},\
  }\href {https://doi.org/10.1103/PhysRevC.107.L011602} {\bibfield  {journal}
  {\bibinfo  {journal} {Phys. Rev. C}\ }\textbf {\bibinfo {volume} {107}},\
  \bibinfo {pages} {L011602} (\bibinfo {year} {2023})}\BibitemShut {NoStop}%
\bibitem [{\citenamefont {Wiedeking}\ \emph {et~al.}(2021)\citenamefont
  {Wiedeking}, \citenamefont {Guttormsen}, \citenamefont {Larsen},
  \citenamefont {Zeiser}, \citenamefont {G{\"{o}}rgen}, \citenamefont
  {Liddick}, \citenamefont {M{\"{u}}cher}, \citenamefont {Siem},\ and\
  \citenamefont {Spyrou}}]{wiedeking2020independent}%
  \BibitemOpen
  \bibfield  {author} {\bibinfo {author} {\bibfnamefont {M.}~\bibnamefont
  {Wiedeking}}, \bibinfo {author} {\bibfnamefont {M.}~\bibnamefont
  {Guttormsen}}, \bibinfo {author} {\bibfnamefont {A.~C.}\ \bibnamefont
  {Larsen}}, \bibinfo {author} {\bibfnamefont {F.}~\bibnamefont {Zeiser}},
  \bibinfo {author} {\bibfnamefont {A.}~\bibnamefont {G{\"{o}}rgen}}, \bibinfo
  {author} {\bibfnamefont {S.~N.}\ \bibnamefont {Liddick}}, \bibinfo {author}
  {\bibfnamefont {D.}~\bibnamefont {M{\"{u}}cher}}, \bibinfo {author}
  {\bibfnamefont {S.}~\bibnamefont {Siem}},\ and\ \bibinfo {author}
  {\bibfnamefont {A.}~\bibnamefont {Spyrou}},\ }\bibfield  {title} {\bibinfo
  {title} {{Independent normalization for {$\gamma$} -ray strength functions:
  The shape method}},\ }\href {https://doi.org/10.1103/PhysRevC.104.014311}
  {\bibfield  {journal} {\bibinfo  {journal} {Physical Review C}\ }\textbf
  {\bibinfo {volume} {104}},\ \bibinfo {pages} {014311} (\bibinfo {year}
  {2021})}\BibitemShut {NoStop}%
\bibitem [{\citenamefont {Brown}\ and\ \citenamefont
  {Larsen}(2014)}]{Brown2014}%
  \BibitemOpen
  \bibfield  {author} {\bibinfo {author} {\bibfnamefont {B.~A.}\ \bibnamefont
  {Brown}}\ and\ \bibinfo {author} {\bibfnamefont {A.~C.}\ \bibnamefont
  {Larsen}},\ }\bibfield  {title} {\bibinfo {title} {{Large low-energy M1
  strength for Fe 56,57 within the nuclear shell model}},\ }\href
  {https://doi.org/10.1103/PhysRevLett.113.252502} {\bibfield  {journal}
  {\bibinfo  {journal} {Physical Review Letters}\ }\textbf {\bibinfo {volume}
  {113}},\ \bibinfo {pages} {252502} (\bibinfo {year} {2014})}\BibitemShut
  {NoStop}%
\bibitem [{\citenamefont {Guttormsen}\ \emph {et~al.}(2015)\citenamefont
  {Guttormsen}, \citenamefont {Aiche}, \citenamefont {Bello~Garrote},
  \citenamefont {Bernstein}, \citenamefont {Bleuel}, \citenamefont {Byun},
  \citenamefont {Ducasse}, \citenamefont {Eriksen}, \citenamefont {Giacoppo},
  \citenamefont {Görgen}, \citenamefont {Gunsing}, \citenamefont {Hagen},
  \citenamefont {Jurado}, \citenamefont {Klintefjord}, \citenamefont {Larsen},
  \citenamefont {Lebois}, \citenamefont {Leniau}, \citenamefont {Nyhus},
  \citenamefont {Renstrøm}, \citenamefont {Rose}, \citenamefont {Sahin},
  \citenamefont {Siem}, \citenamefont {Tornyi}, \citenamefont {Tveten},
  \citenamefont {Voinov}, \citenamefont {Wiedeking},\ and\ \citenamefont
  {Wilson}}]{guttormsen_experimental_2015}%
  \BibitemOpen
  \bibfield  {author} {\bibinfo {author} {\bibfnamefont {M.}~\bibnamefont
  {Guttormsen}}, \bibinfo {author} {\bibfnamefont {M.}~\bibnamefont {Aiche}},
  \bibinfo {author} {\bibfnamefont {F.~L.}\ \bibnamefont {Bello~Garrote}},
  \bibinfo {author} {\bibfnamefont {L.~A.}\ \bibnamefont {Bernstein}}, \bibinfo
  {author} {\bibfnamefont {D.~L.}\ \bibnamefont {Bleuel}}, \bibinfo {author}
  {\bibfnamefont {Y.}~\bibnamefont {Byun}}, \bibinfo {author} {\bibfnamefont
  {Q.}~\bibnamefont {Ducasse}}, \bibinfo {author} {\bibfnamefont {T.~K.}\
  \bibnamefont {Eriksen}}, \bibinfo {author} {\bibfnamefont {F.}~\bibnamefont
  {Giacoppo}}, \bibinfo {author} {\bibfnamefont {A.}~\bibnamefont {Görgen}},
  \bibinfo {author} {\bibfnamefont {F.}~\bibnamefont {Gunsing}}, \bibinfo
  {author} {\bibfnamefont {T.~W.}\ \bibnamefont {Hagen}}, \bibinfo {author}
  {\bibfnamefont {B.}~\bibnamefont {Jurado}}, \bibinfo {author} {\bibfnamefont
  {M.}~\bibnamefont {Klintefjord}}, \bibinfo {author} {\bibfnamefont {A.~C.}\
  \bibnamefont {Larsen}}, \bibinfo {author} {\bibfnamefont {L.}~\bibnamefont
  {Lebois}}, \bibinfo {author} {\bibfnamefont {B.}~\bibnamefont {Leniau}},
  \bibinfo {author} {\bibfnamefont {H.~T.}\ \bibnamefont {Nyhus}}, \bibinfo
  {author} {\bibfnamefont {T.}~\bibnamefont {Renstrøm}}, \bibinfo {author}
  {\bibfnamefont {S.~J.}\ \bibnamefont {Rose}}, \bibinfo {author}
  {\bibfnamefont {E.}~\bibnamefont {Sahin}}, \bibinfo {author} {\bibfnamefont
  {S.}~\bibnamefont {Siem}}, \bibinfo {author} {\bibfnamefont {T.~G.}\
  \bibnamefont {Tornyi}}, \bibinfo {author} {\bibfnamefont {G.~M.}\
  \bibnamefont {Tveten}}, \bibinfo {author} {\bibfnamefont {A.}~\bibnamefont
  {Voinov}}, \bibinfo {author} {\bibfnamefont {M.}~\bibnamefont {Wiedeking}},\
  and\ \bibinfo {author} {\bibfnamefont {J.}~\bibnamefont {Wilson}},\
  }\bibfield  {title} {\bibinfo {title} {Experimental level densities of atomic
  nuclei},\ }\href {https://doi.org/10.1140/epja/i2015-15170-4} {\bibfield
  {journal} {\bibinfo  {journal} {The European Physical Journal A}\ }\textbf
  {\bibinfo {volume} {51}},\ \bibinfo {pages} {170} (\bibinfo {year}
  {2015})}\BibitemShut {NoStop}%
\bibitem [{\citenamefont {Campo}\ \emph {et~al.}(2017)\citenamefont {Campo},
  \citenamefont {Larsen}, \citenamefont {Garrote}, \citenamefont {Eriksen},
  \citenamefont {Giacoppo}, \citenamefont {G\"orgen}, \citenamefont
  {Guttormsen}, \citenamefont {Klintefjord}, \citenamefont {Renstr\o{}m},
  \citenamefont {Sahin}, \citenamefont {Siem}, \citenamefont {Tornyi},\ and\
  \citenamefont {Tveten}}]{PhysRevC.96.014312}%
  \BibitemOpen
  \bibfield  {author} {\bibinfo {author} {\bibfnamefont {L.~C.}\ \bibnamefont
  {Campo}}, \bibinfo {author} {\bibfnamefont {A.~C.}\ \bibnamefont {Larsen}},
  \bibinfo {author} {\bibfnamefont {F.~L.~B.}\ \bibnamefont {Garrote}},
  \bibinfo {author} {\bibfnamefont {T.~K.}\ \bibnamefont {Eriksen}}, \bibinfo
  {author} {\bibfnamefont {F.}~\bibnamefont {Giacoppo}}, \bibinfo {author}
  {\bibfnamefont {A.}~\bibnamefont {G\"orgen}}, \bibinfo {author}
  {\bibfnamefont {M.}~\bibnamefont {Guttormsen}}, \bibinfo {author}
  {\bibfnamefont {M.}~\bibnamefont {Klintefjord}}, \bibinfo {author}
  {\bibfnamefont {T.}~\bibnamefont {Renstr\o{}m}}, \bibinfo {author}
  {\bibfnamefont {E.}~\bibnamefont {Sahin}}, \bibinfo {author} {\bibfnamefont
  {S.}~\bibnamefont {Siem}}, \bibinfo {author} {\bibfnamefont {T.~G.}\
  \bibnamefont {Tornyi}},\ and\ \bibinfo {author} {\bibfnamefont {G.~M.}\
  \bibnamefont {Tveten}},\ }\bibfield  {title} {\bibinfo {title}
  {{Investigating the {$\gamma$} decay of $^{65}$Ni from particle-{$\gamma$}
  coincidence data}},\ }\href {https://doi.org/10.1103/PhysRevC.96.014312}
  {\bibfield  {journal} {\bibinfo  {journal} {Physical Review C}\ }\textbf
  {\bibinfo {volume} {96}},\ \bibinfo {pages} {014312} (\bibinfo {year}
  {2017})}\BibitemShut {NoStop}%
\bibitem [{\citenamefont {Renstr{\o}m}\ \emph {et~al.}(2016)\citenamefont
  {Renstr{\o}m}, \citenamefont {Nyhus}, \citenamefont {Utsunomiya},
  \citenamefont {Schwengner}, \citenamefont {Goriely}, \citenamefont {Larsen},
  \citenamefont {Filipescu}, \citenamefont {Gheorghe}, \citenamefont
  {Bernstein}, \citenamefont {Bleuel}, \citenamefont {Glodariu}, \citenamefont
  {G{\"{o}}rgen}, \citenamefont {Guttormsen}, \citenamefont {Hagen},
  \citenamefont {Kheswa}, \citenamefont {Lui}, \citenamefont {Negi},
  \citenamefont {Ruud}, \citenamefont {Shima}, \citenamefont {Siem},
  \citenamefont {Takahisa}, \citenamefont {Tesileanu}, \citenamefont {Tornyi},
  \citenamefont {Tveten},\ and\ \citenamefont
  {Wiedeking}}]{PhysRevC.93.064302}%
  \BibitemOpen
  \bibfield  {author} {\bibinfo {author} {\bibfnamefont {T.}~\bibnamefont
  {Renstr{\o}m}}, \bibinfo {author} {\bibfnamefont {H.~T.}\ \bibnamefont
  {Nyhus}}, \bibinfo {author} {\bibfnamefont {H.}~\bibnamefont {Utsunomiya}},
  \bibinfo {author} {\bibfnamefont {R.}~\bibnamefont {Schwengner}}, \bibinfo
  {author} {\bibfnamefont {S.}~\bibnamefont {Goriely}}, \bibinfo {author}
  {\bibfnamefont {A.~C.}\ \bibnamefont {Larsen}}, \bibinfo {author}
  {\bibfnamefont {D.~M.}\ \bibnamefont {Filipescu}}, \bibinfo {author}
  {\bibfnamefont {I.}~\bibnamefont {Gheorghe}}, \bibinfo {author}
  {\bibfnamefont {L.~A.}\ \bibnamefont {Bernstein}}, \bibinfo {author}
  {\bibfnamefont {D.~L.}\ \bibnamefont {Bleuel}}, \bibinfo {author}
  {\bibfnamefont {T.}~\bibnamefont {Glodariu}}, \bibinfo {author}
  {\bibfnamefont {A.}~\bibnamefont {G{\"{o}}rgen}}, \bibinfo {author}
  {\bibfnamefont {M.}~\bibnamefont {Guttormsen}}, \bibinfo {author}
  {\bibfnamefont {T.~W.}\ \bibnamefont {Hagen}}, \bibinfo {author}
  {\bibfnamefont {B.~V.}\ \bibnamefont {Kheswa}}, \bibinfo {author}
  {\bibfnamefont {Y.~W.}\ \bibnamefont {Lui}}, \bibinfo {author} {\bibfnamefont
  {D.}~\bibnamefont {Negi}}, \bibinfo {author} {\bibfnamefont {I.~E.}\
  \bibnamefont {Ruud}}, \bibinfo {author} {\bibfnamefont {T.}~\bibnamefont
  {Shima}}, \bibinfo {author} {\bibfnamefont {S.}~\bibnamefont {Siem}},
  \bibinfo {author} {\bibfnamefont {K.}~\bibnamefont {Takahisa}}, \bibinfo
  {author} {\bibfnamefont {O.}~\bibnamefont {Tesileanu}}, \bibinfo {author}
  {\bibfnamefont {T.~G.}\ \bibnamefont {Tornyi}}, \bibinfo {author}
  {\bibfnamefont {G.~M.}\ \bibnamefont {Tveten}},\ and\ \bibinfo {author}
  {\bibfnamefont {M.}~\bibnamefont {Wiedeking}},\ }\bibfield  {title} {\bibinfo
  {title} {{Low-energy enhancement in the {$\gamma$} -ray strength functions of
  Ge 73,74}},\ }\href {https://doi.org/10.1103/PhysRevC.93.064302} {\bibfield
  {journal} {\bibinfo  {journal} {Physical Review C}\ }\textbf {\bibinfo
  {volume} {93}},\ \bibinfo {pages} {64302} (\bibinfo {year}
  {2016})}\BibitemShut {NoStop}%
\bibitem [{\citenamefont {Midtb{\o}}\ \emph {et~al.}(2018)\citenamefont
  {Midtb{\o}}, \citenamefont {Larsen}, \citenamefont {Renstr{\o}m},
  \citenamefont {Bello~Garrote},\ and\ \citenamefont {Lima}}]{Midtbo2018}%
  \BibitemOpen
  \bibfield  {author} {\bibinfo {author} {\bibfnamefont {J.~E.}\ \bibnamefont
  {Midtb{\o}}}, \bibinfo {author} {\bibfnamefont {A.~C.}\ \bibnamefont
  {Larsen}}, \bibinfo {author} {\bibfnamefont {T.}~\bibnamefont {Renstr{\o}m}},
  \bibinfo {author} {\bibfnamefont {F.~L.}\ \bibnamefont {Bello~Garrote}},\
  and\ \bibinfo {author} {\bibfnamefont {E.}~\bibnamefont {Lima}},\ }\bibfield
  {title} {\bibinfo {title} {{Consolidating the concept of low-energy magnetic
  dipole decay radiation}},\ }\href
  {https://doi.org/10.1103/PhysRevC.98.064321} {\bibfield  {journal} {\bibinfo
  {journal} {Physical Review C}\ }\textbf {\bibinfo {volume} {98}},\ \bibinfo
  {pages} {064321} (\bibinfo {year} {2018})}\BibitemShut {NoStop}%
\bibitem [{\citenamefont {Feroz}\ and\ \citenamefont
  {Hobson}(2008)}]{Feroz2008}%
  \BibitemOpen
  \bibfield  {author} {\bibinfo {author} {\bibfnamefont {F.}~\bibnamefont
  {Feroz}}\ and\ \bibinfo {author} {\bibfnamefont {M.~P.}\ \bibnamefont
  {Hobson}},\ }\bibfield  {title} {\bibinfo {title} {{Multimodal nested
  sampling: an efficient and robust alternative to Markov Chain Monte Carlo
  methods for astronomical data analyses}},\ }\href
  {https://doi.org/10.1111/j.1365-2966.2007.12353.x} {\bibfield  {journal}
  {\bibinfo  {journal} {Monthly Notices of the Royal Astronomical Society}\
  }\textbf {\bibinfo {volume} {384}},\ \bibinfo {pages} {449} (\bibinfo {year}
  {2008})}\BibitemShut {NoStop}%
\bibitem [{\citenamefont {Feroz}\ \emph {et~al.}(2009)\citenamefont {Feroz},
  \citenamefont {Hobson},\ and\ \citenamefont {Bridges}}]{Feroz2009}%
  \BibitemOpen
  \bibfield  {author} {\bibinfo {author} {\bibfnamefont {F.}~\bibnamefont
  {Feroz}}, \bibinfo {author} {\bibfnamefont {M.~P.}\ \bibnamefont {Hobson}},\
  and\ \bibinfo {author} {\bibfnamefont {M.}~\bibnamefont {Bridges}},\
  }\bibfield  {title} {\bibinfo {title} {{MultiNest: an efficient and robust
  Bayesian inference tool for cosmology and particle physics}},\ }\href
  {https://doi.org/10.1111/j.1365-2966.2009.14548.x} {\bibfield  {journal}
  {\bibinfo  {journal} {Monthly Notices of the Royal Astronomical Society}\
  }\textbf {\bibinfo {volume} {398}},\ \bibinfo {pages} {1601} (\bibinfo {year}
  {2009})}\BibitemShut {NoStop}%
\bibitem [{\citenamefont {Feroz}\ \emph {et~al.}(2019)\citenamefont {Feroz},
  \citenamefont {Hobson}, \citenamefont {Cameron},\ and\ \citenamefont
  {Pettitt}}]{Feroz2019}%
  \BibitemOpen
  \bibfield  {author} {\bibinfo {author} {\bibfnamefont {F.}~\bibnamefont
  {Feroz}}, \bibinfo {author} {\bibfnamefont {M.~P.}\ \bibnamefont {Hobson}},
  \bibinfo {author} {\bibfnamefont {E.}~\bibnamefont {Cameron}},\ and\ \bibinfo
  {author} {\bibfnamefont {A.~N.}\ \bibnamefont {Pettitt}},\ }\bibfield
  {title} {\bibinfo {title} {{Importance Nested Sampling and the MultiNest
  Algorithm}},\ }\bibfield  {journal} {\bibinfo  {journal} {The Open Journal of
  Astrophysics}\ }\textbf {\bibinfo {volume} {2}},\ \href
  {https://doi.org/10.21105/astro.1306.2144} {10.21105/astro.1306.2144}
  (\bibinfo {year} {2019})\BibitemShut {NoStop}%
\bibitem [{\citenamefont {Buchner}\ \emph {et~al.}(2014)\citenamefont
  {Buchner}, \citenamefont {Georgakakis}, \citenamefont {Nandra}, \citenamefont
  {Hsu}, \citenamefont {Rangel}, \citenamefont {Brightman}, \citenamefont
  {Merloni}, \citenamefont {Salvato}, \citenamefont {Donley},\ and\
  \citenamefont {Kocevski}}]{Buchner2014}%
  \BibitemOpen
  \bibfield  {author} {\bibinfo {author} {\bibfnamefont {J.}~\bibnamefont
  {Buchner}}, \bibinfo {author} {\bibfnamefont {A.}~\bibnamefont
  {Georgakakis}}, \bibinfo {author} {\bibfnamefont {K.}~\bibnamefont {Nandra}},
  \bibinfo {author} {\bibfnamefont {L.}~\bibnamefont {Hsu}}, \bibinfo {author}
  {\bibfnamefont {C.}~\bibnamefont {Rangel}}, \bibinfo {author} {\bibfnamefont
  {M.}~\bibnamefont {Brightman}}, \bibinfo {author} {\bibfnamefont
  {A.}~\bibnamefont {Merloni}}, \bibinfo {author} {\bibfnamefont
  {M.}~\bibnamefont {Salvato}}, \bibinfo {author} {\bibfnamefont
  {J.}~\bibnamefont {Donley}},\ and\ \bibinfo {author} {\bibfnamefont
  {D.}~\bibnamefont {Kocevski}},\ }\bibfield  {title} {\bibinfo {title} {{X-ray
  spectral modelling of the AGN obscuring region in the CDFS: Bayesian model
  selection and catalogue}},\ }\href
  {https://doi.org/10.1051/0004-6361/201322971} {\bibfield  {journal} {\bibinfo
   {journal} {Astronomy {\&} Astrophysics}\ }\textbf {\bibinfo {volume}
  {564}},\ \bibinfo {pages} {A125} (\bibinfo {year} {2014})}\BibitemShut
  {NoStop}%
\bibitem [{\citenamefont {Bethe}(1936)}]{PhysRev.50.332}%
  \BibitemOpen
  \bibfield  {author} {\bibinfo {author} {\bibfnamefont {H.~A.}\ \bibnamefont
  {Bethe}},\ }\bibfield  {title} {\bibinfo {title} {{An Attempt to Calculate
  the Number of Energy Levels of a Heavy Nucleus}},\ }\href
  {https://doi.org/10.1103/PhysRev.50.332} {\bibfield  {journal} {\bibinfo
  {journal} {Physical Review}\ }\textbf {\bibinfo {volume} {50}},\ \bibinfo
  {pages} {332} (\bibinfo {year} {1936})}\BibitemShut {NoStop}%
\bibitem [{\citenamefont {Gilbert}\ and\ \citenamefont
  {Cameron}(1965)}]{doi:10.1139/p65-139}%
  \BibitemOpen
  \bibfield  {author} {\bibinfo {author} {\bibfnamefont {A.}~\bibnamefont
  {Gilbert}}\ and\ \bibinfo {author} {\bibfnamefont {A.~G.~W.}\ \bibnamefont
  {Cameron}},\ }\bibfield  {title} {\bibinfo {title} {{a Composite
  Nuclear-Level Density Formula With Shell Corrections}},\ }\href
  {https://doi.org/10.1139/p65-139} {\bibfield  {journal} {\bibinfo  {journal}
  {Canadian Journal of Physics}\ }\textbf {\bibinfo {volume} {43}},\ \bibinfo
  {pages} {1446} (\bibinfo {year} {1965})}\BibitemShut {NoStop}%
\bibitem [{\citenamefont {Ericson}(1959)}]{ERICSON1959481}%
  \BibitemOpen
  \bibfield  {author} {\bibinfo {author} {\bibfnamefont {T.}~\bibnamefont
  {Ericson}},\ }\bibfield  {title} {\bibinfo {title} {{A statistical analysis
  of excited nuclear states}},\ }\href
  {https://doi.org/10.1016/0029-5582(59)90291-3} {\bibfield  {journal}
  {\bibinfo  {journal} {Nuclear Physics}\ }\textbf {\bibinfo {volume} {11}},\
  \bibinfo {pages} {481} (\bibinfo {year} {1959})}\BibitemShut {NoStop}%
\bibitem [{\citenamefont {Goriely}\ \emph {et~al.}(2008)\citenamefont
  {Goriely}, \citenamefont {Hilaire},\ and\ \citenamefont
  {Koning}}]{PhysRevC.78.064307}%
  \BibitemOpen
  \bibfield  {author} {\bibinfo {author} {\bibfnamefont {S.}~\bibnamefont
  {Goriely}}, \bibinfo {author} {\bibfnamefont {S.}~\bibnamefont {Hilaire}},\
  and\ \bibinfo {author} {\bibfnamefont {A.~J.}\ \bibnamefont {Koning}},\
  }\bibfield  {title} {\bibinfo {title} {{Improved microscopic nuclear level
  densities within the Hartree-Fock-Bogoliubov plus combinatorial method}},\
  }\href {https://doi.org/10.1103/PhysRevC.78.064307} {\bibfield  {journal}
  {\bibinfo  {journal} {Physical Review C - Nuclear Physics}\ }\textbf
  {\bibinfo {volume} {78}},\ \bibinfo {pages} {064307} (\bibinfo {year}
  {2008})}\BibitemShut {NoStop}%
\bibitem [{\citenamefont {Capote}\ \emph {et~al.}(2009)\citenamefont {Capote},
  \citenamefont {Herman}, \citenamefont {Oblo{\v{z}}insk{\'{y}}}, \citenamefont
  {Young}, \citenamefont {Goriely}, \citenamefont {Belgya}, \citenamefont
  {Ignatyuk}, \citenamefont {Koning}, \citenamefont {Hilaire}, \citenamefont
  {Plujko}, \citenamefont {Avrigeanu}, \citenamefont {Bersillon}, \citenamefont
  {Chadwick}, \citenamefont {Fukahori}, \citenamefont {Ge}, \citenamefont
  {Han}, \citenamefont {Kailas}, \citenamefont {Kopecky}, \citenamefont
  {Maslov}, \citenamefont {Reffo}, \citenamefont {Sin}, \citenamefont
  {Soukhovitskii},\ and\ \citenamefont {Talou}}]{Capote2009}%
  \BibitemOpen
  \bibfield  {author} {\bibinfo {author} {\bibfnamefont {R.}~\bibnamefont
  {Capote}}, \bibinfo {author} {\bibfnamefont {M.}~\bibnamefont {Herman}},
  \bibinfo {author} {\bibfnamefont {P.}~\bibnamefont {Oblo{\v{z}}insk{\'{y}}}},
  \bibinfo {author} {\bibfnamefont {P.~G.}\ \bibnamefont {Young}}, \bibinfo
  {author} {\bibfnamefont {S.}~\bibnamefont {Goriely}}, \bibinfo {author}
  {\bibfnamefont {T.}~\bibnamefont {Belgya}}, \bibinfo {author} {\bibfnamefont
  {A.~V.}\ \bibnamefont {Ignatyuk}}, \bibinfo {author} {\bibfnamefont {A.~J.}\
  \bibnamefont {Koning}}, \bibinfo {author} {\bibfnamefont {S.}~\bibnamefont
  {Hilaire}}, \bibinfo {author} {\bibfnamefont {V.~A.}\ \bibnamefont {Plujko}},
  \bibinfo {author} {\bibfnamefont {M.}~\bibnamefont {Avrigeanu}}, \bibinfo
  {author} {\bibfnamefont {O.}~\bibnamefont {Bersillon}}, \bibinfo {author}
  {\bibfnamefont {M.~B.}\ \bibnamefont {Chadwick}}, \bibinfo {author}
  {\bibfnamefont {T.}~\bibnamefont {Fukahori}}, \bibinfo {author}
  {\bibfnamefont {Z.}~\bibnamefont {Ge}}, \bibinfo {author} {\bibfnamefont
  {Y.}~\bibnamefont {Han}}, \bibinfo {author} {\bibfnamefont {S.}~\bibnamefont
  {Kailas}}, \bibinfo {author} {\bibfnamefont {J.}~\bibnamefont {Kopecky}},
  \bibinfo {author} {\bibfnamefont {V.~M.}\ \bibnamefont {Maslov}}, \bibinfo
  {author} {\bibfnamefont {G.}~\bibnamefont {Reffo}}, \bibinfo {author}
  {\bibfnamefont {M.}~\bibnamefont {Sin}}, \bibinfo {author} {\bibfnamefont
  {E.~S.}\ \bibnamefont {Soukhovitskii}},\ and\ \bibinfo {author}
  {\bibfnamefont {P.}~\bibnamefont {Talou}},\ }\bibfield  {title} {\bibinfo
  {title} {{RIPL - Reference Input Parameter Library for Calculation of Nuclear
  Reactions and Nuclear Data Evaluations}},\ }\href
  {https://doi.org/10.1016/j.nds.2009.10.004} {\bibfield  {journal} {\bibinfo
  {journal} {Nuclear Data Sheets}\ }\textbf {\bibinfo {volume} {110}},\
  \bibinfo {pages} {3107} (\bibinfo {year} {2009})}\BibitemShut {NoStop}%
\bibitem [{\citenamefont {Guttormsen}\ \emph {et~al.}(2017)\citenamefont
  {Guttormsen}, \citenamefont {Goriely}, \citenamefont {Larsen}, \citenamefont
  {G\"orgen}, \citenamefont {Hagen}, \citenamefont {Renstr\o{}m}, \citenamefont
  {Siem}, \citenamefont {Syed}, \citenamefont {Tagliente}, \citenamefont
  {Toft}, \citenamefont {Utsunomiya}, \citenamefont {Voinov},\ and\
  \citenamefont {Wikan}}]{PhysRevC.96.024313}%
  \BibitemOpen
  \bibfield  {author} {\bibinfo {author} {\bibfnamefont {M.}~\bibnamefont
  {Guttormsen}}, \bibinfo {author} {\bibfnamefont {S.}~\bibnamefont {Goriely}},
  \bibinfo {author} {\bibfnamefont {A.~C.}\ \bibnamefont {Larsen}}, \bibinfo
  {author} {\bibfnamefont {A.}~\bibnamefont {G\"orgen}}, \bibinfo {author}
  {\bibfnamefont {T.~W.}\ \bibnamefont {Hagen}}, \bibinfo {author}
  {\bibfnamefont {T.}~\bibnamefont {Renstr\o{}m}}, \bibinfo {author}
  {\bibfnamefont {S.}~\bibnamefont {Siem}}, \bibinfo {author} {\bibfnamefont
  {N.~U.~H.}\ \bibnamefont {Syed}}, \bibinfo {author} {\bibfnamefont
  {G.}~\bibnamefont {Tagliente}}, \bibinfo {author} {\bibfnamefont {H.~K.}\
  \bibnamefont {Toft}}, \bibinfo {author} {\bibfnamefont {H.}~\bibnamefont
  {Utsunomiya}}, \bibinfo {author} {\bibfnamefont {A.~V.}\ \bibnamefont
  {Voinov}},\ and\ \bibinfo {author} {\bibfnamefont {K.}~\bibnamefont
  {Wikan}},\ }\bibfield  {title} {\bibinfo {title} {{Quasicontinuum {$\gamma$}
  decay of $^{91,92}$Zr: Benchmarking indirect (n,{$\gamma$}) cross section
  measurements for the s process}},\ }\href
  {https://doi.org/10.1103/PhysRevC.96.024313} {\bibfield  {journal} {\bibinfo
  {journal} {Physical Review C}\ }\textbf {\bibinfo {volume} {96}},\ \bibinfo
  {pages} {024313} (\bibinfo {year} {2017})}\BibitemShut {NoStop}%
\bibitem [{\citenamefont {Goriely}\ and\ \citenamefont
  {Plujko}(2019)}]{PhysRevC.99.014303}%
  \BibitemOpen
  \bibfield  {author} {\bibinfo {author} {\bibfnamefont {S.}~\bibnamefont
  {Goriely}}\ and\ \bibinfo {author} {\bibfnamefont {V.}~\bibnamefont
  {Plujko}},\ }\bibfield  {title} {\bibinfo {title} {{Simple empirical E1 and
  M1 strength functions for practical applications}},\ }\href
  {https://doi.org/10.1103/PhysRevC.99.014303} {\bibfield  {journal} {\bibinfo
  {journal} {Physical Review C}\ }\textbf {\bibinfo {volume} {99}},\ \bibinfo
  {pages} {014303} (\bibinfo {year} {2019})}\BibitemShut {NoStop}%
\bibitem [{\citenamefont {Goriely}\ \emph {et~al.}(2018)\citenamefont
  {Goriely}, \citenamefont {Hilaire}, \citenamefont {P{\'{e}}ru},\ and\
  \citenamefont {Sieja}}]{Goriely2018}%
  \BibitemOpen
  \bibfield  {author} {\bibinfo {author} {\bibfnamefont {S.}~\bibnamefont
  {Goriely}}, \bibinfo {author} {\bibfnamefont {S.}~\bibnamefont {Hilaire}},
  \bibinfo {author} {\bibfnamefont {S.}~\bibnamefont {P{\'{e}}ru}},\ and\
  \bibinfo {author} {\bibfnamefont {K.}~\bibnamefont {Sieja}},\ }\bibfield
  {title} {\bibinfo {title} {{Gogny-HFB+QRPA dipole strength function and its
  application to radiative nucleon capture cross section}},\ }\href
  {https://doi.org/10.1103/PhysRevC.98.014327} {\bibfield  {journal} {\bibinfo
  {journal} {Physical Review C}\ }\textbf {\bibinfo {volume} {98}},\ \bibinfo
  {pages} {014327} (\bibinfo {year} {2018})}\BibitemShut {NoStop}%
\bibitem [{\citenamefont {Hilaire}\ and\ \citenamefont
  {Girod}(2007)}]{Hilaire07}%
  \BibitemOpen
  \bibfield  {author} {\bibinfo {author} {\bibfnamefont {S.}~\bibnamefont
  {Hilaire}}\ and\ \bibinfo {author} {\bibfnamefont {M.}~\bibnamefont
  {Girod}},\ }\bibfield  {title} {\bibinfo {title} {{Large-scale mean-field
  calculations from proton to neutron drip lines using the D1S Gogny force}},\
  }\href {https://doi.org/10.1140/epja/i2007-10450-2} {\bibfield  {journal}
  {\bibinfo  {journal} {The European Physical Journal A}\ }\textbf {\bibinfo
  {volume} {33}},\ \bibinfo {pages} {237} (\bibinfo {year} {2007})}\BibitemShut
  {NoStop}%
\bibitem [{\citenamefont {Schwengner}\ \emph {et~al.}(2013)\citenamefont
  {Schwengner}, \citenamefont {Frauendorf},\ and\ \citenamefont
  {Larsen}}]{PhysRevLett.111.232504}%
  \BibitemOpen
  \bibfield  {author} {\bibinfo {author} {\bibfnamefont {R.}~\bibnamefont
  {Schwengner}}, \bibinfo {author} {\bibfnamefont {S.}~\bibnamefont
  {Frauendorf}},\ and\ \bibinfo {author} {\bibfnamefont {A.~C.}\ \bibnamefont
  {Larsen}},\ }\bibfield  {title} {\bibinfo {title} {{Low-Energy Enhancement of
  Magnetic Dipole Radiation}},\ }\href
  {https://doi.org/10.1103/PhysRevLett.111.232504} {\bibfield  {journal}
  {\bibinfo  {journal} {Physical Review Letters}\ }\textbf {\bibinfo {volume}
  {111}},\ \bibinfo {pages} {232504} (\bibinfo {year} {2013})}\BibitemShut
  {NoStop}%
\bibitem [{\citenamefont {Frauendorf}\ and\ \citenamefont
  {Schwengner}(2022)}]{Frauendorf22}%
  \BibitemOpen
  \bibfield  {author} {\bibinfo {author} {\bibfnamefont {S.}~\bibnamefont
  {Frauendorf}}\ and\ \bibinfo {author} {\bibfnamefont {R.}~\bibnamefont
  {Schwengner}},\ }\bibfield  {title} {\bibinfo {title} {{Evolution of
  low-lying M1 modes in germanium isotopes}},\ }\href
  {https://doi.org/10.1103/PhysRevC.105.034335} {\bibfield  {journal} {\bibinfo
   {journal} {Physical Review C}\ }\textbf {\bibinfo {volume} {105}},\ \bibinfo
  {pages} {034335} (\bibinfo {year} {2022})}\BibitemShut {NoStop}%
\bibitem [{\citenamefont {Sieja}(2017)}]{PhysRevLett.119.052502}%
  \BibitemOpen
  \bibfield  {author} {\bibinfo {author} {\bibfnamefont {K.}~\bibnamefont
  {Sieja}},\ }\bibfield  {title} {\bibinfo {title} {{Electric and Magnetic
  Dipole Strength at Low Energy}},\ }\href
  {https://doi.org/10.1103/PhysRevLett.119.052502} {\bibfield  {journal}
  {\bibinfo  {journal} {Phys. Rev. Lett.}\ }\textbf {\bibinfo {volume} {119}},\
  \bibinfo {pages} {052502} (\bibinfo {year} {2017})}\BibitemShut {NoStop}%
\bibitem [{\citenamefont {Jones}\ \emph {et~al.}(2018)\citenamefont {Jones},
  \citenamefont {Macchiavelli}, \citenamefont {Wiedeking}, \citenamefont
  {Bernstein}, \citenamefont {Crawford}, \citenamefont {Campbell},
  \citenamefont {Clark}, \citenamefont {Cromaz}, \citenamefont {Fallon},
  \citenamefont {Lee}, \citenamefont {Salathe}, \citenamefont {Wiens},
  \citenamefont {Ayangeakaa}, \citenamefont {Bleuel}, \citenamefont {Bottoni},
  \citenamefont {Carpenter}, \citenamefont {Davids}, \citenamefont {Elson},
  \citenamefont {G\"orgen}, \citenamefont {Guttormsen}, \citenamefont
  {Janssens}, \citenamefont {Kinnison}, \citenamefont {Kirsch}, \citenamefont
  {Larsen}, \citenamefont {Lauritsen}, \citenamefont {Reviol}, \citenamefont
  {Sarantites}, \citenamefont {Siem}, \citenamefont {Voinov},\ and\
  \citenamefont {Zhu}}]{Jones2018}%
  \BibitemOpen
  \bibfield  {author} {\bibinfo {author} {\bibfnamefont {M.~D.}\ \bibnamefont
  {Jones}}, \bibinfo {author} {\bibfnamefont {A.~O.}\ \bibnamefont
  {Macchiavelli}}, \bibinfo {author} {\bibfnamefont {M.}~\bibnamefont
  {Wiedeking}}, \bibinfo {author} {\bibfnamefont {L.~A.}\ \bibnamefont
  {Bernstein}}, \bibinfo {author} {\bibfnamefont {H.~L.}\ \bibnamefont
  {Crawford}}, \bibinfo {author} {\bibfnamefont {C.~M.}\ \bibnamefont
  {Campbell}}, \bibinfo {author} {\bibfnamefont {R.~M.}\ \bibnamefont {Clark}},
  \bibinfo {author} {\bibfnamefont {M.}~\bibnamefont {Cromaz}}, \bibinfo
  {author} {\bibfnamefont {P.}~\bibnamefont {Fallon}}, \bibinfo {author}
  {\bibfnamefont {I.~Y.}\ \bibnamefont {Lee}}, \bibinfo {author} {\bibfnamefont
  {M.}~\bibnamefont {Salathe}}, \bibinfo {author} {\bibfnamefont
  {A.}~\bibnamefont {Wiens}}, \bibinfo {author} {\bibfnamefont {A.~D.}\
  \bibnamefont {Ayangeakaa}}, \bibinfo {author} {\bibfnamefont {D.~L.}\
  \bibnamefont {Bleuel}}, \bibinfo {author} {\bibfnamefont {S.}~\bibnamefont
  {Bottoni}}, \bibinfo {author} {\bibfnamefont {M.~P.}\ \bibnamefont
  {Carpenter}}, \bibinfo {author} {\bibfnamefont {H.~M.}\ \bibnamefont
  {Davids}}, \bibinfo {author} {\bibfnamefont {J.}~\bibnamefont {Elson}},
  \bibinfo {author} {\bibfnamefont {A.}~\bibnamefont {G\"orgen}}, \bibinfo
  {author} {\bibfnamefont {M.}~\bibnamefont {Guttormsen}}, \bibinfo {author}
  {\bibfnamefont {R.~V.~F.}\ \bibnamefont {Janssens}}, \bibinfo {author}
  {\bibfnamefont {J.~E.}\ \bibnamefont {Kinnison}}, \bibinfo {author}
  {\bibfnamefont {L.}~\bibnamefont {Kirsch}}, \bibinfo {author} {\bibfnamefont
  {A.~C.}\ \bibnamefont {Larsen}}, \bibinfo {author} {\bibfnamefont
  {T.}~\bibnamefont {Lauritsen}}, \bibinfo {author} {\bibfnamefont
  {W.}~\bibnamefont {Reviol}}, \bibinfo {author} {\bibfnamefont {D.~G.}\
  \bibnamefont {Sarantites}}, \bibinfo {author} {\bibfnamefont
  {S.}~\bibnamefont {Siem}}, \bibinfo {author} {\bibfnamefont {A.~V.}\
  \bibnamefont {Voinov}},\ and\ \bibinfo {author} {\bibfnamefont
  {S.}~\bibnamefont {Zhu}},\ }\bibfield  {title} {\bibinfo {title} {Examination
  of the low-energy enhancement of the $\ensuremath{\gamma}$-ray strength
  function of $^{56}\mathrm{Fe}$},\ }\href
  {https://doi.org/10.1103/PhysRevC.97.024327} {\bibfield  {journal} {\bibinfo
  {journal} {Phys. Rev. C}\ }\textbf {\bibinfo {volume} {97}},\ \bibinfo
  {pages} {024327} (\bibinfo {year} {2018})}\BibitemShut {NoStop}%
\bibitem [{\citenamefont {Rossi}\ \emph {et~al.}(2013)\citenamefont {Rossi},
  \citenamefont {Adrich}, \citenamefont {Aksouh}, \citenamefont {Alvarez-Pol},
  \citenamefont {Aumann}, \citenamefont {Benlliure}, \citenamefont {B\"ohmer},
  \citenamefont {Boretzky}, \citenamefont {Casarejos}, \citenamefont
  {Chartier}, \citenamefont {Chatillon}, \citenamefont {Cortina-Gil},
  \citenamefont {Datta~Pramanik}, \citenamefont {Emling}, \citenamefont
  {Ershova}, \citenamefont {Fernandez-Dominguez}, \citenamefont {Geissel},
  \citenamefont {Gorska}, \citenamefont {Heil}, \citenamefont {Johansson},
  \citenamefont {Junghans}, \citenamefont {Kelic-Heil}, \citenamefont
  {Kiselev}, \citenamefont {Klimkiewicz}, \citenamefont {Kratz}, \citenamefont
  {Kr\"ucken}, \citenamefont {Kurz}, \citenamefont {Labiche}, \citenamefont
  {Le~Bleis}, \citenamefont {Lemmon}, \citenamefont {Litvinov}, \citenamefont
  {Mahata}, \citenamefont {Maierbeck}, \citenamefont {Movsesyan}, \citenamefont
  {Nilsson}, \citenamefont {Nociforo}, \citenamefont {Palit}, \citenamefont
  {Paschalis}, \citenamefont {Plag}, \citenamefont {Reifarth}, \citenamefont
  {Savran}, \citenamefont {Scheit}, \citenamefont {Simon}, \citenamefont
  {S\"ummerer}, \citenamefont {Wagner}, \citenamefont
  {Walu\ifmmode~\acute{s}\else \'{s}\fi{}}, \citenamefont {Weick},\ and\
  \citenamefont {Winkler}}]{PhysRevLett.111.242503}%
  \BibitemOpen
  \bibfield  {author} {\bibinfo {author} {\bibfnamefont {D.~M.}\ \bibnamefont
  {Rossi}}, \bibinfo {author} {\bibfnamefont {P.}~\bibnamefont {Adrich}},
  \bibinfo {author} {\bibfnamefont {F.}~\bibnamefont {Aksouh}}, \bibinfo
  {author} {\bibfnamefont {H.}~\bibnamefont {Alvarez-Pol}}, \bibinfo {author}
  {\bibfnamefont {T.}~\bibnamefont {Aumann}}, \bibinfo {author} {\bibfnamefont
  {J.}~\bibnamefont {Benlliure}}, \bibinfo {author} {\bibfnamefont
  {M.}~\bibnamefont {B\"ohmer}}, \bibinfo {author} {\bibfnamefont
  {K.}~\bibnamefont {Boretzky}}, \bibinfo {author} {\bibfnamefont
  {E.}~\bibnamefont {Casarejos}}, \bibinfo {author} {\bibfnamefont
  {M.}~\bibnamefont {Chartier}}, \bibinfo {author} {\bibfnamefont
  {A.}~\bibnamefont {Chatillon}}, \bibinfo {author} {\bibfnamefont
  {D.}~\bibnamefont {Cortina-Gil}}, \bibinfo {author} {\bibfnamefont
  {U.}~\bibnamefont {Datta~Pramanik}}, \bibinfo {author} {\bibfnamefont
  {H.}~\bibnamefont {Emling}}, \bibinfo {author} {\bibfnamefont
  {O.}~\bibnamefont {Ershova}}, \bibinfo {author} {\bibfnamefont
  {B.}~\bibnamefont {Fernandez-Dominguez}}, \bibinfo {author} {\bibfnamefont
  {H.}~\bibnamefont {Geissel}}, \bibinfo {author} {\bibfnamefont
  {M.}~\bibnamefont {Gorska}}, \bibinfo {author} {\bibfnamefont
  {M.}~\bibnamefont {Heil}}, \bibinfo {author} {\bibfnamefont {H.~T.}\
  \bibnamefont {Johansson}}, \bibinfo {author} {\bibfnamefont {A.}~\bibnamefont
  {Junghans}}, \bibinfo {author} {\bibfnamefont {A.}~\bibnamefont
  {Kelic-Heil}}, \bibinfo {author} {\bibfnamefont {O.}~\bibnamefont {Kiselev}},
  \bibinfo {author} {\bibfnamefont {A.}~\bibnamefont {Klimkiewicz}}, \bibinfo
  {author} {\bibfnamefont {J.~V.}\ \bibnamefont {Kratz}}, \bibinfo {author}
  {\bibfnamefont {R.}~\bibnamefont {Kr\"ucken}}, \bibinfo {author}
  {\bibfnamefont {N.}~\bibnamefont {Kurz}}, \bibinfo {author} {\bibfnamefont
  {M.}~\bibnamefont {Labiche}}, \bibinfo {author} {\bibfnamefont
  {T.}~\bibnamefont {Le~Bleis}}, \bibinfo {author} {\bibfnamefont
  {R.}~\bibnamefont {Lemmon}}, \bibinfo {author} {\bibfnamefont {Y.~A.}\
  \bibnamefont {Litvinov}}, \bibinfo {author} {\bibfnamefont {K.}~\bibnamefont
  {Mahata}}, \bibinfo {author} {\bibfnamefont {P.}~\bibnamefont {Maierbeck}},
  \bibinfo {author} {\bibfnamefont {A.}~\bibnamefont {Movsesyan}}, \bibinfo
  {author} {\bibfnamefont {T.}~\bibnamefont {Nilsson}}, \bibinfo {author}
  {\bibfnamefont {C.}~\bibnamefont {Nociforo}}, \bibinfo {author}
  {\bibfnamefont {R.}~\bibnamefont {Palit}}, \bibinfo {author} {\bibfnamefont
  {S.}~\bibnamefont {Paschalis}}, \bibinfo {author} {\bibfnamefont
  {R.}~\bibnamefont {Plag}}, \bibinfo {author} {\bibfnamefont {R.}~\bibnamefont
  {Reifarth}}, \bibinfo {author} {\bibfnamefont {D.}~\bibnamefont {Savran}},
  \bibinfo {author} {\bibfnamefont {H.}~\bibnamefont {Scheit}}, \bibinfo
  {author} {\bibfnamefont {H.}~\bibnamefont {Simon}}, \bibinfo {author}
  {\bibfnamefont {K.}~\bibnamefont {S\"ummerer}}, \bibinfo {author}
  {\bibfnamefont {A.}~\bibnamefont {Wagner}}, \bibinfo {author} {\bibfnamefont
  {W.}~\bibnamefont {Walu\ifmmode~\acute{s}\else \'{s}\fi{}}}, \bibinfo
  {author} {\bibfnamefont {H.}~\bibnamefont {Weick}},\ and\ \bibinfo {author}
  {\bibfnamefont {M.}~\bibnamefont {Winkler}},\ }\bibfield  {title} {\bibinfo
  {title} {{Measurement of the dipole polarizability of the unstable
  neutron-rich nucleus $^{68}$Ni}},\ }\href
  {https://doi.org/10.1103/PhysRevLett.111.242503} {\bibfield  {journal}
  {\bibinfo  {journal} {Physical Review Letters}\ }\textbf {\bibinfo {volume}
  {111}},\ \bibinfo {pages} {242503} (\bibinfo {year} {2013})}\BibitemShut
  {NoStop}%
\bibitem [{\citenamefont {Von~Egidy}\ \emph {et~al.}(1988)\citenamefont
  {Von~Egidy}, \citenamefont {Schmidt},\ and\ \citenamefont
  {Behkami}}]{VONEGIDY1988189}%
  \BibitemOpen
  \bibfield  {author} {\bibinfo {author} {\bibfnamefont {T.}~\bibnamefont
  {Von~Egidy}}, \bibinfo {author} {\bibfnamefont {H.~H.}\ \bibnamefont
  {Schmidt}},\ and\ \bibinfo {author} {\bibfnamefont {A.~N.}\ \bibnamefont
  {Behkami}},\ }\bibfield  {title} {\bibinfo {title} {{Nuclear level densities
  and level spacing distributions: Part II}},\ }\href
  {https://doi.org/10.1016/0375-9474(88)90491-5} {\bibfield  {journal}
  {\bibinfo  {journal} {Nuclear Physics, Section A}\ }\textbf {\bibinfo
  {volume} {481}},\ \bibinfo {pages} {189} (\bibinfo {year}
  {1988})}\BibitemShut {NoStop}%
\bibitem [{\citenamefont {Egidy}\ and\ \citenamefont
  {Bucurescu}(2005)}]{VonEgidy2005}%
  \BibitemOpen
  \bibfield  {author} {\bibinfo {author} {\bibfnamefont {T.~v.}\ \bibnamefont
  {Egidy}}\ and\ \bibinfo {author} {\bibfnamefont {D.}~\bibnamefont
  {Bucurescu}},\ }\bibfield  {title} {\bibinfo {title} {{Systematics of nuclear
  level density parameters}},\ }\href
  {https://doi.org/10.1103/PhysRevC.72.044311} {\bibfield  {journal} {\bibinfo
  {journal} {Physical Review C}\ }\textbf {\bibinfo {volume} {72}},\ \bibinfo
  {pages} {044311} (\bibinfo {year} {2005})}\BibitemShut {NoStop}%
\bibitem [{\citenamefont {von Egidy}\ and\ \citenamefont
  {Bucurescu}(2009)}]{PhysRevC.80.054310}%
  \BibitemOpen
  \bibfield  {author} {\bibinfo {author} {\bibfnamefont {T.}~\bibnamefont {von
  Egidy}}\ and\ \bibinfo {author} {\bibfnamefont {D.}~\bibnamefont
  {Bucurescu}},\ }\bibfield  {title} {\bibinfo {title} {{Experimental
  energy-dependent nuclear spin distributions}},\ }\href
  {https://doi.org/10.1103/PhysRevC.80.054310} {\bibfield  {journal} {\bibinfo
  {journal} {Physical Review C}\ }\textbf {\bibinfo {volume} {80}},\ \bibinfo
  {pages} {054310} (\bibinfo {year} {2009})}\BibitemShut {NoStop}%
\bibitem [{\citenamefont {Junde}\ \emph {et~al.}(2005)\citenamefont {Junde},
  \citenamefont {Xiaolong},\ and\ \citenamefont {Tuli}}]{JUNDE2005159}%
  \BibitemOpen
  \bibfield  {author} {\bibinfo {author} {\bibfnamefont {H.}~\bibnamefont
  {Junde}}, \bibinfo {author} {\bibfnamefont {H.}~\bibnamefont {Xiaolong}},\
  and\ \bibinfo {author} {\bibfnamefont {J.~K.}\ \bibnamefont {Tuli}},\
  }\bibfield  {title} {\bibinfo {title} {{Nuclear Data Sheets for A = 67}},\
  }\href {https://doi.org/https://doi.org/10.1016/j.nds.2005.10.006} {\bibfield
   {journal} {\bibinfo  {journal} {Nuclear Data Sheets}\ }\textbf {\bibinfo
  {volume} {106}},\ \bibinfo {pages} {159} (\bibinfo {year}
  {2005})}\BibitemShut {NoStop}%
\bibitem [{\citenamefont {Diriken}\ \emph {et~al.}(2015)\citenamefont
  {Diriken}, \citenamefont {Patronis}, \citenamefont {Andreyev}, \citenamefont
  {Antalic}, \citenamefont {Bildstein}, \citenamefont {Blazhev}, \citenamefont
  {Darby}, \citenamefont {De~Witte}, \citenamefont {Eberth}, \citenamefont
  {Elseviers}, \citenamefont {Fedosseev}, \citenamefont {Flavigny},
  \citenamefont {Fransen}, \citenamefont {Georgiev}, \citenamefont
  {Gernhauser}, \citenamefont {Hess}, \citenamefont {Huyse}, \citenamefont
  {Jolie}, \citenamefont {Kr{\"{o}}ll}, \citenamefont {Kr{\"{u}}cken},
  \citenamefont {Lutter}, \citenamefont {Marsh}, \citenamefont {Mertzimekis},
  \citenamefont {Muecher}, \citenamefont {Orlandi}, \citenamefont {Pakou},
  \citenamefont {Raabe}, \citenamefont {Randisi}, \citenamefont {Reiter},
  \citenamefont {Roger}, \citenamefont {Seidlitz}, \citenamefont {Seliverstov},
  \citenamefont {Sotty}, \citenamefont {Tornqvist}, \citenamefont {Van
  De~Walle}, \citenamefont {Van~Duppen}, \citenamefont {Voulot}, \citenamefont
  {Warr}, \citenamefont {Wenander},\ and\ \citenamefont
  {Wimmer}}]{PhysRevC.91.054321}%
  \BibitemOpen
  \bibfield  {author} {\bibinfo {author} {\bibfnamefont {J.}~\bibnamefont
  {Diriken}}, \bibinfo {author} {\bibfnamefont {N.}~\bibnamefont {Patronis}},
  \bibinfo {author} {\bibfnamefont {A.}~\bibnamefont {Andreyev}}, \bibinfo
  {author} {\bibfnamefont {S.}~\bibnamefont {Antalic}}, \bibinfo {author}
  {\bibfnamefont {V.}~\bibnamefont {Bildstein}}, \bibinfo {author}
  {\bibfnamefont {A.}~\bibnamefont {Blazhev}}, \bibinfo {author} {\bibfnamefont
  {I.~G.}\ \bibnamefont {Darby}}, \bibinfo {author} {\bibfnamefont
  {H.}~\bibnamefont {De~Witte}}, \bibinfo {author} {\bibfnamefont
  {J.}~\bibnamefont {Eberth}}, \bibinfo {author} {\bibfnamefont
  {J.}~\bibnamefont {Elseviers}}, \bibinfo {author} {\bibfnamefont {V.~N.}\
  \bibnamefont {Fedosseev}}, \bibinfo {author} {\bibfnamefont {F.}~\bibnamefont
  {Flavigny}}, \bibinfo {author} {\bibfnamefont {C.}~\bibnamefont {Fransen}},
  \bibinfo {author} {\bibfnamefont {G.}~\bibnamefont {Georgiev}}, \bibinfo
  {author} {\bibfnamefont {R.}~\bibnamefont {Gernhauser}}, \bibinfo {author}
  {\bibfnamefont {H.}~\bibnamefont {Hess}}, \bibinfo {author} {\bibfnamefont
  {M.}~\bibnamefont {Huyse}}, \bibinfo {author} {\bibfnamefont
  {J.}~\bibnamefont {Jolie}}, \bibinfo {author} {\bibfnamefont
  {T.}~\bibnamefont {Kr{\"{o}}ll}}, \bibinfo {author} {\bibfnamefont
  {R.}~\bibnamefont {Kr{\"{u}}cken}}, \bibinfo {author} {\bibfnamefont
  {R.}~\bibnamefont {Lutter}}, \bibinfo {author} {\bibfnamefont {B.~A.}\
  \bibnamefont {Marsh}}, \bibinfo {author} {\bibfnamefont {T.}~\bibnamefont
  {Mertzimekis}}, \bibinfo {author} {\bibfnamefont {D.}~\bibnamefont
  {Muecher}}, \bibinfo {author} {\bibfnamefont {R.}~\bibnamefont {Orlandi}},
  \bibinfo {author} {\bibfnamefont {A.}~\bibnamefont {Pakou}}, \bibinfo
  {author} {\bibfnamefont {R.}~\bibnamefont {Raabe}}, \bibinfo {author}
  {\bibfnamefont {G.}~\bibnamefont {Randisi}}, \bibinfo {author} {\bibfnamefont
  {P.}~\bibnamefont {Reiter}}, \bibinfo {author} {\bibfnamefont
  {T.}~\bibnamefont {Roger}}, \bibinfo {author} {\bibfnamefont
  {M.}~\bibnamefont {Seidlitz}}, \bibinfo {author} {\bibfnamefont
  {M.}~\bibnamefont {Seliverstov}}, \bibinfo {author} {\bibfnamefont
  {C.}~\bibnamefont {Sotty}}, \bibinfo {author} {\bibfnamefont
  {H.}~\bibnamefont {Tornqvist}}, \bibinfo {author} {\bibfnamefont
  {J.}~\bibnamefont {Van De~Walle}}, \bibinfo {author} {\bibfnamefont
  {P.}~\bibnamefont {Van~Duppen}}, \bibinfo {author} {\bibfnamefont
  {D.}~\bibnamefont {Voulot}}, \bibinfo {author} {\bibfnamefont
  {N.}~\bibnamefont {Warr}}, \bibinfo {author} {\bibfnamefont {F.}~\bibnamefont
  {Wenander}},\ and\ \bibinfo {author} {\bibfnamefont {K.}~\bibnamefont
  {Wimmer}},\ }\bibfield  {title} {\bibinfo {title} {{Experimental study of the
  $^{66}$Ni(d,p)$^{67}$Ni one-neutron transfer reaction}},\ }\href
  {https://doi.org/10.1103/PhysRevC.91.054321} {\bibfield  {journal} {\bibinfo
  {journal} {Physical Review C}\ }\textbf {\bibinfo {volume} {91}},\ \bibinfo
  {pages} {54321} (\bibinfo {year} {2015})}\BibitemShut {NoStop}%
\bibitem [{\citenamefont {Koning}\ \emph {et~al.}(2007)\citenamefont {Koning},
  \citenamefont {Hilaire},\ and\ \citenamefont {Duijvestijn}}]{Koning2007}%
  \BibitemOpen
  \bibfield  {author} {\bibinfo {author} {\bibfnamefont {A.~J.}\ \bibnamefont
  {Koning}}, \bibinfo {author} {\bibfnamefont {S.}~\bibnamefont {Hilaire}},\
  and\ \bibinfo {author} {\bibfnamefont {M.~C.}\ \bibnamefont {Duijvestijn}},\
  }\bibfield  {title} {\bibinfo {title} {{TALYS-1.0}},\ }in\ \href
  {https://doi.org/10.1051/ndata:07767} {\emph {\bibinfo {booktitle}
  {ND2007}}}\ (\bibinfo  {publisher} {EDP Sciences},\ \bibinfo {address} {Les
  Ulis, France},\ \bibinfo {year} {2007})\ pp.\ \bibinfo {pages}
  {211--214}\BibitemShut {NoStop}%
\bibitem [{\citenamefont {Koning}\ and\ \citenamefont
  {Delaroche}(2003)}]{koning_local_2003}%
  \BibitemOpen
  \bibfield  {author} {\bibinfo {author} {\bibfnamefont {A.}~\bibnamefont
  {Koning}}\ and\ \bibinfo {author} {\bibfnamefont {J.}~\bibnamefont
  {Delaroche}},\ }\bibfield  {title} {\bibinfo {title} {Local and global
  nucleon optical models from 1 $kev$ to 200 $mev$},\ }\href
  {https://doi.org/10.1016/S0375-9474(02)01321-0} {\bibfield  {journal}
  {\bibinfo  {journal} {Nuclear Physics A}\ }\textbf {\bibinfo {volume}
  {713}},\ \bibinfo {pages} {231} (\bibinfo {year} {2003})}\BibitemShut
  {NoStop}%
\bibitem [{\citenamefont {Bauge}\ \emph {et~al.}(2001)\citenamefont {Bauge},
  \citenamefont {Delaroche},\ and\ \citenamefont {Girod}}]{PhysRevC.63.024607}%
  \BibitemOpen
  \bibfield  {author} {\bibinfo {author} {\bibfnamefont {E.}~\bibnamefont
  {Bauge}}, \bibinfo {author} {\bibfnamefont {J.~P.}\ \bibnamefont
  {Delaroche}},\ and\ \bibinfo {author} {\bibfnamefont {M.}~\bibnamefont
  {Girod}},\ }\bibfield  {title} {\bibinfo {title} {{Lane-consistent,
  semimicroscopic nucleon-nucleus optical model}},\ }\href
  {https://doi.org/10.1103/PhysRevC.63.024607} {\bibfield  {journal} {\bibinfo
  {journal} {Phys. Rev. C}\ }\textbf {\bibinfo {volume} {63}},\ \bibinfo
  {pages} {024607} (\bibinfo {year} {2001})}\BibitemShut {NoStop}%
\bibitem [{\citenamefont {Plompen}\ \emph {et~al.}(2020)\citenamefont
  {Plompen}, \citenamefont {Cabellos}, \citenamefont {De~Saint~Jean},
  \citenamefont {Fleming}, \citenamefont {Algora}, \citenamefont {Angelone},
  \citenamefont {Archier}, \citenamefont {Bauge}, \citenamefont {Bersillon},
  \citenamefont {Blokhin}, \citenamefont {Cantargi}, \citenamefont {Chebboubi},
  \citenamefont {Diez}, \citenamefont {Duarte}, \citenamefont {Dupont},
  \citenamefont {Dyrda}, \citenamefont {Erasmus}, \citenamefont {Fiorito},
  \citenamefont {Fischer}, \citenamefont {Flammini}, \citenamefont {Foligno},
  \citenamefont {Gilbert}, \citenamefont {Granada}, \citenamefont {Haeck},
  \citenamefont {Hambsch}, \citenamefont {Helgesson}, \citenamefont {Hilaire},
  \citenamefont {Hill}, \citenamefont {Hursin}, \citenamefont {Ichou},
  \citenamefont {Jacqmin}, \citenamefont {Jansky}, \citenamefont {Jouanne},
  \citenamefont {Kellett}, \citenamefont {Kim}, \citenamefont {Kim},
  \citenamefont {Kodeli}, \citenamefont {Koning}, \citenamefont {Konobeyev},
  \citenamefont {Kopecky}, \citenamefont {Kos}, \citenamefont {Kr{\'{a}}sa},
  \citenamefont {Leal}, \citenamefont {Leclaire}, \citenamefont {Leconte},
  \citenamefont {Lee}, \citenamefont {Leeb}, \citenamefont {Litaize},
  \citenamefont {Majerle}, \citenamefont {M{\'{a}}rquez Dami{\'{a}}n},
  \citenamefont {Michel-Sendis}, \citenamefont {Mills}, \citenamefont
  {Morillon}, \citenamefont {Nogu{\`{e}}re}, \citenamefont {Pecchia},
  \citenamefont {Pelloni}, \citenamefont {Pereslavtsev}, \citenamefont {Perry},
  \citenamefont {Rochman}, \citenamefont {R{\"{o}}hrmoser}, \citenamefont
  {Romain}, \citenamefont {Romojaro}, \citenamefont {Roubtsov}, \citenamefont
  {Sauvan}, \citenamefont {Schillebeeckx}, \citenamefont {Schmidt},
  \citenamefont {Serot}, \citenamefont {Simakov}, \citenamefont {Sirakov},
  \citenamefont {Sj{\"{o}}strand}, \citenamefont {Stankovskiy}, \citenamefont
  {Sublet}, \citenamefont {Tamagno}, \citenamefont {Trkov}, \citenamefont
  {van~der Marck}, \citenamefont {{\'{A}}lvarez-Velarde}, \citenamefont
  {Villari}, \citenamefont {Ware}, \citenamefont {Yokoyama},\ and\
  \citenamefont {{\v{Z}}erovnik}}]{Plompen2020TheJEFF-3.3}%
  \BibitemOpen
  \bibfield  {author} {\bibinfo {author} {\bibfnamefont {A.~J.~M.}\
  \bibnamefont {Plompen}}, \bibinfo {author} {\bibfnamefont {O.}~\bibnamefont
  {Cabellos}}, \bibinfo {author} {\bibfnamefont {C.}~\bibnamefont
  {De~Saint~Jean}}, \bibinfo {author} {\bibfnamefont {M.}~\bibnamefont
  {Fleming}}, \bibinfo {author} {\bibfnamefont {A.}~\bibnamefont {Algora}},
  \bibinfo {author} {\bibfnamefont {M.}~\bibnamefont {Angelone}}, \bibinfo
  {author} {\bibfnamefont {P.}~\bibnamefont {Archier}}, \bibinfo {author}
  {\bibfnamefont {E.}~\bibnamefont {Bauge}}, \bibinfo {author} {\bibfnamefont
  {O.}~\bibnamefont {Bersillon}}, \bibinfo {author} {\bibfnamefont
  {A.}~\bibnamefont {Blokhin}}, \bibinfo {author} {\bibfnamefont
  {F.}~\bibnamefont {Cantargi}}, \bibinfo {author} {\bibfnamefont
  {A.}~\bibnamefont {Chebboubi}}, \bibinfo {author} {\bibfnamefont
  {C.}~\bibnamefont {Diez}}, \bibinfo {author} {\bibfnamefont {H.}~\bibnamefont
  {Duarte}}, \bibinfo {author} {\bibfnamefont {E.}~\bibnamefont {Dupont}},
  \bibinfo {author} {\bibfnamefont {J.}~\bibnamefont {Dyrda}}, \bibinfo
  {author} {\bibfnamefont {B.}~\bibnamefont {Erasmus}}, \bibinfo {author}
  {\bibfnamefont {L.}~\bibnamefont {Fiorito}}, \bibinfo {author} {\bibfnamefont
  {U.}~\bibnamefont {Fischer}}, \bibinfo {author} {\bibfnamefont
  {D.}~\bibnamefont {Flammini}}, \bibinfo {author} {\bibfnamefont
  {D.}~\bibnamefont {Foligno}}, \bibinfo {author} {\bibfnamefont {M.~R.}\
  \bibnamefont {Gilbert}}, \bibinfo {author} {\bibfnamefont {J.~R.}\
  \bibnamefont {Granada}}, \bibinfo {author} {\bibfnamefont {W.}~\bibnamefont
  {Haeck}}, \bibinfo {author} {\bibfnamefont {F.-J.}\ \bibnamefont {Hambsch}},
  \bibinfo {author} {\bibfnamefont {P.}~\bibnamefont {Helgesson}}, \bibinfo
  {author} {\bibfnamefont {S.}~\bibnamefont {Hilaire}}, \bibinfo {author}
  {\bibfnamefont {I.}~\bibnamefont {Hill}}, \bibinfo {author} {\bibfnamefont
  {M.}~\bibnamefont {Hursin}}, \bibinfo {author} {\bibfnamefont
  {R.}~\bibnamefont {Ichou}}, \bibinfo {author} {\bibfnamefont
  {R.}~\bibnamefont {Jacqmin}}, \bibinfo {author} {\bibfnamefont
  {B.}~\bibnamefont {Jansky}}, \bibinfo {author} {\bibfnamefont
  {C.}~\bibnamefont {Jouanne}}, \bibinfo {author} {\bibfnamefont {M.~A.}\
  \bibnamefont {Kellett}}, \bibinfo {author} {\bibfnamefont {D.~H.}\
  \bibnamefont {Kim}}, \bibinfo {author} {\bibfnamefont {H.~I.}\ \bibnamefont
  {Kim}}, \bibinfo {author} {\bibfnamefont {I.}~\bibnamefont {Kodeli}},
  \bibinfo {author} {\bibfnamefont {A.~J.}\ \bibnamefont {Koning}}, \bibinfo
  {author} {\bibfnamefont {A.~Y.}\ \bibnamefont {Konobeyev}}, \bibinfo {author}
  {\bibfnamefont {S.}~\bibnamefont {Kopecky}}, \bibinfo {author} {\bibfnamefont
  {B.}~\bibnamefont {Kos}}, \bibinfo {author} {\bibfnamefont {A.}~\bibnamefont
  {Kr{\'{a}}sa}}, \bibinfo {author} {\bibfnamefont {L.~C.}\ \bibnamefont
  {Leal}}, \bibinfo {author} {\bibfnamefont {N.}~\bibnamefont {Leclaire}},
  \bibinfo {author} {\bibfnamefont {P.}~\bibnamefont {Leconte}}, \bibinfo
  {author} {\bibfnamefont {Y.~O.}\ \bibnamefont {Lee}}, \bibinfo {author}
  {\bibfnamefont {H.}~\bibnamefont {Leeb}}, \bibinfo {author} {\bibfnamefont
  {O.}~\bibnamefont {Litaize}}, \bibinfo {author} {\bibfnamefont
  {M.}~\bibnamefont {Majerle}}, \bibinfo {author} {\bibfnamefont {J.~I.}\
  \bibnamefont {M{\'{a}}rquez Dami{\'{a}}n}}, \bibinfo {author} {\bibfnamefont
  {F.}~\bibnamefont {Michel-Sendis}}, \bibinfo {author} {\bibfnamefont {R.~W.}\
  \bibnamefont {Mills}}, \bibinfo {author} {\bibfnamefont {B.}~\bibnamefont
  {Morillon}}, \bibinfo {author} {\bibfnamefont {G.}~\bibnamefont
  {Nogu{\`{e}}re}}, \bibinfo {author} {\bibfnamefont {M.}~\bibnamefont
  {Pecchia}}, \bibinfo {author} {\bibfnamefont {S.}~\bibnamefont {Pelloni}},
  \bibinfo {author} {\bibfnamefont {P.}~\bibnamefont {Pereslavtsev}}, \bibinfo
  {author} {\bibfnamefont {R.~J.}\ \bibnamefont {Perry}}, \bibinfo {author}
  {\bibfnamefont {D.}~\bibnamefont {Rochman}}, \bibinfo {author} {\bibfnamefont
  {A.}~\bibnamefont {R{\"{o}}hrmoser}}, \bibinfo {author} {\bibfnamefont
  {P.}~\bibnamefont {Romain}}, \bibinfo {author} {\bibfnamefont
  {P.}~\bibnamefont {Romojaro}}, \bibinfo {author} {\bibfnamefont
  {D.}~\bibnamefont {Roubtsov}}, \bibinfo {author} {\bibfnamefont
  {P.}~\bibnamefont {Sauvan}}, \bibinfo {author} {\bibfnamefont
  {P.}~\bibnamefont {Schillebeeckx}}, \bibinfo {author} {\bibfnamefont {K.~H.}\
  \bibnamefont {Schmidt}}, \bibinfo {author} {\bibfnamefont {O.}~\bibnamefont
  {Serot}}, \bibinfo {author} {\bibfnamefont {S.}~\bibnamefont {Simakov}},
  \bibinfo {author} {\bibfnamefont {I.}~\bibnamefont {Sirakov}}, \bibinfo
  {author} {\bibfnamefont {H.}~\bibnamefont {Sj{\"{o}}strand}}, \bibinfo
  {author} {\bibfnamefont {A.}~\bibnamefont {Stankovskiy}}, \bibinfo {author}
  {\bibfnamefont {J.~C.}\ \bibnamefont {Sublet}}, \bibinfo {author}
  {\bibfnamefont {P.}~\bibnamefont {Tamagno}}, \bibinfo {author} {\bibfnamefont
  {A.}~\bibnamefont {Trkov}}, \bibinfo {author} {\bibfnamefont
  {S.}~\bibnamefont {van~der Marck}}, \bibinfo {author} {\bibfnamefont
  {F.}~\bibnamefont {{\'{A}}lvarez-Velarde}}, \bibinfo {author} {\bibfnamefont
  {R.}~\bibnamefont {Villari}}, \bibinfo {author} {\bibfnamefont {T.~C.}\
  \bibnamefont {Ware}}, \bibinfo {author} {\bibfnamefont {K.}~\bibnamefont
  {Yokoyama}},\ and\ \bibinfo {author} {\bibfnamefont {G.}~\bibnamefont
  {{\v{Z}}erovnik}},\ }\bibfield  {title} {\bibinfo {title} {{The joint
  evaluated fission and fusion nuclear data library, JEFF-3.3}},\ }\href
  {https://doi.org/10.1140/epja/s10050-020-00141-9} {\bibfield  {journal}
  {\bibinfo  {journal} {The European Physical Journal A}\ }\textbf {\bibinfo
  {volume} {56}},\ \bibinfo {pages} {181} (\bibinfo {year} {2020})}\BibitemShut
  {NoStop}%
\bibitem [{\citenamefont {Koning}\ \emph {et~al.}(2019)\citenamefont {Koning},
  \citenamefont {Rochman}, \citenamefont {Sublet}, \citenamefont {Dzysiuk},
  \citenamefont {Fleming},\ and\ \citenamefont {van~der
  Marck}}]{Koning2019TENDL:Technology}%
  \BibitemOpen
  \bibfield  {author} {\bibinfo {author} {\bibfnamefont {A.}~\bibnamefont
  {Koning}}, \bibinfo {author} {\bibfnamefont {D.}~\bibnamefont {Rochman}},
  \bibinfo {author} {\bibfnamefont {J.-C.}\ \bibnamefont {Sublet}}, \bibinfo
  {author} {\bibfnamefont {N.}~\bibnamefont {Dzysiuk}}, \bibinfo {author}
  {\bibfnamefont {M.}~\bibnamefont {Fleming}},\ and\ \bibinfo {author}
  {\bibfnamefont {S.}~\bibnamefont {van~der Marck}},\ }\bibfield  {title}
  {\bibinfo {title} {{TENDL: Complete Nuclear Data Library for Innovative
  Nuclear Science and Technology}},\ }\href
  {https://doi.org/10.1016/j.nds.2019.01.002} {\bibfield  {journal} {\bibinfo
  {journal} {Nuclear Data Sheets}\ }\textbf {\bibinfo {volume} {155}},\
  \bibinfo {pages} {1} (\bibinfo {year} {2019})}\BibitemShut {NoStop}%
\bibitem [{\citenamefont {Rochman}\ \emph {et~al.}(2020)\citenamefont
  {Rochman}, \citenamefont {Koning},\ and\ \citenamefont
  {Sublet}}]{Rochman2020}%
  \BibitemOpen
  \bibfield  {author} {\bibinfo {author} {\bibfnamefont {D.}~\bibnamefont
  {Rochman}}, \bibinfo {author} {\bibfnamefont {A.}~\bibnamefont {Koning}},\
  and\ \bibinfo {author} {\bibfnamefont {J.-C.}\ \bibnamefont {Sublet}},\
  }\bibfield  {title} {\bibinfo {title} {A statistical analysis of evaluated
  neutron resonances with tares for jeff-3.3, jendl-4.0, endf/b-viii.0 and
  tendl-2019},\ }\href
  {https://doi.org/https://doi.org/10.1016/j.nds.2019.12.003} {\bibfield
  {journal} {\bibinfo  {journal} {Nuclear Data Sheets}\ }\textbf {\bibinfo
  {volume} {163}},\ \bibinfo {pages} {163} (\bibinfo {year}
  {2020})}\BibitemShut {NoStop}%
\bibitem [{\citenamefont {Cyburt}\ \emph {et~al.}(2010)\citenamefont {Cyburt},
  \citenamefont {Amthor}, \citenamefont {Ferguson}, \citenamefont {Meisel},
  \citenamefont {Smith}, \citenamefont {Warren}, \citenamefont {Heger},
  \citenamefont {Hoffman}, \citenamefont {Rauscher}, \citenamefont {Sakharuk},
  \citenamefont {Schatz}, \citenamefont {Thielemann},\ and\ \citenamefont
  {Wiescher}}]{cyburt_jina_2010}%
  \BibitemOpen
  \bibfield  {author} {\bibinfo {author} {\bibfnamefont {R.~H.}\ \bibnamefont
  {Cyburt}}, \bibinfo {author} {\bibfnamefont {A.~M.}\ \bibnamefont {Amthor}},
  \bibinfo {author} {\bibfnamefont {R.}~\bibnamefont {Ferguson}}, \bibinfo
  {author} {\bibfnamefont {Z.}~\bibnamefont {Meisel}}, \bibinfo {author}
  {\bibfnamefont {K.}~\bibnamefont {Smith}}, \bibinfo {author} {\bibfnamefont
  {S.}~\bibnamefont {Warren}}, \bibinfo {author} {\bibfnamefont
  {A.}~\bibnamefont {Heger}}, \bibinfo {author} {\bibfnamefont {R.~D.}\
  \bibnamefont {Hoffman}}, \bibinfo {author} {\bibfnamefont {T.}~\bibnamefont
  {Rauscher}}, \bibinfo {author} {\bibfnamefont {A.}~\bibnamefont {Sakharuk}},
  \bibinfo {author} {\bibfnamefont {H.}~\bibnamefont {Schatz}}, \bibinfo
  {author} {\bibfnamefont {F.~K.}\ \bibnamefont {Thielemann}},\ and\ \bibinfo
  {author} {\bibfnamefont {M.}~\bibnamefont {Wiescher}},\ }\bibfield  {title}
  {\bibinfo {title} {The jina reaclib database: Its recent updates and impact
  on type-i x-ray bursts},\ }\href
  {https://doi.org/10.1088/0067-0049/189/1/240} {\bibfield  {journal} {\bibinfo
   {journal} {The Astrophysical Journal Supplement Series}\ }\textbf {\bibinfo
  {volume} {189}},\ \bibinfo {pages} {240} (\bibinfo {year}
  {2010})}\BibitemShut {NoStop}%
\bibitem [{\citenamefont {Arnould}\ and\ \citenamefont
  {Goriely}(2006)}]{arnould06}%
  \BibitemOpen
  \bibfield  {author} {\bibinfo {author} {\bibfnamefont {M.}~\bibnamefont
  {Arnould}}\ and\ \bibinfo {author} {\bibfnamefont {S.}~\bibnamefont
  {Goriely}},\ }\bibfield  {title} {\bibinfo {title} {Microscopic nuclear
  models for astrophysics: The brussels bruslib nuclear library and beyond},\
  }\href
  {https://dipot.ulb.ac.be/dspace/bitstream/2013/88909/1/Elsevier\_67323.pdf}
  {\bibfield  {journal} {\bibinfo  {journal} {Nuclear Physics A}\ }\textbf
  {\bibinfo {volume} {777}},\ \bibinfo {pages} {157} (\bibinfo {year}
  {2006})}\BibitemShut {NoStop}%
\bibitem [{\citenamefont {Denissenkov}\ \emph {et~al.}(2018)\citenamefont
  {Denissenkov}, \citenamefont {Perdikakis}, \citenamefont {Herwig},
  \citenamefont {Schatz}, \citenamefont {Ritter}, \citenamefont {Pignatari},
  \citenamefont {Jones}, \citenamefont {Nikas},\ and\ \citenamefont
  {Spyrou}}]{denissenkov_impact_2018}%
  \BibitemOpen
  \bibfield  {author} {\bibinfo {author} {\bibfnamefont {P.}~\bibnamefont
  {Denissenkov}}, \bibinfo {author} {\bibfnamefont {G.}~\bibnamefont
  {Perdikakis}}, \bibinfo {author} {\bibfnamefont {F.}~\bibnamefont {Herwig}},
  \bibinfo {author} {\bibfnamefont {H.}~\bibnamefont {Schatz}}, \bibinfo
  {author} {\bibfnamefont {C.}~\bibnamefont {Ritter}}, \bibinfo {author}
  {\bibfnamefont {M.}~\bibnamefont {Pignatari}}, \bibinfo {author}
  {\bibfnamefont {S.}~\bibnamefont {Jones}}, \bibinfo {author} {\bibfnamefont
  {S.}~\bibnamefont {Nikas}},\ and\ \bibinfo {author} {\bibfnamefont
  {A.}~\bibnamefont {Spyrou}},\ }\bibfield  {title} {\bibinfo {title} {{The
  impact of (n,$\gamma$) reaction rate uncertainties of unstable isotopes near
  \textit{N} = 50 on the i-process nucleosynthesis in {He}-shell flash white
  dwarfs}},\ }\href {https://doi.org/10.1088/1361-6471/aabb6e} {\bibfield
  {journal} {\bibinfo  {journal} {Journal of Physics G: Nuclear and Particle
  Physics}\ }\textbf {\bibinfo {volume} {45}},\ \bibinfo {pages} {055203}
  (\bibinfo {year} {2018})}\BibitemShut {NoStop}%
\bibitem [{\citenamefont {Utsunomiya}\ \emph {et~al.}(2018)\citenamefont
  {Utsunomiya}, \citenamefont {Renstr\o{}m}, \citenamefont {Tveten},
  \citenamefont {Goriely}, \citenamefont {Katayama}, \citenamefont {Ari-izumi},
  \citenamefont {Takenaka}, \citenamefont {Symochko}, \citenamefont {Kheswa},
  \citenamefont {Ingeberg}, \citenamefont {Glodariu}, \citenamefont {Lui},
  \citenamefont {Miyamoto}, \citenamefont {Larsen}, \citenamefont {Midtb\o{}},
  \citenamefont {G\"orgen}, \citenamefont {Siem}, \citenamefont {Campo},
  \citenamefont {Guttormsen}, \citenamefont {Hilaire}, \citenamefont {P\'eru},\
  and\ \citenamefont {Koning}}]{PhysRevC.98.054619}%
  \BibitemOpen
  \bibfield  {author} {\bibinfo {author} {\bibfnamefont {H.}~\bibnamefont
  {Utsunomiya}}, \bibinfo {author} {\bibfnamefont {T.}~\bibnamefont
  {Renstr\o{}m}}, \bibinfo {author} {\bibfnamefont {G.~M.}\ \bibnamefont
  {Tveten}}, \bibinfo {author} {\bibfnamefont {S.}~\bibnamefont {Goriely}},
  \bibinfo {author} {\bibfnamefont {S.}~\bibnamefont {Katayama}}, \bibinfo
  {author} {\bibfnamefont {T.}~\bibnamefont {Ari-izumi}}, \bibinfo {author}
  {\bibfnamefont {D.}~\bibnamefont {Takenaka}}, \bibinfo {author}
  {\bibfnamefont {D.}~\bibnamefont {Symochko}}, \bibinfo {author}
  {\bibfnamefont {B.~V.}\ \bibnamefont {Kheswa}}, \bibinfo {author}
  {\bibfnamefont {V.~W.}\ \bibnamefont {Ingeberg}}, \bibinfo {author}
  {\bibfnamefont {T.}~\bibnamefont {Glodariu}}, \bibinfo {author}
  {\bibfnamefont {Y.-W.}\ \bibnamefont {Lui}}, \bibinfo {author} {\bibfnamefont
  {S.}~\bibnamefont {Miyamoto}}, \bibinfo {author} {\bibfnamefont {A.~C.}\
  \bibnamefont {Larsen}}, \bibinfo {author} {\bibfnamefont {J.~E.}\
  \bibnamefont {Midtb\o{}}}, \bibinfo {author} {\bibfnamefont {A.}~\bibnamefont
  {G\"orgen}}, \bibinfo {author} {\bibfnamefont {S.}~\bibnamefont {Siem}},
  \bibinfo {author} {\bibfnamefont {L.~C.}\ \bibnamefont {Campo}}, \bibinfo
  {author} {\bibfnamefont {M.}~\bibnamefont {Guttormsen}}, \bibinfo {author}
  {\bibfnamefont {S.}~\bibnamefont {Hilaire}}, \bibinfo {author} {\bibfnamefont
  {S.}~\bibnamefont {P\'eru}},\ and\ \bibinfo {author} {\bibfnamefont {A.~J.}\
  \bibnamefont {Koning}},\ }\bibfield  {title} {\bibinfo {title} {{Photoneutron
  cross sections for Ni isotopes: Toward understanding (n,{$\gamma$}) cross
  sections relevant to weak s -process nucleosynthesis}},\ }\href
  {https://doi.org/10.1103/PhysRevC.98.054619} {\bibfield  {journal} {\bibinfo
  {journal} {Physical Review C}\ }\textbf {\bibinfo {volume} {98}},\ \bibinfo
  {pages} {054619} (\bibinfo {year} {2018})}\BibitemShut {NoStop}%
\bibitem [{\citenamefont {Crespo~Campo}\ \emph {et~al.}(2016)\citenamefont
  {Crespo~Campo}, \citenamefont {Bello~Garrote}, \citenamefont {Eriksen},
  \citenamefont {G{\"{o}}rgen}, \citenamefont {Guttormsen}, \citenamefont
  {Hadynska-Klek}, \citenamefont {Klintefjord}, \citenamefont {Larsen},
  \citenamefont {Renstr{\o}m}, \citenamefont {Sahin}, \citenamefont {Siem},
  \citenamefont {Springer}, \citenamefont {Tornyi},\ and\ \citenamefont
  {Tveten}}]{PhysRevC.94.044321}%
  \BibitemOpen
  \bibfield  {author} {\bibinfo {author} {\bibfnamefont {L.}~\bibnamefont
  {Crespo~Campo}}, \bibinfo {author} {\bibfnamefont {F.~L.}\ \bibnamefont
  {Bello~Garrote}}, \bibinfo {author} {\bibfnamefont {T.~K.}\ \bibnamefont
  {Eriksen}}, \bibinfo {author} {\bibfnamefont {A.}~\bibnamefont
  {G{\"{o}}rgen}}, \bibinfo {author} {\bibfnamefont {M.}~\bibnamefont
  {Guttormsen}}, \bibinfo {author} {\bibfnamefont {K.}~\bibnamefont
  {Hadynska-Klek}}, \bibinfo {author} {\bibfnamefont {M.}~\bibnamefont
  {Klintefjord}}, \bibinfo {author} {\bibfnamefont {A.~C.}\ \bibnamefont
  {Larsen}}, \bibinfo {author} {\bibfnamefont {T.}~\bibnamefont {Renstr{\o}m}},
  \bibinfo {author} {\bibfnamefont {E.}~\bibnamefont {Sahin}}, \bibinfo
  {author} {\bibfnamefont {S.}~\bibnamefont {Siem}}, \bibinfo {author}
  {\bibfnamefont {A.}~\bibnamefont {Springer}}, \bibinfo {author}
  {\bibfnamefont {T.~G.}\ \bibnamefont {Tornyi}},\ and\ \bibinfo {author}
  {\bibfnamefont {G.~M.}\ \bibnamefont {Tveten}},\ }\bibfield  {title}
  {\bibinfo {title} {{Statistical {$\gamma$}-decay properties of $^{64}$Ni and
  deduced (n,{$\gamma$}) cross section of the s-process branch-point nucleus
  $^{63}$Ni}},\ }\href {https://doi.org/10.1103/PhysRevC.94.044321} {\bibfield
  {journal} {\bibinfo  {journal} {Physical Review C}\ }\textbf {\bibinfo
  {volume} {94}},\ \bibinfo {pages} {044321} (\bibinfo {year}
  {2016})}\BibitemShut {NoStop}%
\bibitem [{\citenamefont {Spyrou}\ \emph {et~al.}(2017)\citenamefont {Spyrou},
  \citenamefont {Larsen}, \citenamefont {Liddick}, \citenamefont {Naqvi},
  \citenamefont {Crider}, \citenamefont {Dombos}, \citenamefont {Guttormsen},
  \citenamefont {Bleuel}, \citenamefont {Couture}, \citenamefont {Campo},
  \citenamefont {Lewis}, \citenamefont {Mosby}, \citenamefont {Mumpower},
  \citenamefont {Perdikakis}, \citenamefont {Prokop}, \citenamefont {Quinn},
  \citenamefont {Renstrom}, \citenamefont {Siem},\ and\ \citenamefont
  {Surman}}]{Spyrou2017}%
  \BibitemOpen
  \bibfield  {author} {\bibinfo {author} {\bibfnamefont {A.}~\bibnamefont
  {Spyrou}}, \bibinfo {author} {\bibfnamefont {A.~C.}\ \bibnamefont {Larsen}},
  \bibinfo {author} {\bibfnamefont {S.~N.}\ \bibnamefont {Liddick}}, \bibinfo
  {author} {\bibfnamefont {F.}~\bibnamefont {Naqvi}}, \bibinfo {author}
  {\bibfnamefont {B.~P.}\ \bibnamefont {Crider}}, \bibinfo {author}
  {\bibfnamefont {A.~C.}\ \bibnamefont {Dombos}}, \bibinfo {author}
  {\bibfnamefont {M.}~\bibnamefont {Guttormsen}}, \bibinfo {author}
  {\bibfnamefont {D.~L.}\ \bibnamefont {Bleuel}}, \bibinfo {author}
  {\bibfnamefont {A.}~\bibnamefont {Couture}}, \bibinfo {author} {\bibfnamefont
  {L.~C.}\ \bibnamefont {Campo}}, \bibinfo {author} {\bibfnamefont
  {R.}~\bibnamefont {Lewis}}, \bibinfo {author} {\bibfnamefont
  {S.}~\bibnamefont {Mosby}}, \bibinfo {author} {\bibfnamefont {M.~R.}\
  \bibnamefont {Mumpower}}, \bibinfo {author} {\bibfnamefont {G.}~\bibnamefont
  {Perdikakis}}, \bibinfo {author} {\bibfnamefont {C.~J.}\ \bibnamefont
  {Prokop}}, \bibinfo {author} {\bibfnamefont {S.~J.}\ \bibnamefont {Quinn}},
  \bibinfo {author} {\bibfnamefont {T.}~\bibnamefont {Renstrom}}, \bibinfo
  {author} {\bibfnamefont {S.}~\bibnamefont {Siem}},\ and\ \bibinfo {author}
  {\bibfnamefont {R.}~\bibnamefont {Surman}},\ }\bibfield  {title} {\bibinfo
  {title} {{Neutron-capture rates for explosive nucleosynthesis: The case of
  $^{68}$Ni(n,{$\gamma$})$^{69}$Ni}},\ }\href
  {https://doi.org/10.1088/1361-6471/aa5ae7} {\bibfield  {journal} {\bibinfo
  {journal} {Journal of Physics G: Nuclear and Particle Physics}\ }\textbf
  {\bibinfo {volume} {44}},\ \bibinfo {pages} {044002} (\bibinfo {year}
  {2017})}\BibitemShut {NoStop}%
\bibitem [{\citenamefont {Schwengner}\ \emph {et~al.}(2017)\citenamefont
  {Schwengner}, \citenamefont {Frauendorf},\ and\ \citenamefont
  {Brown}}]{Schwengner2017Low-EnergyNuclei}%
  \BibitemOpen
  \bibfield  {author} {\bibinfo {author} {\bibfnamefont {R.}~\bibnamefont
  {Schwengner}}, \bibinfo {author} {\bibfnamefont {S.}~\bibnamefont
  {Frauendorf}},\ and\ \bibinfo {author} {\bibfnamefont {B.~A.}\ \bibnamefont
  {Brown}},\ }\bibfield  {title} {\bibinfo {title} {{Low-Energy Magnetic Dipole
  Radiation in Open-Shell Nuclei}},\ }\href
  {https://doi.org/10.1103/PhysRevLett.118.092502} {\bibfield  {journal}
  {\bibinfo  {journal} {Physical Review Letters}\ }\textbf {\bibinfo {volume}
  {118}},\ \bibinfo {pages} {092502} (\bibinfo {year} {2017})}\BibitemShut
  {NoStop}%
\bibitem [{\citenamefont {Guttormsen}\ \emph {et~al.}(2022)\citenamefont
  {Guttormsen}, \citenamefont {Ay}, \citenamefont {Ozgur}, \citenamefont
  {Algin}, \citenamefont {Larsen}, \citenamefont {Bello~Garrote}, \citenamefont
  {Berg}, \citenamefont {Crespo~Campo}, \citenamefont {Dahl-Jacobsen},
  \citenamefont {Furmyr}, \citenamefont {Gjestvang}, \citenamefont {G\"orgen},
  \citenamefont {Hagen}, \citenamefont {Ingeberg}, \citenamefont {Kheswa},
  \citenamefont {Kullmann}, \citenamefont {Klintefjord}, \citenamefont
  {Markova}, \citenamefont {Midtb\o{}}, \citenamefont {Modamio}, \citenamefont
  {Paulsen}, \citenamefont {Pedersen}, \citenamefont {Renstr\o{}m},
  \citenamefont {Sahin}, \citenamefont {Siem}, \citenamefont {Tveten},\ and\
  \citenamefont {Wiedeking}}]{guttormsen22}%
  \BibitemOpen
  \bibfield  {author} {\bibinfo {author} {\bibfnamefont {M.}~\bibnamefont
  {Guttormsen}}, \bibinfo {author} {\bibfnamefont {K.~O.}\ \bibnamefont {Ay}},
  \bibinfo {author} {\bibfnamefont {M.}~\bibnamefont {Ozgur}}, \bibinfo
  {author} {\bibfnamefont {E.}~\bibnamefont {Algin}}, \bibinfo {author}
  {\bibfnamefont {A.~C.}\ \bibnamefont {Larsen}}, \bibinfo {author}
  {\bibfnamefont {F.~L.}\ \bibnamefont {Bello~Garrote}}, \bibinfo {author}
  {\bibfnamefont {H.~C.}\ \bibnamefont {Berg}}, \bibinfo {author}
  {\bibfnamefont {L.}~\bibnamefont {Crespo~Campo}}, \bibinfo {author}
  {\bibfnamefont {T.}~\bibnamefont {Dahl-Jacobsen}}, \bibinfo {author}
  {\bibfnamefont {F.~W.}\ \bibnamefont {Furmyr}}, \bibinfo {author}
  {\bibfnamefont {D.}~\bibnamefont {Gjestvang}}, \bibinfo {author}
  {\bibfnamefont {A.}~\bibnamefont {G\"orgen}}, \bibinfo {author}
  {\bibfnamefont {T.~W.}\ \bibnamefont {Hagen}}, \bibinfo {author}
  {\bibfnamefont {V.~W.}\ \bibnamefont {Ingeberg}}, \bibinfo {author}
  {\bibfnamefont {B.~V.}\ \bibnamefont {Kheswa}}, \bibinfo {author}
  {\bibfnamefont {I.~K.~B.}\ \bibnamefont {Kullmann}}, \bibinfo {author}
  {\bibfnamefont {M.}~\bibnamefont {Klintefjord}}, \bibinfo {author}
  {\bibfnamefont {M.}~\bibnamefont {Markova}}, \bibinfo {author} {\bibfnamefont
  {J.~E.}\ \bibnamefont {Midtb\o{}}}, \bibinfo {author} {\bibfnamefont
  {V.}~\bibnamefont {Modamio}}, \bibinfo {author} {\bibfnamefont
  {W.}~\bibnamefont {Paulsen}}, \bibinfo {author} {\bibfnamefont {L.~G.}\
  \bibnamefont {Pedersen}}, \bibinfo {author} {\bibfnamefont {T.}~\bibnamefont
  {Renstr\o{}m}}, \bibinfo {author} {\bibfnamefont {E.}~\bibnamefont {Sahin}},
  \bibinfo {author} {\bibfnamefont {S.}~\bibnamefont {Siem}}, \bibinfo {author}
  {\bibfnamefont {G.~M.}\ \bibnamefont {Tveten}},\ and\ \bibinfo {author}
  {\bibfnamefont {M.}~\bibnamefont {Wiedeking}},\ }\bibfield  {title} {\bibinfo
  {title} {Evolution of the $\ensuremath{\gamma}$-ray strength function in
  neodymium isotopes},\ }\href {https://doi.org/10.1103/PhysRevC.106.034314}
  {\bibfield  {journal} {\bibinfo  {journal} {Physical Review C}\ }\textbf
  {\bibinfo {volume} {106}},\ \bibinfo {pages} {034314} (\bibinfo {year}
  {2022})}\BibitemShut {NoStop}%
\bibitem [{\citenamefont {{Iwamoto}}\ \emph {et~al.}(2004)\citenamefont
  {{Iwamoto}}, \citenamefont {{Kajino}}, \citenamefont {{Mathews}},
  \citenamefont {{Fujimoto}},\ and\ \citenamefont {{Aoki}}}]{iwamoto04}%
  \BibitemOpen
  \bibfield  {author} {\bibinfo {author} {\bibfnamefont {N.}~\bibnamefont
  {{Iwamoto}}}, \bibinfo {author} {\bibfnamefont {T.}~\bibnamefont {{Kajino}}},
  \bibinfo {author} {\bibfnamefont {G.~J.}\ \bibnamefont {{Mathews}}}, \bibinfo
  {author} {\bibfnamefont {M.~Y.}\ \bibnamefont {{Fujimoto}}},\ and\ \bibinfo
  {author} {\bibfnamefont {W.}~\bibnamefont {{Aoki}}},\ }\bibfield  {title}
  {\bibinfo {title} {{Flash-Driven Convective Mixing in Low-Mass,
  Metal-deficient Asymptotic Giant Branch Stars: A New Paradigm for Lithium
  Enrichment and a Possible s-Process}},\ }\href
  {https://doi.org/10.1086/380989} {\bibfield  {journal} {\bibinfo  {journal}
  {\apj}\ }\textbf {\bibinfo {volume} {602}},\ \bibinfo {pages} {377} (\bibinfo
  {year} {2004})}\BibitemShut {NoStop}%
\bibitem [{\citenamefont {{Cristallo}}\ \emph {et~al.}(2009)\citenamefont
  {{Cristallo}}, \citenamefont {{Piersanti}}, \citenamefont {{Straniero}},
  \citenamefont {{Gallino}}, \citenamefont {{Dom{\'\i}nguez}},\ and\
  \citenamefont {{K{\"a}ppeler}}}]{cristallo09b}%
  \BibitemOpen
  \bibfield  {author} {\bibinfo {author} {\bibfnamefont {S.}~\bibnamefont
  {{Cristallo}}}, \bibinfo {author} {\bibfnamefont {L.}~\bibnamefont
  {{Piersanti}}}, \bibinfo {author} {\bibfnamefont {O.}~\bibnamefont
  {{Straniero}}}, \bibinfo {author} {\bibfnamefont {R.}~\bibnamefont
  {{Gallino}}}, \bibinfo {author} {\bibfnamefont {I.}~\bibnamefont
  {{Dom{\'\i}nguez}}},\ and\ \bibinfo {author} {\bibfnamefont {F.}~\bibnamefont
  {{K{\"a}ppeler}}},\ }\bibfield  {title} {\bibinfo {title}
  {{Asymptotic-Giant-Branch Models at Very Low Metallicity}},\ }\href
  {https://doi.org/10.1071/AS09003} {\bibfield  {journal} {\bibinfo  {journal}
  {\pasa}\ }\textbf {\bibinfo {volume} {26}},\ \bibinfo {pages} {139} (\bibinfo
  {year} {2009})},\ \Eprint {https://arxiv.org/abs/0904.4173} {arXiv:0904.4173
  [astro-ph.SR]} \BibitemShut {NoStop}%
\bibitem [{\citenamefont {{Suda}}\ and\ \citenamefont
  {{Fujimoto}}(2010)}]{suda10}%
  \BibitemOpen
  \bibfield  {author} {\bibinfo {author} {\bibfnamefont {T.}~\bibnamefont
  {{Suda}}}\ and\ \bibinfo {author} {\bibfnamefont {M.~Y.}\ \bibnamefont
  {{Fujimoto}}},\ }\bibfield  {title} {\bibinfo {title} {{Evolution of low- and
  intermediate-mass stars with [Fe/H] $\leq$ - 2.5}},\ }\href
  {https://doi.org/10.1111/j.1365-2966.2010.16473.x} {\bibfield  {journal}
  {\bibinfo  {journal} {\mnras}\ }\textbf {\bibinfo {volume} {405}},\ \bibinfo
  {pages} {177} (\bibinfo {year} {2010})},\ \Eprint
  {https://arxiv.org/abs/1002.0863} {arXiv:1002.0863 [astro-ph.GA]}
  \BibitemShut {NoStop}%
\bibitem [{\citenamefont {{Stancliffe}}\ \emph {et~al.}(2011)\citenamefont
  {{Stancliffe}}, \citenamefont {{Dearborn}}, \citenamefont {{Lattanzio}},
  \citenamefont {{Heap}},\ and\ \citenamefont {{Campbell}}}]{stancliffe11}%
  \BibitemOpen
  \bibfield  {author} {\bibinfo {author} {\bibfnamefont {R.~J.}\ \bibnamefont
  {{Stancliffe}}}, \bibinfo {author} {\bibfnamefont {D.~S.~P.}\ \bibnamefont
  {{Dearborn}}}, \bibinfo {author} {\bibfnamefont {J.~C.}\ \bibnamefont
  {{Lattanzio}}}, \bibinfo {author} {\bibfnamefont {S.~A.}\ \bibnamefont
  {{Heap}}},\ and\ \bibinfo {author} {\bibfnamefont {S.~W.}\ \bibnamefont
  {{Campbell}}},\ }\bibfield  {title} {\bibinfo {title} {{Three-dimensional
  Hydrodynamical Simulations of a Proton Ingestion Episode in a Low-metallicity
  Asymptotic Giant Branch Star}},\ }\href
  {https://doi.org/10.1088/0004-637X/742/2/121} {\bibfield  {journal} {\bibinfo
   {journal} {\apj}\ }\textbf {\bibinfo {volume} {742}},\ \bibinfo {eid} {121}
  (\bibinfo {year} {2011})},\ \Eprint {https://arxiv.org/abs/1109.1289}
  {arXiv:1109.1289 [astro-ph.SR]} \BibitemShut {NoStop}%
\bibitem [{\citenamefont {{Choplin}}\ \emph {et~al.}(2021)\citenamefont
  {{Choplin}}, \citenamefont {{Siess}},\ and\ \citenamefont
  {{Goriely}}}]{choplin21}%
  \BibitemOpen
  \bibfield  {author} {\bibinfo {author} {\bibfnamefont {A.}~\bibnamefont
  {{Choplin}}}, \bibinfo {author} {\bibfnamefont {L.}~\bibnamefont {{Siess}}},\
  and\ \bibinfo {author} {\bibfnamefont {S.}~\bibnamefont {{Goriely}}},\
  }\bibfield  {title} {\bibinfo {title} {{The intermediate neutron capture
  process. I. Development of the i-process in low-metallicity low-mass AGB
  stars}},\ }\href {https://doi.org/10.1051/0004-6361/202040170} {\bibfield
  {journal} {\bibinfo  {journal} {\aap}\ }\textbf {\bibinfo {volume} {648}},\
  \bibinfo {eid} {A119} (\bibinfo {year} {2021})},\ \Eprint
  {https://arxiv.org/abs/2102.08840} {arXiv:2102.08840 [astro-ph.SR]}
  \BibitemShut {NoStop}%
\bibitem [{\citenamefont {Siess}\ \emph {et~al.}(2000)\citenamefont {Siess},
  \citenamefont {Dufour},\ and\ \citenamefont {Forestini}}]{siess00}%
  \BibitemOpen
  \bibfield  {author} {\bibinfo {author} {\bibfnamefont {L.}~\bibnamefont
  {Siess}}, \bibinfo {author} {\bibfnamefont {E.}~\bibnamefont {Dufour}},\ and\
  \bibinfo {author} {\bibfnamefont {M.}~\bibnamefont {Forestini}},\ }\bibfield
  {title} {\bibinfo {title} {An internet server for pre-main sequence tracks of
  low- and intermediate-mass stars},\ }\href@noop {} {\bibfield  {journal}
  {\bibinfo  {journal} {Astronomy \& Astrophysics}\ }\textbf {\bibinfo {volume}
  {358}},\ \bibinfo {pages} {593} (\bibinfo {year} {2000})}\BibitemShut
  {NoStop}%
\bibitem [{\citenamefont {{Siess}}(2006)}]{siess06}%
  \BibitemOpen
  \bibfield  {author} {\bibinfo {author} {\bibfnamefont {L.}~\bibnamefont
  {{Siess}}},\ }\bibfield  {title} {\bibinfo {title} {{Evolution of massive AGB
  stars. I. Carbon burning phase}},\ }\href
  {https://doi.org/10.1051/0004-6361:20053043} {\bibfield  {journal} {\bibinfo
  {journal} {\aap}\ }\textbf {\bibinfo {volume} {448}},\ \bibinfo {pages} {717}
  (\bibinfo {year} {2006})}\BibitemShut {NoStop}%
\bibitem [{\citenamefont {Goriely}\ and\ \citenamefont
  {Siess}(2018)}]{goriely18}%
  \BibitemOpen
  \bibfield  {author} {\bibinfo {author} {\bibfnamefont {S.}~\bibnamefont
  {Goriely}}\ and\ \bibinfo {author} {\bibfnamefont {L.}~\bibnamefont
  {Siess}},\ }\bibfield  {title} {\bibinfo {title} {Sensitivity of the
  s-process nucleosynthesis in agb stars to the overshoot model},\ }\href@noop
  {} {\bibfield  {journal} {\bibinfo  {journal} {Astronomy \& Astrophysics}\
  }\textbf {\bibinfo {volume} {609}},\ \bibinfo {pages} {A29} (\bibinfo {year}
  {2018})}\BibitemShut {NoStop}%
\bibitem [{\citenamefont {Xu}\ \emph {et~al.}(2013)\citenamefont {Xu},
  \citenamefont {Goriely}, \citenamefont {Jorissen}, \citenamefont {Chen},\
  and\ \citenamefont {Arnould}}]{xu13}%
  \BibitemOpen
  \bibfield  {author} {\bibinfo {author} {\bibfnamefont {Y.}~\bibnamefont
  {Xu}}, \bibinfo {author} {\bibfnamefont {S.}~\bibnamefont {Goriely}},
  \bibinfo {author} {\bibfnamefont {A.}~\bibnamefont {Jorissen}}, \bibinfo
  {author} {\bibfnamefont {G.}~\bibnamefont {Chen}},\ and\ \bibinfo {author}
  {\bibfnamefont {M.}~\bibnamefont {Arnould}},\ }\bibfield  {title} {\bibinfo
  {title} {Databases and tools for nuclear astrophysics applications: Brussels
  nuclear library (bruslib), nuclear astrophysics compilation of reactions ii
  (nacre ii) and nuclear network generator (netgen)},\ }\href
  {https://dipot.ulb.ac.be/dspace/bitstream/2013/138819/1/aa20537-12.pdf}
  {\bibfield  {journal} {\bibinfo  {journal} {Astronomy \& Astrophysics}\
  }\textbf {\bibinfo {volume} {549}},\ \bibinfo {pages} {10} (\bibinfo {year}
  {2013})},\ \Eprint {https://arxiv.org/abs/A106} {A106} \BibitemShut {NoStop}%
\bibitem [{\citenamefont {{Goriely}}\ \emph {et~al.}(2021)\citenamefont
  {{Goriely}}, \citenamefont {{Siess}},\ and\ \citenamefont
  {{Choplin}}}]{goriely21}%
  \BibitemOpen
  \bibfield  {author} {\bibinfo {author} {\bibfnamefont {S.}~\bibnamefont
  {{Goriely}}}, \bibinfo {author} {\bibfnamefont {L.}~\bibnamefont {{Siess}}},\
  and\ \bibinfo {author} {\bibfnamefont {A.}~\bibnamefont {{Choplin}}},\
  }\bibfield  {title} {\bibinfo {title} {{The intermediate neutron capture
  process. II. Nuclear uncertainties}},\ }\href
  {https://doi.org/10.1051/0004-6361/202141575} {\bibfield  {journal} {\bibinfo
   {journal} {\aap}\ }\textbf {\bibinfo {volume} {654}},\ \bibinfo {eid} {A129}
  (\bibinfo {year} {2021})},\ \Eprint {https://arxiv.org/abs/2109.00332}
  {arXiv:2109.00332 [astro-ph.SR]} \BibitemShut {NoStop}%
\bibitem [{\citenamefont {{Choplin}}\ \emph {et~al.}(2022)\citenamefont
  {{Choplin}}, \citenamefont {{Siess}},\ and\ \citenamefont
  {{Goriely}}}]{choplin22}%
  \BibitemOpen
  \bibfield  {author} {\bibinfo {author} {\bibfnamefont {A.}~\bibnamefont
  {{Choplin}}}, \bibinfo {author} {\bibfnamefont {L.}~\bibnamefont {{Siess}}},\
  and\ \bibinfo {author} {\bibfnamefont {S.}~\bibnamefont {{Goriely}}},\
  }\bibfield  {title} {\bibinfo {title} {{The intermediate neutron capture
  process. III. The i-process in AGB stars of different masses and
  metallicities without overshoot}},\ }\href
  {https://doi.org/10.1051/0004-6361/202244360} {\bibfield  {journal} {\bibinfo
   {journal} {\aap}\ }\textbf {\bibinfo {volume} {667}},\ \bibinfo {eid} {A155}
  (\bibinfo {year} {2022})},\ \Eprint {https://arxiv.org/abs/2209.10303}
  {arXiv:2209.10303 [astro-ph.SR]} \BibitemShut {NoStop}%
\bibitem [{\citenamefont {{Denissenkov}}\ \emph {et~al.}(2017)\citenamefont
  {{Denissenkov}}, \citenamefont {{Herwig}}, \citenamefont {{Battino}},
  \citenamefont {{Ritter}}, \citenamefont {{Pignatari}}, \citenamefont
  {{Jones}},\ and\ \citenamefont {{Paxton}}}]{denissenkov17}%
  \BibitemOpen
  \bibfield  {author} {\bibinfo {author} {\bibfnamefont {P.~A.}\ \bibnamefont
  {{Denissenkov}}}, \bibinfo {author} {\bibfnamefont {F.}~\bibnamefont
  {{Herwig}}}, \bibinfo {author} {\bibfnamefont {U.}~\bibnamefont {{Battino}}},
  \bibinfo {author} {\bibfnamefont {C.}~\bibnamefont {{Ritter}}}, \bibinfo
  {author} {\bibfnamefont {M.}~\bibnamefont {{Pignatari}}}, \bibinfo {author}
  {\bibfnamefont {S.}~\bibnamefont {{Jones}}},\ and\ \bibinfo {author}
  {\bibfnamefont {B.}~\bibnamefont {{Paxton}}},\ }\bibfield  {title} {\bibinfo
  {title} {{I-process Nucleosynthesis and Mass Retention Efficiency in He-shell
  Flash Evolution of Rapidly Accreting White Dwarfs}},\ }\href
  {https://doi.org/10.3847/2041-8213/834/2/L10} {\bibfield  {journal} {\bibinfo
   {journal} {\apjl}\ }\textbf {\bibinfo {volume} {834}},\ \bibinfo {eid} {L10}
  (\bibinfo {year} {2017})},\ \Eprint {https://arxiv.org/abs/1610.08541}
  {arXiv:1610.08541 [astro-ph.SR]} \BibitemShut {NoStop}%
\end{thebibliography}%
\end{document}